\documentclass[a4paper,11pt]{article}
\pdfoutput=1  

\usepackage{jheppub}  

\usepackage[T1]{fontenc} 

\usepackage{lipsum}  

\usepackage{multirow}
\usepackage{amsmath,mathrsfs}
\usepackage{amsfonts,amssymb}
\usepackage{accents}

\usepackage{xcolor}

\usepackage{comment}

\usepackage{enumitem}

\usepackage{mwe,tikz}
\usepackage[percent]{overpic}
\usepackage{tikz-cd}

\usepackage{array,ragged2e}

\usepackage{bm}

\usepackage[normalem]{ulem}
\usepackage{caption}
\usepackage{subcaption}

\newcommand{\be}{\begin{equation}}
\newcommand{\ee}{\end{equation}}

\newcommand{\ba}{\begin{aligned}}
\newcommand{\ea}{\end{aligned}}

\newcommand{\nn}{\nonumber}

\newcommand{\cB}{\mathcal{B}}
\newcommand{\cC}{\mathcal{C}}

\newcommand{\cI}{\mathcal{I}}

\newcommand{\cN}{\mathcal{N}}
\newcommand{\cO}{\mathcal{O}}

\newcommand{\cS}{\mathcal{S}}
\newcommand{\cT}{\mathcal{T}}
\newcommand{\cU}{\mathcal{U}}

\newcommand{\cW}{\mathcal{W}}

\newcommand{\AdS}{\mathbf{AdS}}
\newcommand{\sphere}{\mathbf{S}}
\newcommand{\RP}{\mathbf{RP}}

\title{
\boldmath  SymTFTs and Non-Invertible Symmetries \\of 
6d (2,0) SCFTs of Type $D$ from M-theory}

\author[a]{Federico Bonetti,}
\author[b,c,d]{Michele Del Zotto,}
\author[e]{Ruben Minasian}

\affiliation[a]{Departamento de Electromagnetismo y Electr\'onica, Universidad de Murcia, Campus de Espinardo, 30100 Murcia, Spain}
\affiliation[b]{Department of Mathematics, Uppsala University, Box 480, SE-75106, Uppsala, Sweden}
\affiliation[c]{Department of Physics and Astronomy, Uppsala University, Box 516, SE-75120, Uppsala, Sweden}
\affiliation[d]{Center for Geometry and Physics, Uppsala University, Box 480, SE-75106, Uppsala, Sweden}
\affiliation[e]{Institut de Physique Th\'{e}orique, Universit\'{e} Paris Saclay, CNRS, CEA, F-91191, Gif-sur-Yvette, France} 

\emailAdd{f.bonetti@um.es, michele.delzotto@math.uu.se, ruben.minasian@ipht.fr}

\abstract{We  revisit 6d (2,0) SCFTs of type $D_N$
and their  realization in M-theory, focusing on 
absolute  variants of these theories and on their global finite 0- and 2-form symmetries. 
We  derive
the 7d SymTFT capturing these global symmetries from M-theory, both from the point of view of the low-energy supergravity action on $\AdS_7\times \RP^4$
and from M2- and M5-branes giving rise to its topological operators.  
Along the way, results by Gukov, Hsin, and Pei are extended by keeping track of an additional 7d  $\mathbb Z_2$  gauge field, associated to the outer automorphism
of the $D_N$ algebra. In particular, we find an interplay of non-invertible symmetries and mixed anomalies for absolute 6d (2,0) $D_{4k}$ SCFTs with $k\ge 1$.
We  highlight several subtle points related to the non-orientability of $\RP^4$, the half-integral $G_4$-flux that threads it,
and the non-commutativity of fluxes. All these also play an essential role in a holographic derivation of the anomaly polynomial of 6d (2,0) $D_N$ SCFTs.
}

\begin{document} 
\maketitle
\flushbottom

\section{Introduction}

An increasing amount of evidence indicates that the symmetries of a quantum field theory (QFT) are in correspondence with its \textit{topological defects/operators} \cite{Gaiotto:2014kfa}, organized by symmetry categories (see e.g.~\cite{Cordova:2022ruw,McGreevy:2022oyu,Gomes:2023ahz,Schafer-Nameki:2023jdn,Brennan:2023mmt,Bhardwaj:2023kri,Shao:2023gho,Carqueville:2023jhb,Costa:2024wks}
for recent reviews).
This idea has a holographic counterpart, in terms of the existence of topological versions of membranes emerging along the boundary of the holographic dual theory 
\cite{Benini:2022hzx,
Damia:2022bcd,
Apruzzi:2022rei,
GarciaEtxebarria:2022vzq,
Heckman:2022muc,
Heckman:2022xgu,
Antinucci:2022vyk,
Etheredge:2023ler,
Cvetic:2023plv,
Bah:2023ymy,
Apruzzi:2023uma,
Heckman:2024oot,
Heckman:2024obe,
Gutperle:2024vyp,
Bergman:2024aly,
Antinucci:2024bcm,
Cvetic:2024dzu,
Waddleton:2024iiv}, consistent with the analysis of symmetries in holography proposed by Harlow and Ooguri \cite{Harlow:2018tng}. The main purpose of this paper is to revisit the holographic origin of symmetries in the context of 6d (2,0) SCFTs with a holographic dual, with special emphasis on the case of 6d (2,0) SCFTs of type $D$, especially subtle because obtained from a compactification of M-theory on the non-orientable background $\AdS_7 \times \RP^4$ with $G_4$ flux on $\RP^4$.

A convenient way to capture the structure of topological defects/operators of a given $d$-dimensional QFT is to realize it in terms of a bulk-boundary system in one dimension higher exploiting a topological symmetry theory (or SymTFT) \cite{Kapustin:2014gua,Ji:2019jhk,Gaiotto:2020iye,Apruzzi:2021nmk,Freed:2022qnc}. The bulk-boundary system in question consists of a $(d+1)$-dimensional topological field theory---the SymTFT---placed on a finite length slab consisting of an interval times a $d$-dimensional spacetime, completed by the two choices of a $d$-dimensional topological boundary condition $\mathcal B$ on one side of the interval, and of a $d$-dimensional relative field theory $\widehat{\mathcal T}$ on the other side. Since the bulk is topological the system is equivalent to a $d$-dimensional absolute field theory $\mathcal T$. Many examples and applications can be obtained by realizing $d$-dimensional QFTs with known Lagrangian formulations as bulk-boundary systems of this kind.

An interesting application of the above approach is in the context of  
generalized symmetries of quantum fields that do not have a conventional Lagrangian formulation, of which higher dimensional SCFTs are prime examples. Higher dimensional SCFTs can be constructed exploiting stringy techniques, as the worldvolume theories of solitonic membranes in 
string/M-theory, or via geometric engineering. These techniques have been recently extended in various ways to capture the relevant topological symmetry theories, thus determining the generalized symmetries of these quantum fields as well    (see e.g.~\cite{DelZotto:2015isa,
GarciaEtxebarria:2019caf,
Gukov:2020btk,
Albertini:2020mdx,
DelZotto:2020esg,
Closset:2020afy,
Apruzzi:2020zot,
Bah:2020uev,
DelZotto:2020sop,
Apruzzi:2021vcu,
Apruzzi:2021mlh,
Apruzzi:2022dlm,
DelZotto:2022fnw,
DelZotto:2022ras,
Acharya:2023bth,
Chen:2023qnv,
Bashmakov:2023kwo,
Baume:2023kkf,
DelZotto:2024tae,
GarciaEtxebarria:2024fuk,
Braeger:2024jcj,
GarciaEtxebarria:2024jfv,
Tian:2024dgl,
Najjar:2024vmm} for further references). For theories realized as membrane stacks, large $N$ holographic duals are also typically known. This offers a different pathway to the SymTFT, obtained directly from holographic dictionaries, that allows to capture further information about topological symmetries.

The six-dimensional (2,0) SCFTs follow an ADE-type classification arising from their geometric engineering in Type IIB superstrings \cite{Witten:1995zh,Seiberg:1996qx,Strominger:1995ac}. These are among the SCFTs with highest possible dimension according to Nahm theorem and maximal amount of supercharges compatible with a conserved stress energy tensor multiplet
\cite{Nahm:1977tg,Cordova:2016emh}. While the exceptional models of types $E_6$, $E_7$ and $E_8$ do not have a (known) realization as worldvolume theories of fivebranes, the models of types $A_{N-1}$ and $D_N$ do. The models of $A_{N-1}$ type are constructed as the world-volume theories of stacks of $N$ M5-branes in M-theory \cite{Strominger:1995ac,Witten:1995zh}, and therefore have a large $N$ holographic dual $\AdS_7\times \sphere^4$ with $N$ units of $G_4$ flux along the $\sphere^4$, that is obtained from the near-horizon limit of the brane configuration. Similarly, the models of $D_{N}$ type are constructed as the world-volume theories of stacks of $2N$ M5-branes 
in presence of an OM5-plane
(here, we count branes in the ``upstairs'' picture 
with respect to the $\mathbb Z_2$ involution 
that defines the OM5-plane)
\cite{Dasgupta:1995zm,Witten:1995em,Hori:1998iv,Ahn:1998qe,Gimon:1998be,Hanany:2000fq}. Therefore, these models have a large $N$ holographic dual $\AdS_7\times \RP^4$ with $N-\tfrac 12$ units of $G_4$ flux along the $\RP^4$. 
In this paper we are interested in exploiting these correspondences to deepen our understanding of some of 6d (2,0) SCFTs. We are in particular interested in reproducing the field theoretical results obtained about these models from screening \cite{DelZotto:2015isa,GarciaEtxebarria:2019caf,Gukov:2020btk,DelZotto:2022ras} from a holographic perspective. While these results are well-known for 6d (2,0) models of type $A_{N-1}$,\footnote{ \  With some caveats, depending on the value of $N$ that we will discuss.} the corresponding results in type $D_N$ require more care, as the holographic dual for these theories is obtained from M-theory on a background $\AdS_7 \times \RP^4$ with $G_4$ flux on $\RP^4$, which is a non-orientable manifold.\footnote{ 
 \ In particular, $\RP^4$ does not have a spin structure, but rather a Pin$^+$ structure. 
In Appendix \ref{app:killing} we review some properties of Killing pinors of
$\RP^4$.}
The techniques we will develop along the way
allow us to achieve a two-fold goal.
On   one hand, we derive the 
SymTFT for the discrete 0- and 2-form symmetry sector of 6d (2,0) theories of type $D_N$. On the other hand, we 
provide a purely holographic derivation of the anomaly polynomials of these  theories, which (in contrast to the case of type $A$ SCFTs) was not available in the literature.

The SymTFT governing discrete symmetries of 
6d (2,0) $D_N$ theories,
as derived from M-theory, takes the form 
\be \label{eq_preview}
\exp  2\pi i \int_{M_7} \bigg[ 
 \frac 12 \mathsf c_3 \cup \delta \mathsf b_3
- \frac N8 \mathsf c_3 \cup \delta \mathsf c_3
+ \frac 12 \mathsf a_1 \cup \delta \mathsf a_5
+  \frac 14 \mathsf a_1 \cup  \mathsf  c_3  \cup \mathsf c_3 
\bigg] \ , 
\ee 
as discussed in Section \ref{sec_this_is_F7}, to which we refer the reader for
an explanation of our notation.
We confirm 
the analysis of Hsin, Gukov and Pei 
\cite{Gukov:2020btk}
and extend it by keeping track of the 0-form symmetry associated to the $\mathbb Z_2$ outer automorphism of
the $D_N$ algebra (encoded in the fields $\mathsf a_1$, $\mathsf a_5$ in \eqref{eq_preview}).
We find that the topological defects constructed with the fields in \eqref{eq_preview}
support TFTs on their worldvolumes that renders their fusion algebra non-invertible.

Furthermore, our analysis provides
a resolution to the following puzzle.
On one hand, 4d $\cN = 4$ super Yang-Mills theory with gauge group $SO(4N)$ 
has a global $\mathbb Z_2$ 0-form symmetry
(stemming from the outer automorphism of the $D_N$ Dynkin diagram) as well as a global $\mathbb Z_2\times \mathbb Z_2$ center 1-form symmetry.
These symmetries participate in a cubic 't Hooft anomaly, captured by a 5d anomaly action 
of the form 
$\int_{\rm 5d} \frac 12 \mathsf a_1 \cup \mathsf b_2 \cup \mathsf c_2$,
where $\mathsf a_1$, $\mathsf b_2$, $\mathsf c_2$
are background gauge fields for the 0-form and 1-form symmetries \cite{Bhardwaj:2022yxj}.
On the other hand,
4d $\cN = 4$ $SO(4N)$ super Yang-Mills theory   can be obtained by toroidal
reduction of the 6d (2,0) $D_{2N}$ SCFT. 
From this point of view, it is natural to ask if  we can identify the 6d origin
of the mixed 't Hooft anomaly in 4d.
The answer is affirmative. Indeed, the topological action 
\eqref{eq_preview}, reduced on $T^2$,  reproduces the results of \cite{Bergman:2022otk,Etheredge:2023ler}  on the global structures of 4d $\mathcal N=4$ $\mathfrak{so}(4N)$
super Yang-Mills theories, including the aforementioned mixed 't Hooft anomaly,
which originates from the cubic term in \eqref{eq_preview}. This argument highlights the 6d origin for the non-invertible symmetries of this class of models as well.\footnote{ \ This is a first step towards extending our results to 4d $\mathcal N=2$ class $\mathcal S[D_{n}]$ theories building on \cite{Tachikawa:2013hya,Bashmakov:2022jtl,Bashmakov:2022uek,Antinucci:2023ezl}.}

We also apply the 7d topological action \eqref{eq_preview}
to the analysis of   non-invertible symmetries in 6d (2,0) SCFTs of type $D$.
Non-invertible symmetries in 6d have been studied in \cite{Lawrie:2023tdz,Apruzzi:2024cty}. Here, we provide a complementary point of view, 
which applies to 6d theories of type $D_{N=4k}$ for any $k\ge 1$ and hinges on the cubic coupling in \eqref{eq_preview}, building on   
\cite{Tachikawa:2017gyf,Kaidi:2021xfk}.

Finally, we revisit the  gravitational anomaly polynomials for $D$-type 6d SCFTs, originally derived by Intriligator 
\cite{Intriligator:2000eq} 
and Yi \cite{Yi:2001bz}.
In particular, we reproduce 
their results by a direct integration 
of the topological couplings of M-theory on
$\RP^4$. A crucial role is played by a suitable subtraction prescription (see \eqref{eq:subtraction}, \eqref{eq:Msubtraction}) which is necessary to ensure that the M-theory partition function is well-defined, as first pointed out in \cite{Witten:1996hc,Witten:1999vg} and further in \cite{Freed:2019sco}
in the non-orientable setting.

\medskip

The structure of this paper is as follows.
In Section \ref{sec:Atype} we revisit the analysis of the 6d (2,0) SCFTs of type $A$ and their bulk-boundary system.
In Section \ref{sec:results} 
we turn our attention to
6d (2,0) SCFTs of type $D$
and we present their 7d SymTFTs.
Section 
\ref{sec:Mderivation} is devoted to the M-theory
derivation of these SymTFTs.
In Section \ref{sec:application} we apply the SymTFTs to the study of global symmetry structures of absolute variants of 6d (2,0) $D_N$ SCFTs, including non-invertible symmetries.
In Section \ref{sec:DtypeANOMAL}
we revisit the analysis of the anomaly polynomial for these models.
In Section \ref{sec:conclusion} we present our conclusions from this study and outline potential future directions. More technical appendices complement the discussion of the results in the main body of the text.

\section{Warm-up: holography of 6d (2,0) $A_{N-1}$ SCFTs}\label{sec:Atype}

The low energy behavior of a stack of $N$ M-theory fivebranes is captured by an interacting six-dimensional supersymmetric conformal field theory (SCFT) \cite{Strominger:1995ac,Witten:1995zh}. Upon decoupling the center of mass mode, one obtains in the deep IR the irreducible 6d (2,0) SCFT of type $A_{N-1}$. This model is a 6d relative field theory \cite{Freed:2012bs}, meaning that it is a boundary of a 7d bulk theory, that we denote $\mathscr{T}_{\mathrm{7d}}^{A_{N-1}}$ --- see \cite{Monnier:2017klz} for a (conjectural) complete construction. In particular, this implies that a 6d (2,0) $A_{N-1}$ theory assigns to a compact, closed, torsionless, spin manifold, $M^6$, an element of the Hilbert space $\mathscr{T}_{\mathrm{7d}}^{A_{N-1}}(M^6)$. In other words, the 6d (2,0) theory often has a partition vector instead of a partition function \cite{Tachikawa:2013hya}. This fact has a well-known holographic counterpart \cite{Aharony:1998qu}, that we will review in more detail below. Indeed, the holographic dual of a stack of M-theory fivebranes is an $\AdS_7 \times \sphere^4$ background, with $N$ units of $G_4$ flux on the $\sphere^4$,\footnote{ \ Throughout this work, we use a normalization convention for $p$-form fields in which field strengths are quantized in integer units (possibly shifted by 1/2) without $2\pi$ factors. We write the action $S$ in Lorentzian signature. It is defined mod $2\pi \mathbb Z$, so that the path integral weight $e^{iS}$ is well-defined.}
\be
\int_{\sphere^4} G_4 = N\,.
\ee
This implies that on top of the 7d $\mathcal N=2$ supergravity arising from the expansion of M-theory in $\sphere^4$ harmonics, the holographic dual also contains a topological sector of the form \cite{Witten:1998wy,Moore:2004jv} (see also \cite{Gukov:2020btk})
\be\label{eq:7dCS}
2\pi \frac 16 \int_{\AdS_7\times \sphere^4} C_3 \wedge G_4 \wedge  G_4 = 2\pi \frac N2
\int_{\AdS_7 } C_3 \wedge dC_3 \,.
\ee
In other words, the 7d holographic dual of the 6d (2,0) theory contains a (topological) subsector consisting of a dynamical $\mathbb Z_N$ three-form gauge field with a Chern-Simons like coupling. The theory with Lagrangian \eqref{eq:7dCS} has Wilson membranes supported on 3-surfaces~$\Sigma_3$,
\be \label{eq:cWilson}
\mathcal W_q(\Sigma_3) = \text{exp}\, 2\pi\,  i \, q \int_{\Sigma_3} C_3\,.
\ee
These Wilson membranes are topological because $dC_3 = 0$ thanks to the equation of motion for $C_3$ obtained varying the action \eqref{eq:7dCS}. Moreover, it is not hard to show that they satisfy the following identities in correlators
(see e.g.~\cite{Maldacena:2001ss,Banks:2010zn,Kapustin:2014gua}):   
\be
\begin{aligned}
\langle \mathcal W_N(\Sigma_3) \cdots \rangle &= \langle \cdots \rangle \,, \\
\langle\mathcal W_{q_1}(\Sigma_3^{(1)}) \mathcal W_{q_2}(\Sigma_3^{(2)})  \cdots \rangle &= \exp\left( i {2 \pi \over N} \, (q_1 q_2 ) \, \text{Link}(\Sigma_3^{(1)},\Sigma_3^{(2)})\right)\langle \cdots \rangle  \,.
\label{eq:corrA}
\end{aligned}
\ee
In particular, these topological Wilson membranes, can be freely pushed onto the $\AdS_7$ boundary implying that the boundary conditions of the model are in a representation of a Heisenberg algebra
\be\label{eq:Heisenberg}
\mathcal W_{q_1}(\Sigma_3^{(1)}) \mathcal W_{q_2}(\Sigma_3^{(2)}) = \exp\left( i {2 \pi \over N} \, (q_1 q_2 ) \, (\Sigma_3^{(1)} \cdot \Sigma_3^{(2)})\right)\mathcal W_{q_2}(\Sigma_3^{(2)}) \mathcal W_{q_1}(\Sigma_3^{(1)}) \,,
\ee
where $\Sigma_3^{(1)} \cdot \Sigma_3^{(2)}$ is the intersection of the 3-cycles $\Sigma_3^{(1)}$ and $\Sigma_3^{(2)}$ along $\partial \AdS_7$. Following the holographic correspondence, these boundary conditions are the holographic duals for the background fields for the  2-form symmetry $\mathbb Z_N^{(2)}$ of the 6d (2,0) $A_{N-1}$ SCFT, which is captured by its defect group \cite{DelZotto:2015isa}. The non-commutative collection of Wilson surfaces provides the holographic counterpart of the fact that the 6d (2,0) $A_{N-1}$ theory 
has a partition vector rather than a partition function.  
Indeed, radial quantization along the $\AdS_7$ radius produces a Hilbert space along the $\AdS_7$ boundary. Such a Hilbert space is the holographic dual to the Hilbert space of $\mathscr{T}_{\mathrm{7d}}^{A_{N-1}}$, hosting the holographic dual of the partition vector the SCFT.  Because of the 7d CS in \eqref{eq:7dCS} the resulting Hilbert space is not one-dimensional for generic values of 
$N$, which gives the desired result.

\subsection{TQFT correlators from branes at infinity}

The 
M-theory origin of
the
topological operators
of the 7d TQFT
lie in   branes
``wrapping cycles at infinity'' --- see e.g.~\cite{Apruzzi:2022rei,GarciaEtxebarria:2022vzq,Heckman:2022muc,Heckman:2022xgu,Etheredge:2023ler,Bah:2023ymy,Apruzzi:2023uma,Bergman:2024aly,Waddleton:2024iiv}. More precisely,
the branes  wrap cycles
in the compact horizon
geometry ($\sphere^4$ in this case)
while sitting at a fixed
value $r_*$ of the holographic
radial coordinate $r$ in
external spacetime.
We consider the limit in
which $r_*$ goes to infinity,
approaching the holographic
boundary:
this has the effect of freezing out non-topological
modes on the branes, leaving behind only their topological
worldvolume couplings. In the simple setting under consideration, the brane origin of the operators 
 \eqref{eq:cWilson}
is clear: they are   M2-branes sitting at a point (0-cycle) on $\sphere^4$
and extended along   $\Sigma_3$ in external spacetime. 

As a warm-up for the following sections, let us rederive the correlator
\eqref{eq:corrA} from the membrane perspective. Our strategy is as follows. We consider inserting a first M2-brane
on $\Sigma_3$ and determine
its effect on the bulk
3-form of M-theory. Then
we use this information to
detect the extra phase picked
up by the worldvolume
theory on a second M2-brane inserted 
on $\Sigma_3'$. Upon inserting
the first M2-brane on $\Sigma_3 \times {\rm pt}$
(here pt denotes
a point on $\sphere^4$),
the Bianchi identity and equation of motion for the M-theory 3-form
read
\be 
dG_4 = 0 \ , \qquad 
dG_7 = \frac 12 G_4^2
+ X_8 + \delta_8(\Sigma_3 
\times {\rm pt}\subset M_{11}) \  .
\ee 
Here $G_4$ is the 11d 4-form
field strength,
$G_7$ is its Hodge dual,\footnote{\ More precisely,
in our conventions the 
bosonic terms in the
11d low-energy effective action read
\be 
S = 2\pi \int_{M_{11}} 
\bigg[ 
(2\pi \ell_p)^{9} R*1
- \frac 12 (2\pi \ell_p)^6 G_4*G_4
- \frac 16 C_3 G_4 G_4
- C_3 X_8
\bigg] \ , 
\ee 
where $\ell_p$ is the Planck length. We set   $G_7 = - (2\pi \ell_p)^6 *G_4$.
} 
and $X_8$ is the 8-form \cite{Duff:1995wd}
\be \label{eq:X8def}
X_8 = \frac{1}{192} \bigg[
p_1(TM_{11})^2 - 4 p_2(TM_{11})
\bigg] \ ,
\ee 
where $p_i$ are Pontryagin
forms,  polynomials in the curvature
of the 11d metric.
The quantity 
$\delta_8(\Sigma_3 
\times {\rm pt}\subset M_{11})$ is
an 8-form current
(see e.g.~\cite{Cheung:1997az})
describing the insertion
of the M2-brane.
We can write
\be 
\delta_8(\Sigma_3 
\times {\rm pt}\subset M_{11})
= \delta_4(\Sigma_3 \subset M_7)
\wedge v_4 \ , 
\ee 
where $v_4$ denotes the volume form on $\sphere^4$,
normalized to integrate to 1.
The de Rham class of
$v_4$ is the image
in $H^4(\sphere^4;\mathbb R)$ of the generator of
$H^4(\sphere^4;\mathbb Z)$,
the Poincar\'e dual
of ${\rm pt}\in H_0(\sphere^4;\mathbb Z)$. We consider the following ansatz for $G_4$,
\be 
G_4 = N v_4 + g_4 \ , 
\ee 
where $g_4$ is a closed 4-form
with integral periods in external
spacetime, the field strength of the 3-form potential $c_3$. 
This ansatz captures the light modes of $G_4$ associated to cohomology classes on $\sphere^4$.
Upon integrating the
equation of motion
for $G_7$ on $\sphere^4$ we obtain\footnote{\ Here $\textstyle \int_{\sphere^4} G_7$ denotes a 3-form,  the lowest mode in the KK tower of $G_7$ on $\sphere^4$.}
\be \label{eq:g4relation}
g_4 = - \tfrac 1N  \delta_4(\Sigma_3) + \tfrac 1N d \textstyle \int_{\sphere^4} G_7 \ . 
\ee 
This relation demonstrates that
the insertion of the M2-brane 
on $\Sigma_3$ constrains the 
external field strength $g_4$,
which must contain a source
term $- \tfrac 1N  \delta_4(\Sigma_3)$.
Consider now inserting the second M2-brane,
located along a 3-cycle
$\Sigma_3'$ in the external spacetime and sitting at a point on $\sphere^4$.
The effect of the first M2-brane
insertion along $\Sigma_3$
on this second brane is encoded in this source term. The second M2-brane wrapped along $\Sigma_3'$ contributes to the topological defect via 
the topological terms in its
worldvolume action, which indeed gives the holonomy
\be
\textbf{M2}(\Sigma_3') \approx e^{2\pi i \int_{\Sigma_3'} C_3} \cdots
\ee
 Assuming $\Sigma_3'  = \partial B_4'$, we can write
\be 
e^{2\pi i \int_{\Sigma_3'} C_3}
= e^{2\pi i \int_{B_4'} G_4}
= e^{2\pi i \int_{B_4'} g_4}\ . 
\ee 
Since the term $\tfrac 1N d\int_{\sphere^4} G_7$ in \eqref{eq:g4relation} is subleading approaching the holographic boundary,\footnote{\ This follows because by definition $G_7$ is the Hodge star of $G_4$ which contains an inverse metric. Due to the form of the AdS metric the Hodge star along the AdS radius goes as $1/r_*^2$ which in the limit $r_* \to \infty$ vanishes.}  $g_4$ approaches  $ - \tfrac 1N  \delta_4(\Sigma_3)$ in the limit, from which follows that
\be
\langle \textbf{M2}(\Sigma_3')\textbf{M2}(\Sigma_3) \cdots \rangle \approx e^{2\pi i \frac{-1}{N} \int_{B_4'} \delta_4(\Sigma_3)}  \langle \, \dots \rangle
\ee
Making use of 
\be 
\int_{B_4'} \delta_4(\Sigma_3)
= {\rm Link}(\Sigma_3 , \Sigma_3')  \  , 
\ee 
we see that this phase factor 
corresponds
precisely to the one we were after in equation \eqref{eq:corrA}, up to an immaterial sign related to orientation conventions.

\subsection{Comments on Page charges} 
In this section we relate the above discussion to the non-commutative nature of Page charges in M-theory \cite{Moore:2004jv}. The construction of 
Page charges is a bit involved and requires some notation which we now review. The main idea behind this approach is that one can realize (generalized) electric charges for the M-theory $G_4$ field as characters of its \textit{magnetic translation group}. Page charges are
constructed 
in Hamiltonian quantization
on a spacetime of the form
$M_{11} = \mathbb R_t \times X_{10}$.
They can be written 
in terms of the 7-form
\be  \label{eq:P7_def}
P_7 = \frac{1}{2\pi}
\Pi_7
+ \frac 12 
G_4^\bullet \wedge C_3
+ \frac 16 C_3 \wedge dC_3
+ T_7^\bullet \ . 
\ee 
Some comments are in order. Topologically non-trivial $G_4$ configurations can be described by
selecting a topologically non-trivial basepoint, and adding arbitrary
topologically trivial
configurations.
The basepoint
has field strength
$G_4^\bullet \in \Omega^4(X_{10})$,
while the topologically
trivial deviations from
it are
encoded in the \emph{globally defined}
3-form $C_3 \in \Omega^3(X_{10})$.
The 7-form
$T_7^\bullet$ 
is such that
\be \label{eq:T7_def}
dT_7^\bullet = \frac 12 G_4^\bullet \wedge G_4^\bullet + X_8 \ , 
\ee 
where here $X_8$
is constructed with the same
Pontryagin forms of equation \eqref{eq:X8def}, just restricted to the spatial slice
$X_{10}$.
The existence of $T_7^\bullet$
follows from the fact
that 
$\frac 12 G_4^\bullet \wedge G_4^\bullet + X_8$ is trivial
in de Rham cohomology,
which is itself a consequence
of the $G_4$ equation of motion (in the absence of brane sources). 
$T_7^\bullet$ is non-unique;
it is understood
that a definite choice is made in writing $P_7$.
Finally, $\Pi_7$ denotes
the canonical momentum
conjugate to $C_3$.
It is related to $G_7$, the 11d Hodge dual
of $G_4$, as follows.
Let us introduce the quantity 
\be 
\frac{1}{2\pi} \widetilde \Pi_7
= \frac{1}{2\pi} \Pi_7
- \frac 12 G_4^\bullet \wedge C_3
- \frac 13 C_3 \wedge dC_3 \ . 
\ee 
Once can verify that the classical
Gauss law constraint that
follows from varying the action with respect to the temporal component
of the 11d 3-form potential takes the form
\be \label{eq:classical_Gauss}
\frac{1}{2\pi} d\widetilde \Pi_7 
+ \frac 12 (G_4^\bullet + dC_3) \wedge (G_4^\bullet + dC_3) + X_8 = 0 \ .
\ee 
This is nothing but the spatial component of the 11d equation of motion for $G_4$.
We are thus led to the identification
\be 
\frac{1}{2\pi} \widetilde \Pi_7 = - G_7 \ ,
\ee 
where on the RHS we are implicitly restricting $G_7$ to the spatial
slice $X_{10}$. Using \eqref{eq:P7_def}, \eqref{eq:T7_def}, \eqref{eq:classical_Gauss}
it is easy to verify that
$P_7$ is closed. 
Heuristically,  ``$P_7  = dC_6$'', where $C_6$ 
is the electro-magnetic
dual to the 3-form
potential.

\medskip

\noindent Using $P_7$,
one defines the quantities
\be  \label{eq:Page_charges}
\cW(\phi_3) = \exp 2\pi i \int_{X_{10}} \phi_3 \wedge P_7 \ , 
\ee 
where $\phi_3$ is a 
globally-defined
closed 3-form on $X_{10}$,
$\phi_3 \in \Omega^3(X_{10})$. 
Closure of $P_7$
implies that 
$\cW(\phi_3)$
depends on $\phi_3$
only via its de Rham
cohomology class.
The analysis of
\cite{Moore:2004jv} demonstrates
that the 3-form
$\phi_3$ must be chosen
in such a way that\footnote{ \ Here $G_4 = G_4^\bullet + dC_3$ is the total field strength of basepoint plus deviation. We could equally well drop the exact piece $dC_3$
by closure of $\phi_3$, $\omega_3$.}
\be  \label{eq:phi3_condition}
\int_{X_{10}} \phi_3 \wedge \omega_3 \wedge G_4 \in \mathbb Z \ , \qquad
\text{for any $\omega_3 \in \Omega^3_\mathbb Z(X_{10})$} \ . 
\ee 
The symbol 
$\Omega^p_\mathbb Z$
stands for closed
$p$-forms with integral
periods.
The operators
\eqref{eq:Page_charges}
with $\phi_3$
satisfying \eqref{eq:phi3_condition}
generate the so-called
magnetic translation
group; the characters of this group are, by definition, the (topological classes of) electric charges.
Crucially, the magnetic
translation group
needs not be Abelian.  Rather,
we have\footnote{ \ This follows from the canonical commutation relation
\be 
[C_{\mu_1 \dots \mu_3}(x) , \Pi_{\nu_1 \dots \nu_7}(y)] = i \delta^{(10)}(x-y) \epsilon_{\mu_1 \dots \mu_3 \nu_1 \dots \nu_7} \ , 
\ee 
where $\mu_i$, $\nu_j$ are curved
indices on $X_{10}$
and $\epsilon_{\mu_1 \dots \mu_3 \nu_1 \dots \nu_7}$ is a density
taking values in $\{ 0, \pm 1\}$.
}
\be  \label{eq:Page_comm}
\cW(\phi_3)\cW(\phi_3')
= \cW(\phi_3') \cW(\phi_3) \exp \bigg[-2\pi i \int_{X_{10}} 
\phi_3 \wedge \phi_3' \wedge G_4
\bigg] \ .
\ee 

The general analysis
reviewed above readily 
specializes to a 10d spatial slice of the form $M_{10} = M_6 \times \sphere^4$.
Here we are considering
holography in Euclidean signature and,
following 
\cite{Witten:1998wy, Belov:2004ht}, we identify the holographic radial coordinate 
with the time coordinate in Hamiltonian quantization.
It follows that $M_6$ is
identified with the
asymptotic boundary,
i.e.~the 6d spacetime
where the dual 6d SCFT
lives.
The basepoint
flux is identified with the background $G_4$-flux of $N$ units on $\sphere^4$. 

Since $\sphere^4$ does not contribute any non-trivial 3-form
de Rham class,
both $\phi_3$ and 
$\omega_3$ in \eqref{eq:phi3_condition} 
are forms in the external spacetime $M_6$. Integrating
\eqref{eq:phi3_condition} over $\sphere^4$ and using the fact that we have $N$ units of $G_4$ flux,
we find that the
3-form $\phi_3$
must be of the form
\be   \label{eq:phi3_example}
\phi_3 = \frac 1N \Phi_3 \ , \qquad 
\Phi_3  \in \Omega^3_\mathbb Z(M_6) \ . 
\ee 
By Poincar\'e duality
in $M_{6}$
(assumed closed oriented), the closed 3-form
with integral periods
$\Phi_3$ determines
a (nontorsional) 
3-cycle $\Sigma_3$ in $M_6$,  
\be \label{eq:phi3_examplePD}
\Sigma_3 = {\rm PD}_{M_6}[\Phi_3] \ . 
\ee 
Plugging \eqref{eq:phi3_example}
into \eqref{eq:Page_comm}
yields
the phase
\be 
\exp \bigg[ - 2\pi i N \cdot \frac 1N \cdot \frac 1N \int_{X_{10}}
\Phi_3 \wedge \Phi_3'
\bigg]
=\exp \bigg[ - 2\pi i  \frac 1N 
\Sigma_3 \cdot_{M_6} \Sigma_3'
\bigg]  \ ,
\ee 
with the factor $N$ in the numerator
coming from the integral
of $G_4$ on $\sphere^4$.
Here $\cdot_{M_6}$
denotes the intersection number of
cycles in $M_6$.
This phase in the commutation relation of two
$\cW$ operators
in the Hamiltonian formalism
is equivalent to a
non-trivial correlator
of the form
\eqref{eq:corrA}
in a 7d covariant formulation.
We have thus rederived
\eqref{eq:corrA} from yet another perspective,
using non-commutativity of Page charges.\footnote{\ We stress that this phenomenon is distinct
from non-commutativity
of electric/magnetic fluxes associated to torsional cycles.
Torsional phenomena will play an important role in our discussion of 6d (2,0) SCFTs
of type $D_N$.}

Finally,
let us connect the M2-brane perspective
and the Page charge perspective.
To this end, 
we go back to 
\eqref{eq:P7_def}.
We observe that
for $X_{10} = M_6 \times \sphere^4$
we can take $T_7^\bullet = 0$.
If we write the Page charge
in terms of the quantity
$\widetilde \Pi_7$, we have
\begin{align}
\cW(\phi_3)
& = \exp 2\pi i \int_{M_6 \times \sphere^4}
\phi_3 \bigg[  
\frac{1}{2\pi}
\widetilde \Pi_7
+  
G_4^\bullet  C_3
+ \frac 12 C_3  dC_3
\bigg]  \ . 
\end{align}
Since $\phi_3$ 
is purely external,
we need to saturate the $\sphere^4$ integral
using the quantity in bracket.
Moreover, 
as discussed above,
$\Pi_7$ is identified
with the $X_{10}$ component of $G_7$, the 11d Hodge dual of $G_4$.
As a result, the $\widetilde \Pi_7$
term is subleading
in the limit in which
the Hamiltonian slice
approaches the 
holographic boundary.
We are thus led to an expression of the form
\begin{align}
\cW(\phi_3)
& \approx 
e^{2\pi i \int_{M_6}  N \phi_3 C_3}
= e^{2\pi i   \int_{\Sigma_3} C_3}
\ ,
\end{align}
where we have used
\eqref{eq:phi3_example} and \eqref{eq:phi3_examplePD}.
This is identified with the topological
terms on an M2-brane
on $\Sigma_3$.

\section{SymTFT of 6d (2,0) SCFTs of type D}\label{sec:results}

The six-dimensional SCFTs of $A_{N-1}$ type have a family of close cousins which are also realized in M-theory, the 6d (2,0) theories of $D_N$ type. These 6d SCFT arise as the low energy limit for a stack of $N$ pairs of coincident M-theory fivebranes branes on top of an OM5-plane, the M-theory version of a $\mathbb Z_2$ orientifold \cite{Dasgupta:1995zm,Witten:1995em,Hori:1998iv,Ahn:1998qe,Gimon:1998be,Hanany:2000fq}. It is well-known that also the 6d (2,0) theories of $D_N$ type happen to be defined relative to a 7d theory,
which we denote $\mathscr T_{\mathrm{7d}}^{D_N}$.
In the $D_N$ case, however, the 7d theory admits topological gapped boundary conditions corresponding to the existence of
\emph{absolute} versions of these
6d (2,0) SCFT,
as classified in \cite{Gukov:2020btk}.\footnote{\ Here ``absolute'' is used as the opposite of ``relative'' to emphasize that of all 6d (2,0) $D_N$ SCFTs, there are always some which have a partition function, as opposed to a partition vector.} Our main objective in this paper 
is to perform 
a thorough analysis of these TQFTs
$\mathscr T_{\mathrm{7d}}^{D_N}$ and of their
topological operators,
based on the M-theory realization of the
6d (2,0) SCFT of type $D_N$.

In this section we present the result of our analysis. We give the action of the 7d TQFT $\mathscr T_{\mathrm{7d}}^{D_N}$
and we exhibit some of its non-trival
topological operators.
The derivation of these findings is presented in Section \ref{sec:Mderivation} below.

\subsection{The 7d TQFT $\mathscr T_{\mathrm{7d}}^{D_N}$}\label{sec_this_is_F7}

The  TQFT $ \mathscr T_{\mathrm{7d}}^{D_N}$ 
is described by the exponentiated action
\be \label{eq:7d_action_summary}
\mathcal A_{D_N}[M_7] \, \exp  2\pi i \int_{M_7} \bigg[ 
 \frac 12 \mathsf c_3 \cup \delta \mathsf b_3
- \frac N8 \mathsf c_3 \cup \delta \mathsf c_3
+ \frac 12 \mathsf a_1 \cup \delta \mathsf a_5
+  \frac 14 \mathsf a_1 \cup  \mathsf  c_3  \cup \mathsf c_3 
\bigg] \ .
\ee
Some comments are in order. The factor $\mathcal A_{D_N}[M_7]$
describes an invertible 7d field theory,
which encodes the gravitational
and $\mathfrak{so}(5)_R$ R-symmetry
anomalies of the 6d (2,0) $D_N$ SCFT
and has been studied extensively
\cite{Monnier:2013kna,Monnier:2014txa, Monnier:2016jlo,Monnier:2017klz}. (Strictly speaking,
$\mathcal A_{D_N}[M_7]$ is an invertible but not  topological field theory.)
In Section \ref{sec:DtypeANOMAL}, we provide a derivation of the 8-form anomaly polynomial associated to 
$\mathcal A_{D_N}[M_7]$
using an M-theory inflow argument
based on the horizon geometry
$\RP^4$.

In the rest of this section
we set aside $\mathcal A_{D_N}[M_7]$
and focus on the other factor in 
\eqref{eq:7d_action_summary}.
The fields $\mathsf c_3$, $\mathsf b_3$,
$\mathsf a_1$, $\mathsf a_5$
are $\mathbb Z_2$-cochains of
degrees indicated by the subscripts.
The fields $\mathsf c_3$, $\mathsf b_3$
and the quadratic couplings
$\mathsf c_3 \cup \delta \mathsf b_3$,
$\mathsf c_3 \cup \delta \mathsf c_3$ furnish a discrete formulation
of the Abelian 3-form Chern-Simons theory
\be \label{eq:Abelian_3form_CS}
\exp 2\pi i \int_{M_7} \bigg[ 
2 c_3 \wedge db_3 - \frac N2 c_3 \wedge dc_3
\bigg] \,,
\ee 
which has already appeared in \cite{Gukov:2020btk}. 
(The field $c_3$ in the continuum
formulation is a $U(1)$ 3-form gauge field whose field strength is a closed 4-form with integral periods. It is related to the discrete field $\mathsf c_3$ by the formal replacement $c_3 \mapsto \mathsf c_3/2$. Similar remarks apply to the pair $b_3$,
$\mathsf b_3$.)
The discrete $\mathbb Z_2$ field $\mathsf a_1$ is associated to the
$\mathbb Z_2$ outer automorphism
of the Lie algebra~$D_N$.\footnote{\ This $\mathbb Z_2$ action exchanges the spinor (s) and cospinor (c) representations of $D_N=\mathfrak{so}(2N)$, leaving the vector  representation  (v) invariant. If $N=0$ mod 4, (s), (c) are real spinor representations,
if $N=2$ mod 4, they are pseudoreal (quaternionic). If $N$ is odd, 
 (s) and (c) are related by complex conjugation and the $\mathbb Z_2$ automorphism action
 amounts to charge conjugation.
 }  
The field $\mathsf a_5$ is the BF partner
of $\mathsf a_1$.

The equations of motions
for the discrete fields
$\mathsf c_3$, $\mathsf b_3$, $\mathsf a_1$, $\mathsf a_5$ read
\be  \label{eq:higher_group_summary}
\delta \mathsf c_3 = 0 \ , \qquad 
\delta \mathsf a_1 = 0 \ , \qquad
\delta \mathsf b_3 = \mathsf a_1 \cup \mathsf c_3   \ , \qquad
\delta \mathsf a_5 = -\frac 12 \mathsf c_3 \cup \mathsf c_3 \ . 
\ee 
We highlight that 
the coupling
$\mathsf a_1 \cup \mathsf c_3 \cup \mathsf c_3$ 
induces non-trivial RHSs 
for $\delta \mathsf b_3$,
$\delta \mathsf a_5$ on-shell.

In the exponentiated action 
\eqref{eq:7d_action_summary} the gravitational and R-symmetry contributions are decoupled
from the discrete sector described by
$\mathsf c_3$, $\mathsf b_3$,
$\mathsf a_1$, $\mathsf a_5$.
We leave the  exploration of  possible additional terms that can mix these two sectors to future work.

\subsubsection*{Remarks on the cubic term $\mathsf a_1 \cup \mathsf c_3 \cup \mathsf c_3$}

The term $\tfrac 14 \mathsf a_1 \cup \mathsf c_3 \cup \mathsf c_3$ in the topological action
\eqref{eq:7d_action_summary}
deserves further comments. 
In fact, it requires a quadratic refinement of the cup product $\mathsf c_3 \cup \mathsf c_3$.
More precisely, we 
propose the following interpretation for 
$\tfrac 14 \mathsf a_1 \cup \mathsf c_3 \cup \mathsf c_3$. We refer the reader to Appendix
\ref{app_refinement} for further details and references.

Let $X_6$ be a (non-necessarily orientable)
compact 6-manifold. A Wu structure on $X_6$ is a trivialization $\rho_3$ of the fourth Wu class $\nu_4(X_6)$ of $X_6$,
$d\rho_3 = \nu_4(X_6)$.\footnote{\ Following \cite{Hsin:2021qiy}, we can interpret this relation as
encoding a higher-group structure 
mixing a $\mathbb Z_2$ 2-form symmetry with background field $\rho_3$ and Lorentz symmetry.}
Such a $\rho_3$ exists because $\nu_4(X_6)$ vanishes in cohomology. (More generally, $\nu_{p+1}(X_n)$ is always zero in cohomology on a manifold $X_n$ of dimension $n\le 2p+1$, see Appendix \ref{app_refinement}.)
Wu structures on $X_6$ 
form a torsor over $H^3(X_6;\mathbb Z_2)$ and
are in 1-to-1 correspondence
with $\mathbb Z_4$-valued refinements
of the cup product 
in $H^3(X_6;\mathbb Z_2)$
 \cite{brown_math, browder_math}.
For a given Wu structure
$\rho_3$, let us denote the associated
quadratic refinement
as $\widetilde q_{\rho_3}: H^3(X_6;\mathbb Z_2) \rightarrow \mathbb Z_4$. It satisfies
\be \label{eq_refinement_key}
\widetilde q_{\rho_3}(\mathsf x_3 + \mathsf y_3)  = \widetilde q_{\rho_3}(\mathsf x_3 )
+ \widetilde q_{\rho_3}( \mathsf y_3)
+ \theta \int_{X_{6}} \mathsf x_3 \cup \mathsf y_3 \mod 4 \ ,
\ee 
where  $\mathsf x_3$,  $\mathsf y_3 \in H^3(X_{6};\mathbb Z_2)$ and $\theta: H^*(X_{6} ; \mathbb Z_2) \rightarrow H^*(X_{6} ; \mathbb Z_4)$
is the cohomology  map induced by the inclusion
$\mathbb Z_2 \hookrightarrow \mathbb Z_4$
that sends (1 mod 2) to (2 mod 4).

After these preliminaries,
let us now consider the 7d spacetime $M_7$  
where the topological action 
\eqref{eq:7d_action_summary} is formulated.
The 
 $\mathbb Z_2$ gauge field on $M_7$ 
 corresponds to a class $\mathsf a_1 \in H^1(M_7;\mathbb Z_2)$.
By applying Poincar\'e duality 
to $\mathsf a_1$ we get an element
$H_6(M_7;\mathbb Z_2)$. Geometrically, 
we think of this homology class as  represented by a submanifold $X_6$ (which may or may not be orientable).\footnote{\ In general, 
it is a non-trivial question whether  
a  homology class of degree $k$
of a manifold $M_n$ 
can be realized as fundamental class of an embedded submanifold $X_k$, or not. The answer to this question
turns out to be affirmative 
for any homology class in
 $H_{n-1}(M_n;\mathbb Z_2)$, for any $n$, for compact $M_n$,  see e.g.~\cite[Theorem II.26]{manturov2007topological}.
}
We observe that $\nu_4$ is automatically trivial in dimension 7. 
Thus, there exists a Wu structure $d\rho_3 = \nu_4$ on $M_7$.
It induces a Wu structure on $X_6$.
We are now in a position to 
propose a prescription to 
make more rigorous the cubic coupling $\mathsf a_1 \mathsf c_3 \mathsf c_3$.
We claim
\be 
\exp 2\pi i \int_{M_7} \frac 14 \mathsf a_1 \cup \mathsf c_3 \cup \mathsf c_3
= \exp 2\pi i \frac 14 \widetilde q_{\rho_3}^{\; X_6 }(\mathsf c_3) \ ,
\qquad 
X_6 = {\rm PD}_{M_7}[\mathsf a_1] \ .
\ee 
On the RHS  
$\widetilde q_{\rho_3}^{\; X_6 }$ is the $\mathbb Z_4$-valued refinement is computed
in the 6d space $X_6$, with the $\rho_3$ induced from $M_7$.

Next, we argue that the space 
$M_7$ should have
$w_2 = 0= w_1$, i.e.~be a spin manifold.
One way to see it is to use the M-theory
holographic picture,
in which 11d spacetime is of the form $M_{11} = M_7 \times \RP^4$.
The total space $M_{11}$ must be a 
Pin$^+$ manifold
(more on this in \S 
\ref{sec:mc_and_RP4} below)
hence $w_2(M_{11}) = 0$.
We observe that
\be 
w_2(M_{11}) = 
w_2(M_7) + w_2(\RP^4)
+ w_1(M_7) \cup w_1(\RP^4) \ , \quad 
w_1(\RP^4) = \alpha_1  \ , \quad 
w_2(\RP^4) = 0  \ ,
\ee 
where $\alpha_1$ denotes the generator of
$H^1(\RP^4;\mathbb Z_2) \cong \mathbb Z_2$.
The vanishing of
$w_2(M_{11})$
implies 
$w_2(M_7) = 0 =w_1(M_7)$.\footnote{\ A possible loophole in the above argument is
to consider a total space that is not a direct product $M_7 \times \RP^4$, but rather a non-trivial
$\RP^4$ fibration over $M_7$,
captured by an $SO(5)$ bundle
with non-zero $w_2(SO(5))$
to cancel $w_2(M_7)$.
See e.g.~\cite{Cordova:2018acb,Brennan:2022tyl,Brennan:2023vsa,Aspman:2022sfj,Moore:2024vsd}
for analogous setups in 4d in which
theories with fermions can be formulated
on non-Spin manifolds.}
We remark that, if $M_7$ is a spin manifold,
the spin  structure induces a Wu structure
$\rho_3$ \cite{miller1987some,Monnier:2016jlo}.\footnote{\ We also refer the reader to   \cite[sec.~4.4]{Hsieh:2020jpj}
for an argument relating spin and Wu structures on 7-manifolds, based on 
 $E_7$ index theory and motivated physically from E-string theories.
 }

\subsection{Some topological operators in $\mathscr T_{\mathrm{7d}}^{D_N}$}\label{sec_this_is_F7_ops}

The topological operators of the theory $\mathscr T_{\mathrm{7d}}^{D_N}$ include operators corresponding to
holonomies of the fields 
$\mathsf c_3$, $\mathsf b_3$,
$\mathsf a_1$, $\mathsf a_5$.
(They can be extracted systematically
by studying the Gauss law constraints of the 7d action \cite{Apruzzi:2022rei,Bah:2023ymy}.)\footnote{\ We refrain from a study of condensation defects and codimension-one topological operators in 
$\mathscr T_{\mathrm{7d}}^{D_N}$.
}
More precisely, we have the operators 
\be \label{eq:top_ops_all}
\ba
\bm Q_3(\Sigma_3) &= 
e^{i \pi \int_{\Sigma_3} \mathsf c_3} 
\ ,  & 
\qquad \qquad 
\widehat{\bm Q}_3(\Sigma_3) &= 
e^{i \pi \int_{\Sigma_3} \mathsf b_3} \otimes
\, T_3(\Sigma_3; \mathsf a_1, \mathsf c_3)
\ , \\
\bm Q_1(\Sigma_1) &= e^{i \pi \int_{\Sigma_1} \mathsf a_1}  \ ,  & 
\widehat{\bm Q}_5(\Sigma_5) &= 
e^{i \pi \int_{\Sigma_5} \mathsf a_5}
\, \otimes T_5(\Sigma_5; \mathsf c_3) \ .
\ea
\ee 
The notation $T_3(\Sigma_3; \mathsf a_1, \mathsf c_3)$ indicates that the
holonomy of $\mathsf b_3$ has to be dressed with a suitable 3d TQFT, coupled to the 7d field $\mathsf a_1$, $\mathsf c_3$; similarly
for $T_5(\Sigma_5; \mathsf c_3)$.
The inclusion of $T_3(\Sigma_3; \mathsf a_1, \mathsf c_3)$,
$T_5(\Sigma_5; \mathsf c_3)$ is required because   the cubic coupling
$\mathsf a_1 \cup \mathsf c_3 \cup \mathsf c_3$ induces a modification of the standard
 gauge transformations of $\mathsf b_3$,
 $\mathsf a_5$,
which in turn renders the naive
holonomies $e^{i \pi \int_{\Sigma_3} \mathsf b_3}$, $e^{i \pi \int_{\Sigma_5} \mathsf a_5}$
not gauge invariant \cite{Kaidi:2023maf}.
Equivalently, the 7d coupling
$\mathsf a_1 \cup \mathsf c_3 \cup \mathsf c_3$
induces an anomaly 
$\mathsf a_1 \cup \mathsf c_3$
on the 3d worldvolume of 
$e^{i \pi \int_{\Sigma_3} \mathsf b_3}$,
and also an anomaly $\mathsf c_3 \cup \mathsf c_3$
on the 5d worldvolume of 
$e^{i \pi \int_{\Sigma_5} \mathsf a_5}$,
and the topological dressings are required
to absorb these anomalies \cite{Kaidi:2021xfk,Bhardwaj:2022yxj}.
More explicitly,
the topological actions
describing the dressing factors $T_3(\Sigma_3; \mathsf a_1, \mathsf c_3)$,
$T_5(\Sigma_5;  \mathsf c_3)$
are as follows,
\be  \label{eq:T3T5dressing_def}
\ba 
T_3(\Sigma_3; \mathsf a_1, \mathsf c_3) \; :  \quad 
&
\exp 2\pi i  
\int_{\Sigma_3} 
\bigg[ 
\frac 12 \phi_0 \cup \delta \phi_2
 + \frac 12 \phi_2 \cup \mathsf a_1 
 - \frac 12 \phi_0 \cup \mathsf c_3
\bigg]  \ ,
\\
T_5(\Sigma_5; \mathsf c_3) \;:
\quad 
&
\exp 2\pi i  
\int_{\Sigma_5} 
\bigg[ 
-\frac 14 \varphi_2 \cup \delta \varphi_2
 + \frac 12 \varphi_2 \cup \mathsf c_3 
\bigg] \ .
\ea 
\ee 
The fields $\phi_0$, $\phi_2$ are
$\mathbb Z_2$ cochains of degrees 0, 2 living on $\Sigma_3$, while $\varphi_2$ is a 
$\mathbb Z_2$ cochain of degree 2 living on $\Sigma_5$. They couple to the (restrictions to $\Sigma_3$, $\Sigma_5$ of) the bulk fields
$\mathsf a_1$, $\mathsf c_3$.
Due to the $T_3$, $T_5$ factors,
the topological operators
$\widehat{\bm Q}_3(\Sigma_3)$,
$\widehat{\bm Q}_5(\Sigma_5)$
are non-invertible.
We discuss further $T_3$, $T_5$
in Section 
\ref{sec:noninvertiblah}, where we also see how these  topological actions
 emerge from M-theory.

\section{Holographic derivation in M-theory }\label{sec:Mderivation}

We now discuss how the 7d TQFT 
$\mathscr T_{\mathrm{7d}}^{D_N}$
can be derived from M-theory.

Firstly, let us recall that the
6d (2,0) SCFT of type $D_N$ 
emerge at long wavelengths
in the worldvolume
theory of a stack of $N$ pairs of coincident M5-branes on top of an
$\mathbb R^5/\mathbb Z_2$ orbifold singularity. Crucially,
the geometric $\mathbb Z_2$ action
by reflection in $\mathbb R^5$ is
accompanied by a 
flip in sign of 
the M-theory 3-form.
The fixed locus of the $\mathbb Z_2$
action is
sometimes referred to as an OM5-plane
\cite{Dasgupta:1995zm,Witten:1995em,Hori:1998iv,Ahn:1998qe,Gimon:1998be,Hanany:2000fq}.
The associated
holographic
setup in the large $N$ limit is
$\AdS_7 \times \RP^4$,
with 
$\RP^4$ obtained from $\sphere^4$ by modding out by antipodal identification.

Our approach to the study
of $ \mathscr T_{\mathrm{7d}}^{D_N}$ is twofold.
On the one hand, the fields and couplings in the  action for $ \mathscr T_{\mathrm{7d}}^{D_N}$ can be derived from the M-theory low-energy effective action reduced
on the horizon geometry $\RP^4$.
On the other hand,
the topological operators
of $ \mathscr T_{\mathrm{7d}}^{D_N}$
can be realized using M2-branes and
M5-branes wrapping cycles in $\RP^4$ and approaching
the holographic boundary of 
external spacetime. According to the paradigm of 
``branes at infinity''
\cite{Apruzzi:2022rei,GarciaEtxebarria:2022vzq,Heckman:2022muc,Heckman:2022xgu,Etheredge:2023ler,Bah:2023ymy,Apruzzi:2023uma,Bergman:2024aly,Waddleton:2024iiv},
sending the branes towards the holographic boundary 
 has the effect of freezing out their non-topological modes, leaving
behind only their topological worldvolume couplings.

The rest of this section is organized as follows.
In \S\ref{sec:mc_and_RP4} we review  several aspects of M-theory on $\RP^4$, which in particular clarifies various properties of the background of interest.
We proceed in \S\ref{sec_Moverview} with a brief overview of the salient points of the M-theory derivation of $\mathscr T_{\mathrm{7d}}^{D_N}$.
After these preliminaries, 
in
\S.\ref{sec_3fCS_from_M} we turn to the explicit derivation of the 3-form Abelian Chern-Simons theory
\eqref{eq:Abelian_3form_CS}, based
of the non-commutativity of 
torsional fluxes \cite{Freed:2006ya,Freed:2006yc,Albertini:2020mdx}.  
In \S\ref{sec_cubic_derivation}
we present the M-theory origin of the cubic coupling in the action for
$\mathscr T_{\mathrm{7d}}^{D_N}$.
In \S\ref{sec:noninvertiblah}
we
show how a class of non-trivial extended  operators of $\mathscr T_{\mathrm{7d}}^{D_N}$
can be derived from M2- and M5-branes
and in \S\ref{sec_HW} we show that they exhibit a version of 
the Hanany-Witten effect \cite{Hanany:1996ie}.

\subsection{Aspects of M-theory on $\RP^4$
} \label{sec:mc_and_RP4}

As we reviewed in the introduction of this manuscript, a holographic realization of the 6d (2,0) of type $D_N$ is obtained from M-theory on $\AdS_7 \times \RP^4$. The four dimensional real projective space is not orientable, and therefore not spin. There are some features of M-theory on non-orientable manifold that will be important in characterizing holographically the 7d theory $\mathscr T_{\mathrm{7d}}^{D_N}$

\subsubsection{M-theory on
non-orientable spaces
and $m_{\rm c}$ structures}
\label{sec:mc_structure}

M-theory can be consistently
formulated on 
spacetimes that are not necessarily orientable. The presence of the gravitino requires
a suitable  tangential structure, analogous to
a spin structure in the orientable case: the correct choice is a 
${\rm Pin}^+$ structure.\footnote{\ The 
obstruction for existence of a ${\rm Pin}^+$ structure on
a manifold $M$ is the Stiefel-Whitney  class $w_2(TM)$. If $M$ does admit ${\rm Pin}^+$ structures, then the set of isomorphism classes of ${\rm Pin}^+$ structures on $M$ is acted upon freely and transitively by $H^1(M; \mathbb Z_2)$.
}
The 3-form
potential of 11d supergravity requires
additional structure,
on top of the 
${\rm Pin}^+$ structure,
dubbed an $m_{\rm c}$
structure \cite{Witten:2016cio,Freed:2019sco}.
By definition,
an $m_{\rm c}$ structure
on a ${\rm Pin}^+$ manifold $M$
is a choice of 
\emph{twisted} integer
lift of $w_4(TM) \in H^4(M;\mathbb Z_2)$, the fourth
Stiefel-Whitney class of $M$.
A twisted
integer cohomology class
is an element of
$H^\bullet(M;\widetilde{\mathbb Z})$ 
where $\widetilde{\mathbb Z}$
denotes the constant sheaf
of integers on $M$ twisted by the 
orientation bundle
of $M$.
In what follows we will also encounter twisted real
cohomology classes, \emph{i.e.}~elements in
$H^\bullet(M;\widetilde{\mathbb R})$, and twisted
differential forms
$\widetilde \Omega^\bullet(M)$;
in all cases ``twisted'' refers to twisting by the orientation bundle of $M$.
A necessary and sufficient
condition
for a ${\rm Pin}^+$ manifold $M$ to admit
an $m_{\rm c}$ structure
is $\tilde \beta w_4(TM) = 0$,
where $\tilde \beta$ is
the Bockstein homomorphism
$H^4(M;\mathbb Z_2) \rightarrow  H^5(M; \tilde {\mathbb Z})$
associated to the short exact sequence
$0 \rightarrow \widetilde{\mathbb Z} \xrightarrow{2} \widetilde{\mathbb Z} \rightarrow \mathbb Z_2 \rightarrow 0$.

The $m_{\rm c}$ structure
on 11d spacetime
$M_{11}$
enters the formulation
of the quantization condition
for the 4-form flux.
The field strength $G_4$
is a closed twisted 4-form
on $M_{11}$; we use the 
notation $G_4 \in \widetilde {\Omega}^4_d(M_{11})$.
The de Rham class of $G_4$
is an element 
$[G_4]_{\rm dR} \in H^4(M_{11} ; \widetilde{\mathbb R})$.
Now, there exists 
an \emph{integral} twisted
class   $a_4 \in H^4(M_{11} ; \widetilde{\mathbb Z})$ such that 
\be 
2[G_4]_{\rm dR} = \widetilde \varrho(a_4) \ , 
\ee 
where 
$\widetilde \varrho:
H^\bullet(M_{11};\widetilde {\mathbb Z}) \rightarrow H^\bullet(M_{11};\widetilde {\mathbb R})$
is the natural map induced by
$\widetilde{\mathbb Z }
\rightarrow \widetilde {\mathbb R}$.
The shifted quantization
condition on $G_4$
then states that
the class $a_4$ is a
twisted integral lift of
$w_4(TM_{11})$,
\be 
a_4 = w_4(TM_{11}) \mod 2 \ .
\ee 
The class $a_4$ encodes the
$m_{\rm c}$ structure on $M_{11}$.
Suppose $\mathcal C_4$
is a 4-cycle in $M_{11}$.
Since $G_4$ is a twisted 4-form,
in order to define 
$\int_{\mathcal C_4} G_4$
we require the normal
bundle to $\mathcal C_4$
inside $M_{11}$
to be orientable
and oriented. 
We then have
\be
\int_{\mathcal C_4} G_4 = 
\int_{\mathcal C_4}   
 \frac 12 a_4
\stackrel{\text{mod $\mathbb Z$}}{=}
\frac 12 \int_{\mathcal C_4} w_4(TM_{11}) \ .
\ee

Let us briefly 
describe how the general discussion
above specializes to the more
familiar case of orientable
$M_{11}$ \cite{Witten:1996md}.
An orientable manifold $M_{11}$
with a ${\rm Pin}^+$ structure
is a  manifold with a Spin structure.
Twisted cohomology classes 
and forms on $M_{11}$ reduce to untwisted ones.
On a Spin manifold $M_{11}$
there exists an integral class
$\lambda(TM_{11}) \in H^4(M_{11};\mathbb Z)$
that satisfies
$2\lambda(TM_{11}) = p_1(TM_{11})$,
the first Pontrygin class
of $M_{11}$. The class $\lambda$
is a canonical integral
lift of $w_4(TM_{11})$;
thus, every Spin manifold $M_{11}$
has a canonical
$m_{\rm c}$ structure.
The quantization condition
can be phrased as
\be
\int_{\cC_4} G_4 \stackrel{\text{mod $\mathbb Z$}}{=}
\frac 12 \int_{\cC_4} \lambda(TM_{11}) \ .
\ee

\paragraph{Differential cohomology description.}
For later purposes, it is convenient to describe the 3-form gauge field of 11d supergravity using differential cohomology. 
Reviews on differential cohomology aimed at physicists include 
\cite{Freed:2000ta,Hopkins:2002rd,Freed:2006ya,Freed:2006yc,Freed:2006mx,Hsieh:2020jpj}.

The (gauge equivalence class of the) \emph{difference} between
two  3-form gauge field configurations can be described
in terms of a twisted analog
of an ordinary differential cohomology group.
Indeed, the shifted flux quantization condition drops away in taking a difference, leaving us with a standard flux
quantization condition.  
We denote the relevant twisted
differential cohomology group
of degree $p$ on a manifold $M$ as $\breve H^p(M;\widetilde{\mathbb Z})$.
The case relevant for the 3-form gauge field is $p=4$, but we keep $p$ generic in this presentation. 
The group $\breve H^p(M;\widetilde{\mathbb Z})$
can be characterized by the 
commutative diagram
\be 
\label{eq:diff_coho_hexagon}
\small
\begin{tikzcd}
 & H^{p-1}(M;\widetilde{\mathbb R/\mathbb Z}) \ar[rr, "- \widetilde \beta"] \ar[dr, hook, "i"]&   & H^p(M;\widetilde{\mathbb Z}) \ar[dr, "\widetilde \varrho"] &  
\\
H^{p-1}(M;\widetilde{\mathbb R}) \ar[ur] \ar[dr] &   & 
\breve H^p(M;\widetilde{\mathbb Z}) \ar[ur, two heads, "I"] \ar[dr, two heads, "R"] &   & H^p(M;\widetilde{\mathbb R}) 
\\
  & \displaystyle\frac{\widetilde \Omega^{p-1}(M)}{\widetilde \Omega^{p-1}_{\mathbb Z}(M)} \ar[rr] \ar[ur,hook, "\tau"] &   & \widetilde \Omega^p_{\mathbb Z}(M) \ar[ur] &  
\end{tikzcd}
\ee
The two diagonals are exact at the center.
The tilde on various objects
are a reminder that we are considering cohomology
groups and differential forms
that are twisted by the orientation bundle of $M$.
The subscript $\mathbb Z$
on $\widetilde \Omega^p$
indicates integral periods.
We refer to $I$ as the characteristic class map,
and to $R$ as the field strength map. An element $\breve x \in \breve H^p(M;\widetilde{\mathbb Z})$
with $I(\breve a)=0$ is called 
topologically trivial;
an element with $R(\breve a)=0$
is called flat. The map
$\widetilde \beta$
is the Bockstein homomorphism
associated to the short exact
sequence  $0 \rightarrow \widetilde{\mathbb Z} 
\rightarrow \widetilde {\mathbb R} \rightarrow \widetilde {\mathbb R / \mathbb Z} \rightarrow 0$ of  twisted 
coefficients.

Modeling the difference of
3-form gauge field configurations in M-theory
is sufficient to perform
the torsional analog of a Kaluza-Klein expansion,
along the lines of \cite{Apruzzi:2021nmk}.
We refer the reader to Appendix
\ref{app:diff_cochains} for further comments
on how to model the
3-form gauge field itself
using differential cochains.

\subsubsection{The case $M_{11}= M_7 \times \RP^4$}
\label{sec:RPcoho}

We are interested in studying M-theory on a spacetime of the form
$M_{11}= M_7 \times \RP^4$. We assume that $M_7$ is orientable and Spin.
In what follows, we need
some properties of
$\RP^4$.
The relevant cohomology groups are
\begin{align} \label{eq:RP4groups}
H^\bullet(\RP^4;\mathbb Z) &= \{ \mathbb Z, 0, \mathbb Z_2, 0, \mathbb Z_2 \} \ , \nn \\
H^\bullet(\RP^4;\widetilde{\mathbb Z}) &= \{ 0, \mathbb Z_2, 0, \mathbb Z_2, \mathbb Z \} \ , \nn \\ 
H^\bullet(\RP^4; \mathbb Z_2) &= \{ \mathbb Z_2, \mathbb Z_2, \mathbb Z_2, \mathbb Z_2, \mathbb Z_2 \} \ .
\end{align}
Let $t_1$ denote a generator
of $H^1(\RP^4;\widetilde{\mathbb Z})$. We introduce the quantities
\be 
t_2 = t_1 \cup t_1 \equiv  t_1^2 \ , \qquad 
t_3 = t_1 \cup t_1 \cup t_1 \equiv t_1^3 \ , \qquad 
t_4 = t_1 \cup t_1 \cup t_1 \cup t_1 \equiv t_1^4  \ ,
\ee 
which can be taken as generators of $H^2(\RP^4;{\mathbb Z})$,
$H^3(\RP^4;{\mathbb Z})$,
$H^4(\RP^4;\mathbb Z)$,
respectively.
We use $v_4$ to denote
a generator of 
$H^4(\RP^4;\widetilde{\mathbb Z})$,
which integrates to one
on $\RP^4$,
\be  \label{eq:v4normalization}
\int_{\RP^4} v_4 = 1 \ .
\ee 
The integral homology groups
can be obtained by Poincar\'e duality,
which exchanges
twisted and untwisted coefficients,
\be  \label{eq:RP4Poincare}
H_p(\RP^4 ; \mathbb Z) \cong 
H^{4-p}(\RP^4 ; \widetilde{ \mathbb Z}) \ , \quad 
H_p(\RP^4 ; \widetilde{\mathbb Z}) \cong 
H^{4-p}(\RP^4 ; { \mathbb Z}) \ , \quad 
p=0,\dots,4 \ .
\ee 
The total Stiefel-Whitney class of  
$\RP^4$
reads
\be 
w(T \RP^4) = (1 + \alpha_1)^5 = 1 + \alpha_1 + \alpha_1^4  \ , 
\ee 
where $\alpha_1 \in H^1(\RP^4 ; \mathbb Z_2) \cong \mathbb Z_2$ is a generator. Thus
$\RP^4$ is 
non-orientable and
the obstruction 
$w_2(T \RP^4)$ to a ${\rm Pin}^+$
structure vanishes.
Since $H^1(\RP^4 ; \mathbb Z_2) \cong \mathbb Z_2$,
$\RP^4$ admits two (isomorphism
classes of) ${\rm Pin}^+$
structures.\footnote{\ The two Pin$^+$ structures 
are \emph{complementary}
in the sense that 
one can be reached from the other by tensoring with orientation bundle of $\RP^4$. In terms of the
oriented double cover $\sphere^4$,
a section $\psi$ of the spinor
bundle on $\sphere^4$ descends
to a section of a Pin$^+$ bundle
on $\RP^4$
if it satisfies $(\sigma \psi) = \pm \psi$, where $\sigma$ is the antipodal involution acting
on spinors (see Appendix \ref{app:killing}), and the two signs
correspond to the two Pin$^+$ structures on~$\RP^4$~\cite{Witten:2015aba}.
}
We also have
a non-zero $w_4(T\RP^4)$.
We fix an $m_{\rm c}$
structure on $\RP^4$ by 
choosing $v_4$ as 
twisted integral lift of  $w_4(T\RP^4)$,
\be \label{eq:v4_is_lift}
v_4 = w_4(T \RP^4) \mod 2  \ . 
\ee

\paragraph{Background flux.}
The background 4-form is purely internal,
\emph{i.e.}~proportional
to $v_4$. 
Using \eqref{eq:v4_is_lift},
the quantization condition
states that the proportionality constant between $G_4$ and $v_4$ must be a half integer.
We write
\be  \label{eq:flux_bk}
\text{background:} \qquad G_4 = \left( N - \frac 12 \right) v_4 \ . 
\ee 
In the probe brane
picture,
we start from flat 11d space $\mathbb R^{1,5} \times \mathbb R^5$, we place $N$
pairs of M5-branes
at the origin of $\mathbb R^5$,
and we mod out by the
$\mathbb Z_2$ reflection
in $\mathbb R^5$.
Since the fixed point
of the $\mathbb Z_2$ action carries $-1$ 
units of M5-brane charge \cite{Dasgupta:1995zm,Witten:1995em,Witten:1996md}, the total
$G_4$ flux linking
the origin of 
$\mathbb R^5$
in the ``upstairs'' picture
is $2N-1$,
which leads to \eqref{eq:flux_bk}
after taking the quotient.

\subsubsection{Comments on supersymmetry}

As we have seen above, 
M-theory can be consistently formulated
on the internal space $\RP^4$, because
$\RP^4$ admits
an $m_{\rm c}$ structure
(hence in particular a ${\rm Pin}^+$ 
structure).
The holographic dual
of the 6d (2,0) SCFT of type $D_N$
is $\AdS_7 \times \RP^4$,
with $N-1/2$ units of $G_4$ flux
on $\RP^4$.
Since the SCFT has maximal supersymmetry, the same must hold
for the supergravity solution.
Since $\AdS_7 \times \RP^4$ is not Spin, however,
it cannot admit Killing spinors.
Rather, it must admit what we might refer to as Killing
${\rm Pin}^+$-pinors.
A general study of such objects
in 11d supergravity lies beyond the
scope of this work, see however \cite{Lazaroiu:2016vov,Lazaroiu:2016nbq}.
For the case at hand,
we may exploit the fact that
$\RP^4$ is realized
from $\sphere^4$ via antipodal
identification.
In Appendix \ref{app:killing}
we study the behavior of the Killing
spinors on $\sphere^4$ under this map,
following \cite{Witten:2015aba,Metlitski:2015yqa,Wang:2020jgh}.
In particular, we verify that all of them
transform in the correct way
to survive the quotient as well-defined sections of the 
${\rm Pin}^+$ bundle on $\RP^4$.
In external spacetime we still have
standard Killing spinors in $\AdS_7$,
corresponding as usual to the $Q$
and $S$ supercharges of the 6d SCFT.

\subsection{Overview of the M-theory analysis}\label{sec_Moverview}

\begin{table}
\centering
\begingroup
\begin{tabular}{| c || m{41mm} |}
\hline
 \rule[-2.5mm]{0mm}{7.5mm}Field & 
\Centering Supergravity origin \\ \hline \hline 
\rule[-4mm]{0mm}{11mm}$\mathsf c_3$ & $G_4$ on $t_1 = {\rm PD}_{\RP^4}[\RP^3]$
\\ \hline 
\rule[-4mm]{0mm}{11mm}$\mathsf b_3$ & $G_7$  on $t_4 = {\rm PD}_{\RP^4}[\widetilde{\rm pt}]$
\\ \hline 
\rule[-4mm]{0mm}{11mm}$\mathsf a_1$ & $G_4$ on $t_3 = {\rm PD}_{\RP^4}[\RP^1]$ 
\\ \hline 
\rule[-4mm]{0mm}{11mm}$\mathsf a_5$ & $G_7$ on $t_2 = {\rm PD}_{\RP^4}[\RP^2]$ 
\\ \hline 
\end{tabular}
\begin{tabular}{| m{53mm} || m{27mm} |}
\hline
\Centering\rule[-2.5mm]{0mm}{7.5mm}Topological operator & 
\Centering \rule[-2.5mm]{0mm}{7.5mm}Brane origin \\ \hline \hline 
\rule[-4mm]{0mm}{11mm}$\bm Q_3(\Sigma_3)
= e^{i \pi  \int_{\Sigma_3} \mathsf c_3}$ & \rule[-4mm]{0mm}{11mm}M2 on $\Sigma_3 \times \widetilde{{\rm pt}}$
\\ \hline 
\rule[-4mm]{0mm}{11mm}$\widehat {\bm  Q}_3(\Sigma_3)
= e^{i \pi  \int_{\Sigma_3} \mathsf b_3} T_3(\Sigma_3;
\mathsf a_1 , \mathsf c_3)$ & \rule[-4mm]{0mm}{11mm}M5  on $\Sigma_3 \times \RP^3$
\\ \hline 
\rule[-4mm]{0mm}{11mm}$\bm Q_1(\Sigma_1)
= e^{i \pi  \int_{\Sigma_1} \mathsf a_1}$ & \rule[-4mm]{0mm}{11mm}M2 on $\Sigma_1 \times \RP^2$ 
\\ \hline 
\rule[-4mm]{0mm}{11mm}$\widehat {\bm  Q}_5(\Sigma_5)
= e^{i \pi  \int_{\Sigma_5} \mathsf a_5}T_5(\Sigma_5;
\mathsf c_3)$ & \rule[-4mm]{0mm}{11mm}M5 on $\Sigma_5 \times \RP^1$ 
\\ \hline 
\end{tabular}
\endgroup
\caption{On the left,
discrete $\mathbb Z_2$ fields
in  the  TQFT $\mathscr T_{\mathrm{7d}}^{D_N}$ and their origins in terms of 
expansion of $G_4$ and its Hodge dual
$G_7$ onto torsional cohomology classes on $\RP^4$.
On the right, some topological operators in the  TQFT $\mathscr T_{\mathrm{7d}}^{D_N}$ and their origin from
branes in M-theory on $\RP^4$.
The notation $T_p(\Sigma_p)$
denotes a dressing by a $p$-dimensional
TQFT localized on the support
$\Sigma_p$ of the operator, as described in Section \ref{sec_this_is_F7_ops}.
The (co)homology of $\RP^4$ with integral coefficients (twisted and untwisted) is reviewed in Section 
\ref{sec:RPcoho}.
\label{table:summary}}
\end{table}

In the remaining of this section
we demonstrate how the TQFT 
$\mathscr T_{\mathrm{7d}}^{D_N}$
and its operators
can be derived from M-theory on 
$\RP^4$. Before entering the details of the derivation,
let us summarize its main points:
\begin{itemize}
\item The 
fields $\mathsf c_3$, $\mathsf b_3$,
$\mathsf a_1$, $\mathsf a_5$
come from expansion of the M-theory
3-form gauge field $C_3$ and its dual
$C_6$ onto torsional cohomology 
classes in the horizon geometry
$\RP^4$.
Correspondigly, 
the operators
$\bm Q_3(\Sigma_3)$,
$\widehat{\bm Q}_3(\Sigma_3)$,
$\bm Q_1(\Sigma_1)$,
$\widehat{\bm Q}_5(\Sigma_5)$
are identified with M2-branes
and M5-branes wrapping torsional cycles in $\RP^4$.
This is  summarized in Table \ref{table:summary}.

\item 
The 
quadratic terms
$\mathsf c_3 \cup \delta \mathsf b_3$,
$\mathsf c_3 \cup \delta \mathsf c_3$,
$\mathsf a_1 \cup \delta \mathsf a_5$
in the action 
\eqref{eq:7d_action_summary}
originate from flux non-commutativity in
11d supergravity. They are discussed in Section \ref{sec_3fCS_from_M}, where we use an argument based on brane physics to capture flux non-commutativity effects in the presence of a non-zero $G_4$-flux background on
the horizon geometry $\RP^4$.
\item The cubic coupling $\mathsf a_1 \cup \mathsf c_3 \cup \mathsf c_3$ is derived in Section \ref{sec_cubic_derivation},
 from
reduction of the 11d M-theory effective action.

\item 
The realization of the operators
$\widehat{\bm Q}_3(\Sigma_3)$,
$\widehat{\bm Q}_5(\Sigma_5)$
via wrapped M5-branes
allows for a  
derivation 
of the topological dressing factors
$T_3$, $T_5$
from the M5-brane worldvolume theory,
see Section \ref{sec:noninvertiblah}.

\end{itemize}

\subsection{Derivation of 3-form Chern-Simons theory}\label{sec_3fCS_from_M}

In this section we provide an M-theory derivation of the 3-form Abelian Chern-Simons theory
\eqref{eq:Abelian_3form_CS}, based
of the non-commutativity of 
torsional fluxes \cite{Freed:2006ya,Freed:2006yc,Albertini:2020mdx}.
Our argument hinges on the interplay between
pairing of torsional cycles in $\RP^4$,
and the background $G_4$-flux
threading $\RP^4$.

\subsubsection{Topological operators and correlators from branes}

We are interested in studying
topological operators supported
on 3-cycles in external
spacetime $M_{7}$. Following
the approach of 
\cite{Apruzzi:2022rei,GarciaEtxebarria:2022vzq,Heckman:2022muc,Heckman:2022xgu,Etheredge:2023ler,Bah:2023ymy,Apruzzi:2023uma},
these can be realized using branes
wrapping cycles in the internal geometry $\RP^4$
and extending along three directions
externally. Throughout this section,
we work in the limit in which the
external worldvolume of the branes approaches the holographic boundary of $M_7$; as a result, we only keep
track of topological terms
in the brane effective actions.

An M2-brane couples electrically to
the 3-form potential $C_3$,
whose field strength $G_4$
is a twisted differential form.
It follows that we can only consider
M2-branes wrapping twisted
cycles in $\RP^4$,
i.e.~elements of $H_\bullet(\RP^4 ; \widetilde{\mathbb Z})$. By a similar token,
we can only consider M5-branes
 wrapping elements of 
$H_\bullet(\RP^4 ; {\mathbb Z})$, because the dual
6-form potential to $C_3$
has untwisted field strength $*G_4$.
Inspection of \eqref{eq:RP4groups}, \eqref{eq:RP4Poincare}
reveals that we have two ways
of obtaining 3d operators
in external spacetime:
\be  \label{eq:Ws_from_branes}
\text{$\bm Q_3(\Sigma_3)$ = M2-brane on $\Sigma_3 \times \widetilde{{\rm pt}}$} \ ; \qquad 
\text{$\widehat {\bm Q}_3(\widehat \Sigma_3)$  = M5-brane on $\widehat \Sigma_3 \times \RP^3$} \  .
\ee 
Here $\widetilde{{\rm pt}}$
denotes a generator
of $H_0(\RP^4 ; 
\widetilde{\mathbb Z})$,
while
$\RP^3$
denotes a generator
of $H_3(\RP^4 ; 
{\mathbb Z})$.
They are Poincar\'e duals in $\RP^4$ to $t_1^4$,
$t_1$, respectively,
\be 
\widetilde{{\rm pt}} = {\rm PD}_{\RP^4}[t_1^4]
\ , \qquad 
\RP^3 = {\rm PD}_{\RP^4}[t_1] \ . 
\ee

Let us adopt a Hamiltonian
viewpoint,
with the holographic radial coordinate
of $M_7$ treated as Euclidean time \cite{Witten:1998wy,Belov:2004ht}. The supports
$\Sigma_3$, $\widehat \Sigma_3$
of ${\bm Q_3}(\Sigma_3)$, $\widehat {\bm Q}_3(\widehat \Sigma_3)$ are taken to lie inside a 6d fixed-time  slice of $M_7$, which we denote $M_6$.
We write $M_{10} = M_6 \times \RP^4$
for the 10d fixed-time slice
in the total 11d spacetime $M_{11}$.
We aim to compute the
commutators among  the operators
${\bm Q}_3(\Sigma_3)$ and $\widehat {\bm Q}_3(\widehat \Sigma_3)$.
To this end, we proceed as follows.

Building on \cite{Freed:2006ya,Freed:2006yc,Etheredge:2023ler},
we observe that inserting an M5-brane on a 6-cycle in
$M_{10}$
implements an operator
that translates the
electric 3-form potential.
This operator acts on the wavefunctional
of the system,
which 
is a functional of the 3-form potential.
In the present setup,
the M5-brane wraps
$\widehat \Sigma_3 \times \RP^3 \subset M_6 \times \RP^4$, whose Poincar\'e dual 
in $M_{10}$ reads
\be 
{\rm PD}_{M_{10}}[\widehat \Sigma_3 \times \RP^3]
= {\rm PD}_{M_{6}}[\widehat \Sigma_3]
\cup t_1  \ .
\ee 
Since $t_1\in H^1(\RP^4 ; \widetilde{\mathbb Z})$ is   torsional, there exists a class $u_0 \in H^0(\RP^4 ; \widetilde{
{\mathbb R} /\mathbb Z})$
such that
\be
t_1 = \widetilde \beta(u_0) \ , 
\ee 
where $\widetilde\beta$ denotes the Bockstein
homomorphism associated
to the short exact sequence
$0 \rightarrow \widetilde{\mathbb Z}
\rightarrow \widetilde{\mathbb R}
\rightarrow \widetilde{\mathbb R / \mathbb Z}$.
With this notation,
the translation of the 3-form potential implemented by
the M5-brane is\footnote{\ 
Heuristically, we can provide further intuition for 
\eqref{eq:G4shift} as follows.
The M5-brane couples electrically to $C_6$, the 11d electro-magnetic dual to $C_3$. Inserting an M5-brane on
a 6-cycle $\cC_6$ gives a factor
$\exp 2\pi i \int_{\cC_6}C_6
= \exp 2\pi i \int_{M_{10}}{\rm PD}[\cC_6] C_6$, where PD denotes Poincar\'e duality on the spatial slice $M_{10}$.
We are interested in the case in which $\cC_6$ is a torsional cycle. Hence, ${\rm PD}[\cC_6] = \beta(u_3)$ for some $u_3$ class with $\widetilde{\mathbb R/\mathbb Z}$ coefficients.
After a formal integration by parts, we get a factor
$\exp 2\pi i \int_{M_{10}} u_3 P_7$, where $P_7$ is the Page charge, schematically $P_7 =dC_6$. The operator $P_7$ implements translations on $C_3$, so we have formally
$C_3 \mapsto C_3 + u_3$,
which is best understood as
\eqref{eq:G4shift}.
}
\be \label{eq:G4shift}
\breve G_4 \mapsto \breve G_4 + i(u_3) \ , \qquad 
u_3 = {\rm PD}_{M_6}[\widehat \Sigma_3] \cup u_0 
\in H^3(M_{10} ; \widetilde{\mathbb R / \mathbb Z}) \ . 
\ee 
Here we are using the fact that the (gauge equivalence class of the)
translation in the 3-form potential
can be modeled by an element in
$\breve H^4(M_{10};\widetilde{\mathbb Z})$,
as explained in section \ref{sec:mc_structure}.
The map $i$ was introduced in the commutative
diagram \eqref{eq:diff_coho_hexagon}.
The quantity
$i(u_3)$ is a twisted, unshifted, flat
class; it is topologically non-trivial.

\paragraph{$ \bm Q_3 \widehat {\bm Q}_3$ commutator.}
Let us now consider
the effect of the shift
\eqref{eq:G4shift} on
an M2-brane on
$\Sigma_3 \times \widetilde{\rm pt} \subset M_6 \times \RP^4$.
The relevant coupling in the M2-brane worldvolume action is
the standard topological coupling to the 3-form potential.
In the language of differential cohomology, it can be written as
\be 
\exp 2\pi i \int_{\Sigma_3 \times \widetilde{\rm pt}}
\breve G_4 \ . 
\ee 
Under the shift 
\eqref{eq:G4shift},
this coupling induces the following phase factor,
\begin{align} \label{eq:computephase}
\exp 2 \pi i
\int_{\Sigma_3 \times \widetilde{\rm pt}}
i(u_3)
&= \exp 2\pi i \int_{M_{10}}
{\rm PD}_{M_6}[\Sigma_3] \cup t_1^4 \cup u_3
\nn \\
& = \exp \bigg[ -2\pi i \int_{M_{6}}
{\rm PD}_{M_6}[\Sigma_3] \cup 
{\rm PD}_{M_6}[\widehat \Sigma_3] 
\int_{\RP^4} t_1^4 \cup  \beta^{-1}(t_1)
\bigg] \nn \\
& = \exp \bigg[ - 2\pi i  (\Sigma_3 \cdot_{M_6} \widehat \Sigma_3)
\cT_{\RP^4}(t_1^4, t_1) 
\bigg] \ . 
\end{align}
In the previous expression
$(\Sigma_3 \cdot_{M_6} \widehat \Sigma_3)$ denotes the $\mathbb Z$-valued intersection
pairing of non-torsional 3-cycles in $M_6$,
while $\cT_{\RP^4}(t_1^4,t_1)$ denotes the
$\mathbb R/\mathbb Z$-valued pairing of torsional cycles on $\RP^4$.
Making use of\footnote{ \ 
Since $t_1$ is torsional
of degree 2
and the pairing $\cT_{\RP^4}$ is bilinear,
$2\cT_{\RP^4}(t_1^4, t_1)=\cT_{\RP^4}(t_1^4, 2t_1)=0$ mod $\mathbb Z$,
so that we must have
$\cT_{\RP^4}(t_1^4, t_1)=0$
or $\frac 12$ mod $\mathbb Z$.
The option 
$\cT_{\RP^4}(t_1^4, t_1)=0$
 mod $\mathbb Z$
 is ruled out because
 $\cT_{\RP^4}: H^4(\RP^4 ; \mathbb Z) \times H^1(\RP^4 ; \widetilde{\mathbb Z}) \rightarrow 
\mathbb R/\mathbb Z$ is a perfect pairing.
}
\be 
\cT_{\RP^4}(t_1^4, t_1) = \frac 12 \mod \mathbb Z \ ,
\ee 
we can write the phase factor of interest as
\be 
\exp \bigg[ - 2\pi i  \frac{1}{2}   (\Sigma_3 \cdot_{M_6} \widehat \Sigma_3) \bigg]  \ . 
\ee 
This translates to the non-trivial commutator
of ${\bm Q}_3(\Sigma_3)$
and $\widehat {\bm Q}_3(\widehat \Sigma_3)$
operators,
\be  \label{eq:WWhatcomm}
{\bm Q}_3(\Sigma_3)\widehat {\bm Q}_3(\widehat \Sigma_3)=
\widehat {\bm Q}_3(\widehat \Sigma_3 )
{\bm Q}_3(\Sigma_3) \exp \bigg[ - 2\pi i  \frac{1}{2}   (\Sigma_3 \cdot_{M_6} \widehat \Sigma_3) \bigg] \ . 
\ee 
The above considerations are
a reformulation
of the  
non-commutativity
argument of \cite{Freed:2006ya,Freed:2006yc}.

\paragraph{M5-brane coupling to bulk $C_3$.}
Before turning to the analysis
of the commutator
of two $\widehat{\bm Q}_3$ operators,
it is useful to recall
some facts on the coupling of
the worldvolume theory of an M5-brane to the bulk 3-form potential.

The M5-brane supports a chiral 2-form, i.e.~a $U(1)$
2-form gauge field
whose 3-form field strength
sartisfies a self-duality condition.
The partition function of the M5-brane coupled to the bulk $C_3$ can be defined 
by resorting to a 7d 3-form Chern-Simons theory \cite{Witten:1996hc},
formulated on a manifold
$M_7$, with $\partial M_7 = M_6$ the worldvolume of the M5-brane. 
The form of the 7d 3-form
Chern-Simons theory
can be inferred from the topological couplings
of the 11d M-theory effective action. We follow here the exposition in \cite{Witten:1999vg}\footnote{ \ 
The combination
$\pi^*(z_4) + g_4$ introduced below is denoted $w$ in \cite{Witten:1999vg}. There, $v_4 = \lambda$, which is the canonical integral lift of the fourth Stiefel-Whitney class in the Spin setting.}.
A neighborhood of the M5-brane can be captured by the normal
bundle to $M_6$ inside the 11d spacetime. The unit spheres
in the fibers of the normal bundle determine an $\sphere^4$ bundle over $M_6$.
To determine the CS theory on
$M_7$ we consider an 11d spacetime given by an $\sphere^4$ bundle over $M_7$,
$\sphere^4 \hookrightarrow M_{11} \xrightarrow{\pi} M_7$.
The relevant topological couplings on $M_{11}$ can be encoded in the following 12-form,
\be \label{eq:I12_with_subtraction}
I_{12} = \left[ \frac 16 G_4 \wedge G_4 \wedge G_4
+ G_4 \wedge X_8
\right]
 - (G_4 \rightarrow -\tfrac 12 v_4) \ . 
\ee 
In the above expression
$X_8$ denotes a combination of Pontryagin classes of the tangent bundle,
see \eqref{eq:X8def}.
The total $G_4$ is written as the sum of three contributions,
\be \label{eq:G4_three_terms}
G_4 = \pi^*(z_4) + g_4  -\frac 12 v_4  \ .
\ee 
The quantity $v_4$ is a fixed twisted integral lift of $w_4$ that 
specifies the $m_{\rm c}$ structure of spacetime.
The quantity $g_4$ captures
the flux generated by the M5-brane itself:
 $g_4$ integrates to one on the $\sphere^4$ fibers of the
fibration over $M_7$. Finally,
the quantity $\pi^*(z_4)$ captures the  contributions
to $G_4$ that are pulled back
to $M_{11}$ from $M_7$.
The notation $-(G_4 \rightarrow -\tfrac 12 v_4)$
in \eqref{eq:I12_with_subtraction}
means to subtract the previous
quantity with $G_4$ replaced by $-\tfrac 12 v_4$.
To proceed, we fiber-integrate
$I_{12}$ over $\sphere^4$. 
As shown in \cite{Witten:1999vg},
after expressing everything
in terms of quantities on the base $M_7$
one gets the
 8-form
\be \label{eq:8form_def}
I_8 = \frac 12 z_4 \wedge z_4 
+ \frac 12 \nu_4 \wedge z_4 
= \frac 12 \left( z_4 + \frac 12 \nu_4 \right)^2 - \frac 18 \nu_4^2 \ .
\ee 
Here $\nu_4$ denotes a twisted integral
lift of the fourth Wu class of the tangent bundle to $M_7$.
The combination 
$z_4 + \frac 12 \nu_4$
is a 4-form whose periods
satisfy an integrality condition
shifted by the fourth Wu class;
it 
provides the appropriate definition of the restriction
of $G_4$ to $M_7$ \cite{Monnier:2013rpa}.
The 8-form \eqref{eq:8form_def}
furnishes the sought-for
7d Chern-Simons theory.
In the notation of differential cohomology,
the action on $M_7$ can be written as
\be \label{eq:I8_breve}
S = \int_{W_7} \bigg[ 
\frac 12 \breve z_4 \cdot \breve z_4
+ \frac 12 \breve \nu_4 \cdot
\breve z_4
\bigg] \ . 
\ee

\paragraph{$\widehat {\bm Q}_3 \widehat {\bm Q}_3$ commutator.}
We can now study the commutator
of two $\widehat {\bm Q}_3$
operators. Our strategy is
to 
 examine the effect of the shift
\eqref{eq:G4shift},
induced by an M5-brane
on $\widehat \Sigma_3 \times \RP^3$,
on \emph{another}
M5-brane, wrapping
$\widehat \Sigma_3' \times \RP^3$.
To this end, we make use of 
\eqref{eq:I8_breve}, with $W_7$
extending $\widehat \Sigma_3' \times \RP^3$.
More precisely, we consider the quantity 
\be  \label{eq:z_squared}
\frac 12 \int_{\widehat \Sigma_3' \times \mathcal B_4} 
\bigg[ 
\frac 12 \breve z_4 \cdot \breve z_4
+ \frac 12 \breve \nu_4 \cdot
\breve z_4
\bigg] \ .
\ee 
Here
 $\cB_4$ is a 4-chain,  analogous to 
the 5-chain 
governing the annihilation
of Witten's fat strings in $\AdS_5 \times \RP^5$ \cite{Witten:1998xy}.
In that reference,
Witten considers type IIB string theory on $\AdS_5 \times \RP^5$ and studies 5-branes wrapped on a copy of $\RP^4$ inside $\RP^5$, dubbed fat strings.
He constructs a 5-chain that can be regarded as a bordism between 
two copies of $\RP^4$---both ``incoming'' with the same (twisted) orientation---and the empty manifold. Furthermore, he argues that this 5-chain can be identified with $\RP^5$, and in particular is threaded by the non-vanishing $F_5$ 
background flux.
The arguments of \cite{Witten:1998xy} can be adapted to M-theory on $\AdS_7 \times \RP^4$. We   conclude that a 4-chain
$\cB_4$ can be constructed,
which 
can be regarded as a bordism between 
two copies of $\RP^3$---both ``incoming'' with the same orientation---and the empty manifold.  
The extra prefactor $1/2$ in 
\eqref{eq:z_squared} accounts
for the fact that
we have two copies of 
$\RP^3$
``entering'' the 4-chain
$\cB_4$.
Moreover, $\cB_4$ is 
identified with $\RP^4$ and thus threaded by the background $G_4$ flux.

Our task is to identify the modification in \eqref{eq:z_squared}
induced by the shift
$\breve G_4 \mapsto
\breve G_4 + i(u_3)$, see \eqref{eq:G4shift}.
Retracing the relation between $G_4$ and $z_4$ \eqref{eq:G4_three_terms}, we see that this shift of $\breve G_4$ corresponds to the shift
\be
\breve z_4 \mapsto \breve z_4 + i(u_3) \ . 
\ee 
The quantity of interest is therefore
\be    \label{eq:zsq_difference}
\frac 12 \int_{\widehat \Sigma_3' \times \mathcal B_4} 
\bigg[ 
\frac 12 (\breve z_4 + i(u_3))^2 
- \frac 12 \breve z_4^2
\bigg] \ . 
\ee 
We have dropped the contributions proportional to the fourth Wu class
(the
fourth Wu class $\RP^4$ is zero,
and the fourth Wu class of
$\Sigma_3$  vanishes for dimensional reasons).

The only non-zero contribution in 
\eqref{eq:zsq_difference}
is 
\be \label{eq:factorize}
\frac 12 \int_{\widehat \Sigma_3' \times \cB_4}
\breve z_4 \cdot i(u_3)
= 
\bigg( \int_{\RP^4} \breve z_4 \bigg)
\bigg( \frac 12 \int_{\widehat \Sigma_3' \times  \widetilde{\rm pt} } i(u_3) \bigg) \ . 
\ee 
The integral of $\breve z_4$ on $\RP^4$ picks up a non-zero contribution from the background flux \eqref{eq:flux_bk},
\be
\int_{\RP^4} \breve z_4 = N \ . 
\ee
Notice that the result is $N$, and not $N-\frac 12$:
this is due to the fact that, in the relation 
\eqref{eq:G4_three_terms} between $G_4$
and $z_4$, we have to take into account the subtraction
of the $v_4$ term, which 
implements the half-integral quantization of $G_4$ and is removed when considering $z_4$.
The other integral in 
\eqref{eq:factorize} is evaluated following the same steps as in
\eqref{eq:computephase},
\be 
\frac 12 \int_{
\widehat \Sigma_3' \times 
\widetilde{\rm pt}
} i(u_3) = 
-(\widehat \Sigma_3' \cdot_{ M_{6} } \widehat \Sigma_3) 
\int_{\RP^4} \frac 12 t_1^4 \beta^{-1}(t_1) \ .
\ee 
The integral on the RHS
is a \emph{refinement}
of the torsional pairing
$\cT_{\RP^4}(t_1^4, t_1) = \frac 12$
mod $\mathbb Z$, 
\be  \label{eq:quad_refin}
\int_{\RP^4} \frac 12 t_1^4 \beta^{-1}(t_1) =  \frac s4 \mod \mathbb Z \ , \qquad s\in \{ \pm 1 \} \ .  
\ee 
In summary, 
the phase factor of interest takes the form
\be 
\exp 2 \pi i  \frac 12  \int_{\widehat \Sigma_3' \times \mathcal B_4}    \breve z_4 \cdot i(u_3) = \exp    2\pi i \frac{-sN}{4} (\widehat \Sigma_3 '\cdot_{M_6} \widehat \Sigma_3) \ .
\ee 
It corresponds to
the non-trivial commutator
\be  \label{eq:WhatWhatcomm}
\widehat {\bm Q}_3(\widehat \Sigma_3')\widehat {\bm Q}_3(\widehat \Sigma_3)=
\widehat {\bm Q}_3(\widehat \Sigma_3 )
\widehat {\bm Q}_3(\widehat \Sigma_3') \exp 2\pi i \frac{-sN}{4}   (\widehat \Sigma_3' \cdot_{M_6} \widehat \Sigma_3) \ . 
\ee
A more detailed analysis would be required to determine the sign $s$ in the quadratic refinement
of $\cT_{\RP^4}(t_1^4, t_1)$. It is not needed, however, 
to extract the sought-for 7d TQFT, up to sign ambiguities that can be reabsorbed by field redefinitions,
as discussed below in Section~\ref{sec:extractCS}.

\paragraph{${\bm Q}_3 {\bm Q}_3$ commutator.} 
Finally, the commutator of
two ${\bm Q}_3$ operators is trivial. In terms of branes, this can be seen as follows. Heuristically,
inserting an M2-brane
on a torsional 3-cycle induces
a shift in $G_7$ of the form
$\breve G_7 \mapsto \breve G_7 + i(u_6)$.
The second M2-brane, however, does not have any
topological worldvolume
coupling to $\breve G_7$,
and therefore does not pick up any additional phase.

\subsubsection{Extracting the 7d 3-form Chern-Simons terms} \label{sec:extractCS}

There is a standard way of translating commutation relations such as 
\eqref{eq:WWhatcomm}, \eqref{eq:WhatWhatcomm}
in Hamiltonian language into
a Lagrangian for an Abelian
TQFT in 7d written in terms of a collection of 3-form potentials
\cite{Maldacena:2001ss,Banks:2010zn,Gukov:2020btk}.
Quite generally, we can consider the 7d action
\be
S = 2\pi \int_{M_7}   \frac 12 k_{ij} c_3^{(i)} \wedge dc_3^{(j)}
 \ ,
\ee 
where the matrix $k_{ij}$ is
integral and non-degenerate. 
The classical equations
of motion yield 
the correlators\footnote{ \ The supports
$\Sigma_3$, $\Sigma_3'$ are assumed to be distinct and disjoint.}
\be 
\langle 
e^{2\pi i \int_{\Sigma_3} n_{(i)} c_3^{(i)}}
e^{2\pi i \int_{\Sigma'_3} n'_{(j)} c_3^{(j)}}
\rangle
= \exp \bigg[ -2\pi i (k^{-1})^{ij} n_{(i)} n'_{(j)} L_{M_7}(\Sigma_3, \Sigma_3') \bigg] \ ,
\ee 
where $n_{(i)}$, $n_{(j)}' \in \mathbb Z$ and 
$L_{M_7}(\Sigma_3, \Sigma_3')$
denotes the integer linking number of $\Sigma_3$, $\Sigma_3'$ inside $M_7$.
This is mapped to the integer
intersection number of $\Sigma_3$ and $\Sigma_3'$
in the Hamiltonian formulation,
when $\Sigma_3$, $\Sigma_3'$ are both taken to lie inside a spatial surface.

Taking $s=1$ in \eqref{eq:quad_refin},
the commutators computed above correspond to 
\be 
(k^{-1})^{ij} = \begin{pmatrix}
0 & \frac 12 \\
 \frac 12 & \frac {N}4
\end{pmatrix} \ , \qquad 
k_{ij} = \begin{pmatrix}
-N & 2 \\
2 & 0
\end{pmatrix} \ ,
\ee
where rows/columns of these matrices
are ordered as $(\bm Q_3,\widehat {\bm Q}_3)$.
The 7d TQFT action reads 
\be  \label{eq:3formCS}
S_\text{7d TQFT}^\text{quadratic} = 2\pi \int_{M_7} \bigg[
- \frac 12 N c_3 \wedge dc_3
+ 2 c_3 \wedge db_3
\bigg] \ , 
\qquad c_3 \equiv c_3^{(1)} \ , 
\qquad b_3 \equiv c_3^{(2)}
\ee 
(If we had taken $s=-1$ in \eqref{eq:quad_refin}, we would have obtained
the same action up to a redefinition of $c_3^{(2)}$ and an overall flip of orientation.) We also establish the correspondences\footnote{\ In this section we neglect the effects of the cubic coupling $\mathsf a_1 \cup \mathsf c_3 \cup \mathsf c_3$ in the total 7d TQFT,
which are discussed in Sections \ref{sec_cubic_derivation}, \ref{sec:noninvertiblah}.}
\begin{align} 
e^{2\pi i \int_{\Sigma_3} c_3} &\leftrightarrow
{\bm Q}_3(\Sigma_3) 
 \  , & 
e^{2\pi i \int_{\widehat \Sigma_3} b_3}
 &\leftrightarrow 
 \widehat {\bm Q}_3(\widehat \Sigma_3)
 \ , \\ 
c_3 &\leftrightarrow \text{$\breve G_4$ expanded onto $\breve t_1$} \ ,
& 
b_3 &\leftrightarrow \text{$\breve G_7$ expanded onto $\breve t_1^4$} 
\ . \label{eq:origin_of_c3b3}
\end{align}
The identifications on the second line
follow from the fact that 
 ${\bm Q}_3(\Sigma_3)$ is realized by an M2-brane on $\widetilde {\rm pt}$, 
while  
 $\widehat {\bm Q}_3(\widehat \Sigma_3)$ 
is realized by an M5-brane
on $\RP^3$,
see \eqref{eq:Ws_from_branes}.

\subsubsection{Analysis 
depending on the value of $N$ mod 4} 

Depending on the value of $N$ mod
4 we can perform an integral change of basis in the 3-form potentials $c_3$,
$b_3$ in \eqref{eq:3formCS} that 
simplifies the form of the action.
It also allows us to check
\eqref{eq:3formCS} against the results of 
\cite{GarciaEtxebarria:2019caf} based on the analysis of asymptotic non-commuting torsional fluxes
in the type IIB realization of
the 6d (2,0) SCFT of type $D_N$.

\paragraph{$N = 0$ mod 4.}
We write $N = 4m$ and 
perform the 
redefinition
\be 
\begin{pmatrix}
c_3 \\ b_3
\end{pmatrix}
= \begin{pmatrix}
1 & 0 \\ m & 1
\end{pmatrix}
\begin{pmatrix}
x_3 \\ y_3
\end{pmatrix} \ ,
\ee 
which is implemented by a matrix
in $SL(2,\mathbb Z)$. The 7d action reads
\be 
S_\text{7d TQFT}^\text{quadratic} =2\pi \int_{M_7} 2    
 x_3 \wedge dy_3
\ .
\ee 
The correlators of holonomies
of $x_3$, $y_3$ are
\begin{align}
\langle e^{2\pi i \int_{\Sigma_3} x_3} 
e^{2\pi i \int_{\Sigma'_3} x_3}
\rangle &= 
\langle e^{2\pi i \int_{\Sigma_3} y_3} 
e^{2\pi i \int_{\Sigma'_3} y_3}
\rangle
=1 \ , \\
\langle e^{2\pi i \int_{\Sigma_3} x_3} 
e^{2\pi i \int_{\Sigma'_3} y_3}
\rangle &= e^{ -\frac 12 \, 2\pi i L_{M_7}(\Sigma_3, \Sigma_3')} \ ,
\end{align} 
where $L_{M_7}(\Sigma_3, \Sigma_3')$ is the integer
linking number of $\Sigma_3$,
$\Sigma'_3$ in $M_7$.

\paragraph{$N = 1$ mod 4.}
We write $N = 4m + 1$ and 
perform the 
redefinition
\be 
\begin{pmatrix}
c_3 \\ b_3
\end{pmatrix}
= \begin{pmatrix}
1 & 2 \\ m & 2m+1
\end{pmatrix}
\begin{pmatrix}
x_3 \\ y_3
\end{pmatrix} \ .
\ee 
The 7d action reads
\be 
S_\text{7d TQFT}^\text{quadratic} =2\pi  \int_{W_7} \bigg[ 
2 y_3 \wedge dy_3
- \frac 12 x_3  \wedge dx_3 
\bigg] \ .
\ee 
The correlators of holonomies
of $x_3$, $y_3$ are
\begin{align}
\langle e^{2\pi i \int_{\Sigma_3} x_3} 
e^{2\pi i \int_{\Sigma'_3} x_3}
\rangle &= 
\langle e^{2\pi i \int_{\Sigma_3} x_3} 
e^{2\pi i \int_{\Sigma'_3} y_3}
\rangle
=1 \ , \\
\langle e^{2\pi i \int_{\Sigma_3} y_3} 
e^{2\pi i \int_{\Sigma'_3} y_3}
\rangle &= e^{- \frac 14 \, 2\pi i L_{M_7}(\Sigma_3, \Sigma_3')} \ .
\end{align} 
The term $-\frac 12 x_3 dx_3$
describes an invertible factor that decouples from the system
(the holonomy $e^{2\pi i \int_{\Sigma_3}x_3}$ has trivial
correlators with all holonomies).
We are left with a 3-form
Chern-Simons action at level 4.

\paragraph{$N = 2$ mod 4.}
We write $N = 4m+2$ and perform the redefinition
\be 
\begin{pmatrix}
c_3 \\ b_3
\end{pmatrix}
= \begin{pmatrix}
1 & 1 \\ m & m+1
\end{pmatrix}
\begin{pmatrix}
x_3 \\ y_3
\end{pmatrix} \ . 
\ee 
The 7d action reads
\be 
S_\text{7d TQFT}^\text{quadratic} = 2\pi \int_{M_7} \bigg[ 
 y_3 \wedge dy_3
-   x_3 \wedge d x_3
\bigg] \ .
\ee 
The correlators of holonomies
of $x_3$, $y_3$ are
\begin{align}
\langle e^{2\pi i \int_{\Sigma_3} x_3} 
e^{2\pi i \int_{\Sigma'_3} x_3}
\rangle &=
e^{ \frac 12 \, 2\pi i L_{M_7}(\Sigma_3, \Sigma_3')} \ , \quad 
\langle e^{2\pi i \int_{\Sigma_3} y_3} 
e^{2\pi i \int_{\Sigma'_3} y_3}
\rangle
= e^{- \frac 12 \, 2\pi i L_{M_7}(\Sigma_3, \Sigma_3')} \ , \nn \\
\langle e^{2\pi i \int_{\Sigma_3} x_3} 
e^{2\pi i \int_{\Sigma'_3} y_3}
\rangle &= 1 \ .
\end{align} 
We have two decoupled 3-form
Chern-Simons theories at levels $\pm 2$.

\paragraph{$N = 3$ mod 4.}
We write $N = 4m + 3$ and perform the redefinition
\be 
\begin{pmatrix}
c_3 \\ b_3
\end{pmatrix}
= \begin{pmatrix}
2 & 1 \\ 2m+1 & m+1
\end{pmatrix}
\begin{pmatrix}
x_3\\ y_3
\end{pmatrix} \ . 
\ee 
The 7d action reads
\be 
S_\text{7d TQFT}^\text{quadratic} = 2\pi  \int_{W_7} \bigg[ 
 \frac 12 y_3 \wedge d y_3
-  2 x_3 \wedge d x_3
\bigg] \ .
\ee 
The correlators of holonomies
of $x_3$, $y_3$ are
\begin{align}
\langle e^{2\pi i \int_{\Sigma_3} y_3} 
e^{2\pi i \int_{\Sigma'_3} y_3}
\rangle &= 
\langle e^{2\pi i \int_{\Sigma_3} x_3} 
e^{2\pi i \int_{\Sigma'_3} y_3}
\rangle
=1 \ , \\
\langle e^{2\pi i \int_{\Sigma_3} x_3} 
e^{2\pi i \int_{\Sigma'_3} x_3}
\rangle &= e^{ \frac 14 \, 2\pi i L_{M_7}(\Sigma_3, \Sigma_3')} \ .
\end{align} 
The term $\frac 12 y_3 dy_3$
describes an invertible factor that decouples from the system
(the holonomy $e^{2\pi i \int_{\Sigma_3}y_3}$ has trivial
correlators with all holonomies).
We are left with a 3-form
Chern-Simons action at level $-4$.

\paragraph{Comparison with linking pairing computations in Type IIB.}
The above results
are consistent with the
Type IIB analysis of 
\cite{GarciaEtxebarria:2019caf}
based on the torsional
linking pairing
in $\sphere^3/\Gamma$,
where $\Gamma$ is the discrete subgroup of $SU(2)$
associated to the Lie algebra $D_N$.
The linking pairing,
denoted $\mathsf L_\Gamma$
and taking values in $\mathbb R /\mathbb Z$, was found to be 
\begin{align}
\mathsf L_\Gamma & = 
\begin{pmatrix}
0 & \tfrac 12  \\ 
\tfrac 12 & 0
\end{pmatrix} & 
&\text{for $N = 0$ mod 4} \ , \\ 
\mathsf L_\Gamma & = 
\begin{pmatrix}
\tfrac 12 & 0  \\ 
0 & \tfrac 12 
\end{pmatrix} & 
&\text{for $N = 2$ mod 4} \ , \\ 
\mathsf L_\Gamma & = 
- \tfrac 14  & 
&\text{for $N = 1$ mod 4} \ , \\ 
\mathsf L_\Gamma & = 
 \tfrac 14  & 
&\text{for $N = 3$ mod 4} \ .
\end{align}

\subsubsection{Fat strings in $\AdS_5$ and fat membranes
in $\AdS_7$}

The physics of the quadratic terms \eqref{eq:3formCS} in the 7d TQFT can also be analyzed 
from the point of view of ``fat membranes'', i.e.~M5-branes wrapped on   $\RP^3 \subset \RP^4$ with three
non-compact directions.
Indeed, we can follow closely 
the discussion of ``fat strings'' in $\AdS_5 \times \RP^5$
\cite{Witten:1998xy}, which has been used in \cite{Bergman:2022otk} to
study the 5d topological action
associated to the 1-form symmetry sector of 4d $\mathcal N =4$ $\mathfrak{so}(2N)$ super Yang-Mills theory.

\paragraph{Fat strings in $\AdS_5 \times \RP^5$.}

4d $\cN = 4$ SYM with gauge algebra $\mathfrak{so}(2N)$
is dual to $\AdS_5 \times \RP^5$ with $N$ units of
$F_5$ flux through $\RP^5$.
The relevant 2d extended operators in the 5d 
external spacetime are
\begin{align}
\cU_{\rm F1}(\cC_2) & = \text{$\rm F1$ on $\cC_2 \times \widetilde{\rm pt}$} \ , \nn \\
\cU_{\rm D1}(\cC_2) & = \text{$\rm D1$ on $\cC_2 \times \widetilde{\rm pt}$} \ , \nn \\
\cU_{\rm NS5}(\cC_2) & = \text{$\rm NS5$ on $\cC_2 \times \RP^4$} \ , \nn \\
\cU_{\rm D5}(\cC_2) & = \text{$\rm D5$ on $\cC_2 \times \RP^4$} \ .
\end{align}
The last two are dubbed fat strings and they originate
from five-branes in ten dimensions.

\begin{figure}
  \centering
   \includegraphics[width=8.5cm]{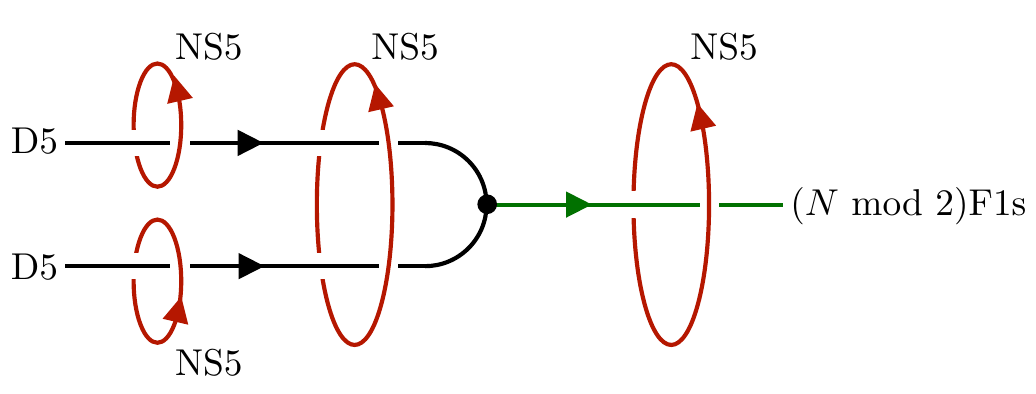}
\caption{Two fat strings in $\AdS_5 \times \RP^5$ (i.e.~D5-branes wrapping
$\RP^4 \subset \RP^5$)
with the same orientation can annihilate 
into $N$ mod 2 fundamental strings.
This imposes constraints on the correlators
among wrapped D5-branes, wrapped NS5-branes,
and F1 strings. On the right hand side,
we probe the F1s with an NS5-brane. Internally, the latter
 wraps $\RP^4 \subset \RP^5$. Externally, it is
supported on a surface linking once with the
support of the F1s. We can slide the NS5-brane
towards the left, and get to a configuration
in which we have twice the correlator
between a wrapped NS5-brane and a wrapped D5-brane
that link once.
}
        \label{fig_strings_annihilate}
\end{figure}

As explained in \cite{Witten:1998xy}, one can construct
a 5-chain $\cB_5$ with the following properties:
$\cB_5$ is a bordism between two copies of $\RP^4$
with the same orientation, and the empty set;
$\cB_5$ is threaded by $N$ units of $F_5$ flux.
The existence of $\cB_5$ implies the existence
of a `vertex' among 2d surface operators,
as depicted schematically in Figure \ref{fig_strings_annihilate}.\footnote{ \ The Wess-Zumino couplings on a D5-brane include a term of the form $a_1 \wedge F_5$, where $a_1$ is the $U(1)$ gauge field on the brane.
Considering this coupling on 
a 6d space of the form
$\Sigma_1 \times \cB_5$, and recalling that $\cB_5$ has $N$ units of $F_5$ flux,
we infer an induced coupling $N\int_{\Sigma_1}a_1$. The endpoints of  F1-strings on a D-brane are charged under $a_1$ and are needed to cancel the induced charge. We consider $N$ mod 2 because $\RP^3$ is a torsional cycle of degree 2.}

In turn, the existence of this `vertex'
imposes  constraints on the correlators
among surface operators in 5d. 
A pictorial derivation of these constraints
is furnished in Figure~\ref{fig_strings_annihilate}
for the case of fat strings from D5-branes.
An analogous argument can be made for 
fat strings from NS5-branes.
We must have
\begin{align}
\langle \cU_{\rm D5}(\cC_2) \cU_{\rm NS5}(\cC_2') 
\rangle ^2 
&= \langle \cU_{\rm F1}(\cC_2) \cU_{\rm NS5}(\cC_2') 
\rangle ^{(N \ {\rm mod} \ 2)}    \ ,  \nn \\
\langle \cU_{\rm NS5}(\cC_2) \cU_{\rm D5}(\cC_2') 
\rangle ^2 
&= \langle \cU_{\rm D1}(\cC_2) \cU_{\rm D5}(\cC_2') 
\rangle ^{(N \ {\rm mod} \ 2)}\ . 
\end{align}
In particular, for odd $N$ we see that
the correlator 
$\langle \cU_{\rm D5}(\cC_2) \cU_{\rm NS5}(\cC_2') 
\rangle$ cannot be trivial.

\begin{figure}
  \centering
\includegraphics[width=9cm]{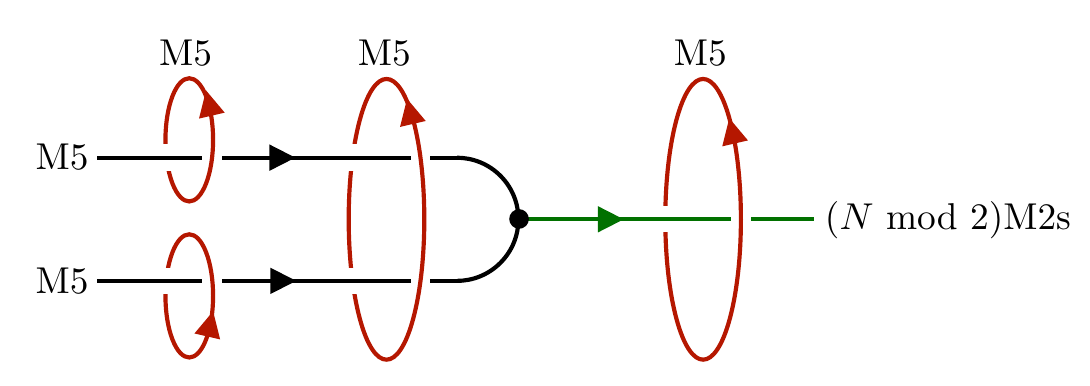}
\caption{ The analog of Figure \ref{fig_strings_annihilate} in $\AdS_7 \times \RP^4$.
All M5-branes in the figure wrap $\RP^3 \subset \RP^4$
and extend along some 3-cycle in external 7d spacetime. The 3-cycles represented by the horizontal black and green lines link with the 3-cycles represented by red loops.
}
        \label{fig_membranes_annihilate}
\end{figure}

\paragraph{Fat membranes in $\AdS_7 \times \RP^4$.}
By ``fat membrane'' in $\AdS_7 \times \RP^4$ we mean an M5-brane wrapping $\RP^3 \subset \RP^4$.
The discussion here is analogous to the previous paragraphs.
Following steps similar to those in \cite{Witten:1998xy}, we can argue that a 4-chain $\cB_4$ exists
with the following properties:
$\cB_4$ is a bordism between two copies of $\RP^3$
with the same orientation, and the empty set;
$\cB_4$ is threaded by the background $G_4$ flux.
In fact, we have already encountered
$\cB_4$ earlier in this section,
in the discussion of the
$\widehat {\bm Q}_3 \widehat {\bm Q}_3$ commutator. As remarked around
\eqref{eq:flux_bk}, the effective flux on 
$\cB_4$ is $N$. Thus, the existence of $\cB_4$ implies the existence
of a `vertex' among 3d  operators in 7d external spacetime,
as depicted schematically in Figure~\ref{fig_membranes_annihilate}.\footnote{ \ Similarly to the fat string case, this is due to the fact that a flux for $G_4$ induces 
a coupling $N \int b_2$, where $b_2$ is the chiral 2-form on the M5-brane worldvolume. Open  M2-branes that end on 2d surfaces inside the M5-brane
are also charged under $b_2$ in the same way, and are needed to cancel the induced charge.} 

Similarly to  the $\AdS_5 \times \RP^5$ case,
the existence of the `vertex'
between fat membranes and M2-branes imposes  constraints on the correlators
among 3d operators in 7d. 
A pictorial derivation of these constraints
is furnished in Figure \ref{fig_membranes_annihilate}.
In particular, for $N$ odd
we see that the 
$\langle \text{M5} \text{M5} \rangle$
correlator cannot be trivial.
This is reflected in the fact that 
\eqref{eq:3formCS} does not simply consist of a single BF term.

\subsection{Cubic coupling from M-theory effective action}
\label{sec_cubic_derivation}

In this section we discuss the M-theory origin of the cubic coupling
$\mathsf a_1 \cup \mathsf c_3 \cup \mathsf c_3$
 in the 7d action
\eqref{eq:7d_action_summary}. 
Our starting point is the following
expansion of $\breve G_4$ onto
twisted
torsional cohomology classes on 
$\RP^4$.
The relevant terms are
\be \label{eq:torsional_ansatz}
\breve G_4 = \left( N - \tfrac 12 \right) \breve v_4
+ \breve c_3 \cdot \breve t_1
+ \breve a_1 \cdot \breve t_1^3 + \dots \ . 
\ee 
Throughout this section, we turn off
the $SO(5)$ gauge fields
associated to the R-symmetry
of the 6d (2,0) SCFT.
The term $( N - \tfrac 12 ) \breve v_4$ encodes the background flux.
The objects $\breve c_3$, $\breve a_1$ are untwisted differential
cohomology classes in external spacetime, with degrees given 
by their subscripts.
The corresponding internal classes
$\breve t_1$, $\breve t_1^3$
are both torsional of degree 2.
As a result,
$\breve a_1$ models a $\mathbb Z_2$
1-form gauge field,
and $\breve c_3$ models
a $\mathbb Z_2$ 3-form gauge field.
We have already encountered
$\breve c_3$ in section \ref{sec:extractCS},
see \eqref{eq:origin_of_c3b3}, where it was denoted $c_3$ because
we were adopting the language of differential forms.
The external gauge field
$\breve a_1$,
 on the other hand,
is associated to the
$\mathbb Z_2$ outer automorphism
of the  Lie algebra $D_N$.

Plugging \eqref{eq:torsional_ansatz} into $- \tfrac 16 \breve G_4^3$
and collecting terms with total internal degree equal to 5,
we obtain
\be 
- \frac 16 \int_{M_7 \times \RP^4} \breve G_4^3
=   \bigg[ \int_{M_7} \breve a_1  \cdot \breve c_3 \cdot  \breve c_3  \bigg] 
\bigg[ \frac 12 \int_{\RP^4}
\breve t_1^5 \bigg] \ . 
\ee 
The integral over $\RP^4$ is a secondary invariant
valued in $\mathbb R/ \mathbb Z$.
In fact, it is equal to the same
quadratic refinement of the linking
pairing between $t_1$ and $t_1^4$
as in \eqref{eq:quad_refin}.  
Selecting the sign $s=1$,
we have the sought-form cubic coupling in the 7d TQFT,
\be  \label{eq:result_cubic}
S_\text{7d TQFT}^\text{cubic} =2\pi\int_{M_7} 
 \frac 14 \breve a_1 \cdot  \breve  c_3  \cdot \breve c_3 
=2\pi\int_{M_7} 
 \frac 14 \mathsf a_1 \cup  \mathsf  c_3  \cup \mathsf  c_3   
 \ . 
\ee 
In the second step, we 
have observed that we are computing a primary invariant in differential cohomology
and we have rephrased it in terms
of ordinary cohomology classes
$\mathsf a_1\in H^2(M_7;\mathbb Z_2)$, $\mathsf c_3 \in H^3(M_7;\mathbb Z_2)$.

Let us remark that 
the quantity
$- \tfrac 16 \int_{M_{11}} \breve G_4^3$, 
which is written in the language of differential
cohomology, captures the familiar
topological coupling $-\tfrac 16 \int_{M_{11}} C_3 \wedge G_4 \wedge G_4$ in the low-energy M-theory effective action.
This coupling 
in general is not well-defined by itself. Rather, the consistency of the 11d effective action relies on a subtle interplay between the two-derivative
coupling $- \tfrac 16 \int_{M_{11}} C_3 \wedge G_4 \wedge G_4$,
the higher-derivative coupling
$-  \int_{M_{11}} C_3 \wedge X_8$
(see \eqref{eq:X8def} for the definition of $X_8$),
and the gravitino measure in the path integral
\cite{Witten:1996md,Freed:2019sco}.
In the derivation
given above, leading to \eqref{eq:result_cubic},
we have only taken into account the two-derivative term. We believe that 
\eqref{eq:result_cubic} captures non-trivial
physics of the 7d topological theory we want to study, but we also expect that a more refined derivation should be possible. In particular,
it should be feasible to give a first-principle
argument for the   quadratic refinement 
underpinning the interpretation of the
$\mathsf a_1 \mathsf c_3 \mathsf c_3$ coupling,
as per our proposal discussed in Section \ref{sec_this_is_F7}. We leave a systematic
investigation of these points for future research.

\subsection{Branes and non-invertible topological operators}\label{sec:noninvertiblah}

As already noted in Section \ref{sec_this_is_F7},
the cubic coupling
$\mathsf a_1 \cup \mathsf c_3 \cup \mathsf \mathsf c_3$
in \eqref{eq:7d_action_summary}
implies that the naive holonomies
$e^{i \pi \int_{\Sigma_3} \mathsf  b_3}$
and $e^{i \pi \int_{\Sigma_5} \mathsf  a_5}$
are not well-defined topological
operators of the 7d TQFT.
Field-theoretically, this can be seen  from
two equivalent points of view, cfr.~\cite{Kaidi:2021xfk,Kaidi:2023maf}:
the cubic coupling requires a modification of the gauge transformations of $\mathsf b_3$, $\mathsf a_5$;
the cubic coupling induces anomalies
on the worldvolumes of the naive
holonomy defects $e^{i \pi \int_{\Sigma_3} \mathsf  b_3}$
and $e^{i \pi \int_{\Sigma_5} \mathsf  a_5}$.
In this section, we discuss 
the derivation
of the topological dressing
factors $T_3(\Sigma_3; \mathsf a_1, \mathsf c_3)$, $T_5(\Sigma_5 ; \mathsf c_3)$
in \eqref{eq:top_ops_all} from the M5-brane
worldvolume theory.

The brane origin of the operators
$\bm Q_3(\Sigma_3)$, $\widehat{\bm Q}_3(\Sigma_3)$, $\bm Q_1(\Sigma_1)$,
$\widehat {\bm Q}_5(\Sigma_5)$
is summarized in Table \ref{table:summary}.
We observe that the operators requiring a topological dressing are precisely those originating from M5-branes. In fact, we can identify the origin  of 
the topological dressing in
the chiral tensor living on the worldvolume of the M5-brane, and its coupling to the bulk~$G_4$.

It is notoriously challenging to write down an effective action for the chiral 2-form and its coupling to $G_4$. In what follows, we adopt a strategy building on \cite{Bah:2020jas,Apruzzi:2023uma} which aims at capture some robust topological aspects of the physical system at hand.
The usefulness of this approach for concrete computations has been tested
in several examples in 
\cite{Bah:2020jas,Apruzzi:2023uma}.\footnote{ \ We refer the reader to \cite{Belov:2006jd,Belov:2006xj,Hsieh:2020jpj}, building on \cite{Witten:1996hc,Witten:1999vg},
for a rigorous 
treatment of chiral fields
via an auxiliary topological action in one dimension higher.
}
The main point of this approach is to use an auxiliary topological action in one dimension higher, understood as a functional of the field strengths in the problem, whose variation formally reproduces the relevant Bianchi identities.
In the present context, the key observation is that 
the field strength $h_3$ of the chiral 2-form on the M5-brane satisfies
$dh_3 = G_4$
\cite{Townsend:1995af}, \cite[around eq.~(2.8)]{Witten:1995em}, \cite{Witten:1996hc}.
This modified Bianchi identity for $h_3$ can be
obtained from the following auxiliary topological 7d action,
\be 
\label{eq_aux_7d_action}
\int_{W_7} \bigg[ 
\frac 12 h_3 \wedge dh_3 - h_3 \wedge G_4 \bigg] \ , 
\ee 
upon variation of the action with respect to $h_3$. Here we assume for simplicity
that the worldvolume of the M5-brane can be written as $\partial W_7$ for $W_7$ an auxiliary manifold. 
Heuristically,
the term $\frac 12 h_3 dh_3$ is a proxy for the 6d kinetic term
of the chiral 2-form.
It serves a function analogous to the quantity
$\frac 12 E_5 dE_5$ in \cite{Bah:2020jas}, which captures the chiral 4-form of Type IIB string theory.
The second term $h_3G_4$ is a proxy for a 6d coupling of the schematic form
$h_3C_3$.

We can further motivate  
\eqref{eq_aux_7d_action} building on the analysis of \cite{Hsieh:2020jpj}.
As stated in Section 5.1 of that reference, 
the relevant 7d action for studying a 6d chiral 2-form 
can be written as (we identify their $\mathsf A_A$ with $h_3$)
\be 
\int_{W_7} \bigg[ \frac 12 h_3 \wedge dh_3 + \mathsf w \wedge h_3 \bigg] \ , \qquad 
\mathsf w = - \frac 14 p_1(W_7) \ .
\ee 
As noted in Section \ref{sec:mc_and_RP4}, in the  spin setting
$\frac 12 \lambda = \frac 14 p_1$ is the canonical integral lift of $w_4$.
The integral lift of $w_4$ governs the shifted quantization condition for $G_4$. Thus, modulo $\mathbb Z$, $G_4$ and $-\mathsf w$ are equal.
We notice that we are working in the background
$\AdS_7 \times \RP^4$,
which is not spin, but that here the focus is on the manifold $W_7$.
As explained in greater detail below, we are interested in $W_7$ of the form $W_4 \times \RP^3$ or
$W_6 \times \RP^1$.
We observe that both $\RP^3$ and $\RP^1$ are spin
(see Appendix \ref{app:killing}).
Thus, assuming $W_4$, $W_6$ are spin, we can work in the setting in which $W_7$ is spin.

For our discussion below, we find it useful to recast 
\eqref{eq_aux_7d_action} in the notation of differential cohomology.
In particular, 
since the term $\frac 12 h_3 dh_3$ has the structure of a Chern-Simons term for a 3-form potential, we consider 
a differential cohomology class $\breve h_4$ of degree 4. The field strength of $\breve h_4$ is $dh_3$. We rewrite 
\eqref{eq_aux_7d_action} as
\be 
\int_{W_7} \bigg[ 
\frac 12 \breve h_4 \cdot \breve h_4
- \breve h_4 \cdot \breve G_4
\bigg] \ .
\ee 
This formulation is best suited for expansion onto torsional cycles, to which we now turn.\footnote{ \ The term $\frac 12 \breve h_4 \cdot \breve h_4$ is more precisely  understood as a quadratic refinement. For our purposes, we do not need to study the latter closely. In the expansion below, we are only interested in cross-terms originating from $\frac 12 \breve h_4 \cdot \breve h_4$.}

Let us consider the operator
$\widehat{\bm Q}_3(\Sigma_3)$,
originating from an M5-brane on 
$\Sigma_3 \times \RP^3$.
We are led to consider
$W_7 = W_4 \times \RP^3$ where
$\partial W_4 = \Sigma_3$.
The twisted cohomology classes
on $\RP^3$
that
we can used to expand $\breve h_4$
are $t_1$ and $t_1^3$.
We can then write
\be 
\breve h_4 = \breve \phi_3 \cdot \breve t_1 + \breve \phi_1 \cdot
\breve t_1^3 \  . 
\ee 
The expansion of the relevant terms in $\breve G_4$ 
is as in \eqref{eq:torsional_ansatz}, repeated here for convenience,
\be 
\breve G_4 = \breve c_3 \cdot t_1
+ \breve a_1 \cdot \breve t_1^3 \ . 
\ee 
The integral over $\RP^3$ is saturated by collecting terms with a $\breve t_1^4$ factor
and using $\int_{\RP^3} \breve t_1^4 = \frac 12$ mod $\mathbb Z$. We obtain
\be 
\int_{W_7} \bigg[ 
\frac 12 \breve h_4 \cdot \breve h_4
- \breve h_4 \cdot \breve G_4
\bigg]
= \int_{W_4} \bigg[ 
 \frac 12 \breve \phi_1 \cdot \breve \phi_3
+ \frac 12 \breve \phi_3 \cdot \breve a_1
 -  \frac 12 \breve \phi_1 \cdot 
 \breve c_3
\bigg] \ . 
\ee 
Recall $\partial W_4 = \Sigma_3$.
We can write the final action directly as an integral over
$\Sigma_3$, as a localized BF
term coupled to the bulk fields
$a_1$, $c_3$, arriving at\footnote{ \ Here we are dealing with discrete gauge fields originating from expansion of $\breve h_4$, $\breve G_4$ onto torsional elements in cohomology of degree 2. In this case, there is no 1/2 factor when we translate into expressions written with cocycles/cochains.
We refer the reader to \cite[sec.~2.3]{Apruzzi:2021nmk} for further explanations. This is to be contrasted with the case of a BF coupling at level $k$ written with continuous $U(1)$ fields, and subsequently translated into cochains. In this case, we have $1/k$ factors. For instance, we have seen this above in  \eqref{eq:Abelian_3form_CS} compared to   \eqref{eq:7d_action_summary}.}
 \be  \label{eq:T3dressing}
T_3(\Sigma_3; \mathsf a_1, \mathsf c_3) \sim \sum_{\phi_0, \phi_2} \exp 2\pi i  
\int_{\Sigma_3} 
\bigg[ 
\frac 12 \phi_0 \cup \delta \phi_2
+ \frac 12 \phi_2 \cup \mathsf a_1 
 - \frac 12  \phi_0 \cup \mathsf c_3
\bigg]  \ .
\ee 
The sum on $\phi_p$
is over $\mathbb Z_2$ $p$-cochains
on $\Sigma_3$.

By a similar token we can
analyze the operator
$\widehat {\bm Q}_5(\Sigma_5)$,
originating from an M5-brane on $\RP^1$. 
Thus, we consider
$W_7 = W_6 \times \RP^1$
with $\partial W_6= \Sigma_5$.
In this case, the only twisted cohomology class that we can use to expand $\breve h_4$ is $\breve t_1$, thus
we write
\be 
\breve h_4 = \breve \varphi_3 \cdot \breve t_1 \ . 
\ee 
Our task is to collect terms
with $\breve t_1^2$ to saturate the
$\RP^1$ integral,
using $\int_{\RP^1} \breve t_1^2 = \frac 12$ mod $\mathbb Z$.
With the same $\breve G_4$ expansion as before, we compute 
\be 
\int_{W_7} \bigg[ 
\frac 12 \breve h_4 \cdot \breve h_4
- \breve h_4 \cdot \breve G_4
\bigg] 
= \int_{W_6} \bigg[ 
- \frac 14 \breve \varphi_3 \cdot \breve \varphi_3
+ \frac 12 \breve \varphi_3 \cdot 
\breve c_3
\bigg] \ . 
\ee 
(The $\breve h_4$ squared term
involves a quadratic refinement of the secondary invariant $\int_{\RP^1} \breve t_1^2 = \frac 12$ mod $\mathbb Z$,
which is $\pm \frac 14$ mod $\mathbb Z$. A more careful analysis would be needed to fix the sign, but we do not need it for the present purposes. We proceed selecting the plus sign.)
Upon rewriting the action on $\Sigma_5$, we arrive at 
 \be \label{eq:T5dressing}
T_5(\Sigma_5; \mathsf c_3) \sim \sum_{ \varphi_2} \exp 2\pi i  
\int_{\Sigma_5} 
\bigg[ 
-\frac 14 \varphi_2 \cup \delta \varphi_2
 + \frac 12 \varphi_2 \cup \mathsf c_3 
\bigg]  \ .
\ee 
The field $\varphi_2$ is localized on $\Sigma_5$.

\paragraph{Comments on $\frac 14  \varphi_2 \cup \delta \varphi_2$.}
In writing the term $\frac 14 \varphi_2 \cup \delta \varphi_2$, we are using a schematic notation.
More properly, it is understood as in \cite[app.~C]{Gukov:2020btk}, see also \cite{Kapustin:2017jrc}. 
Below, we consider $\mathbb Z_N$ cochain for generic even $N$.  
The case of interest is $N=2$.

The formula for the Pontryagin square at the level of cochains reads
\be \label{eq_Pontryagin}
\mathfrak P(\varphi_2) = \varphi_2 \cup \varphi_2 - \delta \varphi_2 \cup_1 \varphi_2 \ . 
\ee 
The Pontryagin square of a $\mathbb Z_N$ cochain
is a $\mathbb Z_{2N}$ cochain. The quantity $\varphi_2$ is understood as an integral lift of a $\mathbb Z_N$ chain,
so that $\delta \varphi_2$ is generically non-zero but it is divisible by $N$.
Here $\cup_1$ is the first   Steenrod's higher cup product
\cite{steenrod1947products}
(see e.g.~\cite{Kapustin:2014gua,Benini:2018reh} for expositions aimed at physicists).
From \eqref{eq_Pontryagin}, acting with $\delta$, we can write
\be \label{eq_rewriting}
\varphi_2 \cup \frac{\delta \varphi_2}{N}
=\frac{1}{2N} \delta \mathfrak P(\varphi_2)
- \frac N2  \frac{\delta \varphi_2}{N} \cup_1 
\frac{\delta \varphi_2}{N}
\ .
\ee 
In the above expression, $\delta/N$ on the RHS is interpreted as  the Bockstein homomorphism
 associated to the short exact sequence $0 \rightarrow \mathbb Z \xrightarrow{N} \mathbb Z \rightarrow \mathbb Z_N \rightarrow 0$.
The advantage of the rewriting \eqref{eq_rewriting} stems from the fact that the RHS consists of an exact piece, plus a quantity that can be expressed using Bocksteins and is well-defined modulo $N$. Thus, 
we use the integral of the RHS on a 5-cycle to define the corresponding integral of the LHS,
\be \label{eq_better_def}
\exp 2\pi i \frac 1N \int_{\Sigma_5} \varphi_2 \cup \frac{\delta \varphi_2}{N}
= \exp \pi i \int_{\Sigma_5} 
\frac{\delta \varphi_2}{N} \cup_1 
\frac{\delta \varphi_2}{N} 
\ . 
\ee 
Specializing to $N=2$, we have the proper definition
of $\exp 2\pi i \int_{\Sigma_5} \frac 14 \varphi_2 \cup \delta \varphi_2$ that enters
the action
\eqref{eq:T5dressing}
of $T_5$.

\subsection{Hanany-Witten effect} \label{sec_HW}

The M2- and M5-branes in the setup 
we are studying can undergo non-trivial
Hanany-Witten transitions \cite{Hanany:1996ie}.
In this section, we describe this effect and   its implications for the topological operators
$\bm Q_3(\Sigma_3)$, $\widehat{\bm Q}_3(\Sigma_3)$, $\bm Q_1(\Sigma_1)$,
$\widehat {\bm Q}_5(\Sigma_5)$
of the 7d TQFT~$\mathscr T_{\mathrm{7d}}^{D_N}$.
The Hanany-Witten effect has been shown
to play a crucial role in many string/M-theory constructions
where ``branes at infinity'' 
realize generalized symmetries, see e.g.~\cite{Apruzzi:2022rei,Heckman:2022xgu,Apruzzi:2023uma}.

\begin{figure} \centering
\includegraphics[width=5cm]{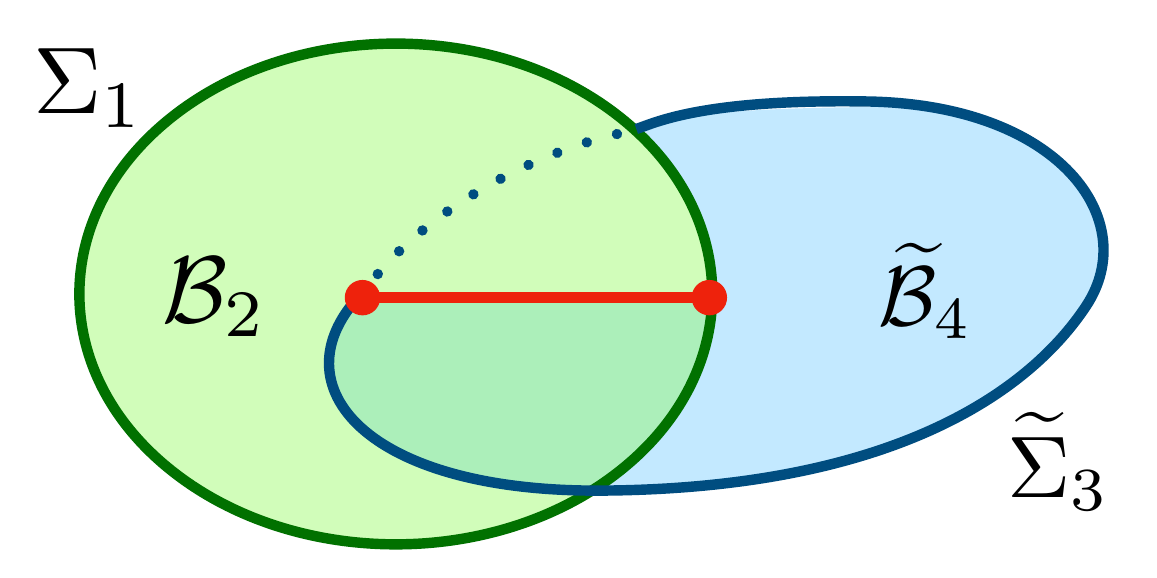}
\caption{Configuration of M5-branes undergoing a Hanany-Witten transition as in \eqref{eq_HW}. Both branes share two common spacetime directions, suppressed in the figure. In the remaining five external directions $M_5$, one M5-brane wraps $\Sigma_1 = \partial \cB_2$, the second wraps $\widetilde \Sigma_3 = \partial \widetilde{\cB}_4$. We imagine that, at first, the two branes were un-linked (far apart with disjoint $\cB_2$, $\widetilde \cB_4$).
Upon linking them,
an M2-brane is created, which lies along the red segment, i.e.~the intersection of $\cB_2$ and $\widetilde {\cB}_4$ in $M_5$.
}
\label{fig_HW}
\end{figure}

Table \ref{table:summary} shows that the relevant M5-branes for the problem
at hand are those 
wrapping  $\RP ^3$ and $\RP ^1$ inside
$\RP^4$. We claim that such M5-branes
can undergo a Hanany-Witten transition,
as specified by the brane configuration
described in the table below,
\begin{equation} \label{eq_HW}
    \begin{tabular}{|c|| >{\centering}p{4mm} | >{\centering}p{4mm} | >{\centering}p{4mm} | 
    >{\centering}p{4mm}
    | >{\centering}p{4mm} |
    >{\centering}p{4mm}
    |
    >{\centering}p{4mm}
    |
    >{\centering\arraybackslash}p{24mm}
    |}
       \hline 
       \rule{0pt}{9pt}Brane & $x^0$ & $x^1$ &$x^2$ &$x^3$ & $x^4$ & $x^5$ & $x^6$ & $\RP^4$    \\\hline \hline
       \rule{0pt}{10.pt}M5  & $\mathsf X$ &$\mathsf X$  &$\mathsf X$ &  &   &   &   & $\RP^3$   \\   \hline
       \rule{0pt}{10.pt} M5'  & $\mathsf X$ &$\mathsf X$  &   &$\mathsf X$ & $\mathsf X$&$\mathsf X$ &  & $\RP^1$    \\ \hline
       \rule{0pt}{10.pt}M2  & $\mathsf X$ & $\mathsf X$ &  &    &  & & $\mathsf X$ & $\widetilde{\mathrm{pt}}$ \\ \hline
    \end{tabular}
\end{equation}
This version of the Hanany-Witten effect is slighly non-standard, in that the branes involved wrap torsional cycles in the internal geometry $\RP^4$. Nonetheless, we can argue for
\eqref{eq_HW} using an argument based on linking as in the original paper \cite{Hanany:1996ie}. More precisely, we find it convenient to make use of 
the Bianchi identities of $G_4$ and its dual $G_7$, treating an M5-brane as a magnetic source for the former, and an M2-brane as a magnetic source for the latter
\cite{Marolf:2000cb,Apruzzi:2023uma}.
The Bianchi identities read
\be 
dG_4= J_5^{\rm M5} \ , \qquad 
dG_7 = \frac 12 G_4^2 + X_8 + J_8^{\rm M2} \ .
\ee 
Here $J_5^{\rm M5}$ is a delta-function supported 5-form
 that describes
the insertion of an M5-brane. Similarly $J_8^{\rm M2}$ describes the insertion of an M2-brane. 
The quantity $X_8$ is a closed 8-form constructed with the Pontryagin classes of the tangent bundle to 11d spacetime,
see \eqref{eq:X8def} and Section \ref{sec:DtypeANOMAL} below.
From $ddG_4 =0 =ddG_7$
we see that
\be \label{eq_current_Bianchi}
dJ_5^{\rm M5} = 0 \ , \qquad 
dJ_8^{\rm M2} = - G_4 \wedge J_5^{\rm M5} \ .
\ee 
Let us write 11d spacetime as $M_{11} = M_2 \times M_5 \times \RP^4$, where we have singled out two directions in external 7d spacetime. We consider an M5-brane with worldvolume
$M_2 \times \Sigma_1 \times \RP^3$,
where $\Sigma_1$ is a 1-cycle in $M_5$.
The associated 5-form current reads\footnote{ \ The quantity $t_1$ is best understood here as a differential cohomology class, but for notational simplicity we use the wedge product of forms.}
\be 
J_5^{\rm M5} = \delta_4(\Sigma_1 \subset M_5) \wedge t_1 \ , \qquad
t_1 = {\rm PD}_{\RP^4}[\RP^3] \ . 
\ee 
We assume that $\Sigma_1$ is homologically trivial in $M_5$,
$\Sigma_1  = \partial \cB_2$, so that 
\be 
\delta_4(\Sigma_1 \subset M_5) =  d \delta_3(\cB_2 \subset M_5) \ .
\ee 
This allows us to write an expression for the $G_4$ flux generated by this M5-brane  (up to closed terms, i.e.~a homogeneous solution)
\be \label{eq_G4_sourced}
G_4 = \delta_3(\cB_2 \subset M_5) \wedge t_1 \ . 
\ee 
Next, we focus on the other M5-brane, which has worldvolume
$M_2 \times \widetilde \Sigma_3 \times \RP^1$,
with $\widetilde \Sigma_3$ a 3-cycle in $M_5$.
Its current reads
\be 
J_5^{\rm M5'} = \delta_2(\widetilde \Sigma_3 \subset M_5) \wedge t_3 \ , \qquad 
t_3 = {\rm PD}_{\RP^4}[\RP^1] \ . 
\ee 
The product of $G_4$ in \eqref{eq_G4_sourced} (the flux sourced by the first M5-brane) and $J_5^{\rm M5'}$
is a measure of the linking between the two M5-branes, 
\be  \label{eq_flux_and_link}
G_4 \wedge J_5^{\rm M5'} = {\rm Link}_{M_5}(\Sigma_1 , \widetilde \Sigma_3) {\rm vol}_{M_5} \wedge {\rm PD}_{\RP^4}[\widetilde {\mathrm {pt}}] \ .
\ee 
Here, we have used
\be 
\ba 
\delta_3(\cB_2 \subset M_5) \wedge \delta_2(\widetilde \Sigma_3 \subset M_5)
&= (\cB_2 \cdot _{M_5} \widetilde \Sigma_3) {\rm vol}_{M_5}
= {\rm Link}_{M_5}(\Sigma_1 , \widetilde \Sigma_3) {\rm vol}_{M_5} \ , \\ 
t_1 \wedge t_3 & = t_4 = {\rm PD}_{\RP^4}[\widetilde {\mathrm {pt}}] \ .
\ea 
\ee 
From \eqref{eq_flux_and_link} and \eqref{eq_current_Bianchi} we see that, if we change the linking of the M5-branes in the $M_5$ directions,
the quantity $dJ_8^{\rm M2}$ has to change, too. This encodes the fact that an M2-brane is generated, stretching between the two M5-branes. 
We can see this explicitly, as follows.
Let us write $\widetilde \Sigma_3 = \partial \widetilde \cB_4$
inside $M_5$. Then,
\be 
J_5^{\rm M5'} = d\delta_1(\widetilde \cB_4 \subset M_5) \wedge t_3 \ , 
\ee 
and as a result (we do not keep track of signs)
\be 
G_4 \wedge J_5^{\rm M5'}= d\Big( \delta_3(\cB_2\subset M_5) \wedge \delta_1(\widetilde \cB_4 \subset M_5) \wedge t_4  \Big) \ . 
\ee 
This is compatible with
\be 
J_8^{\rm M2} = \delta_3(\cB_2\subset M_5) \wedge \delta_1(\widetilde \cB_4 \subset M_5) \wedge t_4  \ , 
\ee 
which indicates that an M2-brane is created, which is supported on the intersection of $\cB_2$ and $\widetilde \cB_4$ in $M_5$. As depicted in Figure \ref{fig_HW}, this intersection
is a segment connecting $\Sigma_1$ and $\widetilde \Sigma_3$.
We also see   that the internal profile of the M2-brane that is 
$\widetilde {\mathrm{pt}}$, Poincar\'e dual in $\RP^4$ to $t_4$.
In summary, we have confirmed the process
 described in 
\eqref{eq_HW}.

Now, let us interpret \eqref{eq_HW} in light of the
topological operators of the 7d TQFT
$\mathscr T_{\mathrm{7d}}^{D_N}$. Making use of Table \ref{table:summary},
it is straightforward to translate \eqref{eq_HW} into 
\begin{equation} \label{eq_7d_moves}
    \begin{tabular}{|c|| >{\centering}p{4mm} | >{\centering}p{4mm} | >{\centering}p{4mm} | 
    >{\centering}p{4mm}
    | >{\centering}p{4mm} |
    >{\centering}p{4mm}
    |
    >{\centering\arraybackslash}p{4mm}
    |}
       \hline 
       \rule{0pt}{9pt}Top.~op. & $x^0$ & $x^1$ &$x^2$ &$x^3$ & $x^4$ & $x^5$ & $x^6$    \\\hline \hline
       \rule{0pt}{10.pt} \rule[-2mm]{0mm}{6.5mm} $\widehat{\bm Q}_3$  & $\mathsf X$ &$\mathsf X$  &$\mathsf X$ &  &   &   &      \\   \hline
       \rule{0pt}{10.pt} \rule[-2mm]{0mm}{6.5mm} $\widehat{\bm Q}_5$  & $\mathsf X$ &$\mathsf X$  &   &$\mathsf X$ & $\mathsf X$&$\mathsf X$ &       \\ \hline
       \rule{0pt}{10.pt} \rule[-2mm]{0mm}{6.5mm} ${\bm Q}_3$  & $\mathsf X$ & $\mathsf X$ &  &    &  & & $\mathsf X$   \\ \hline
    \end{tabular}
\end{equation}
In other words, if two topological operators
$\widehat{\bm Q}_3$ and $\widehat{\bm Q}_5$ share two
spacetime directions, and we change their linking in the remaining five directions,
 then an operator
${\bm Q}_3$ is created/annihilated, stretching between them.

Next, we describe another kind of Hanany-Witten effect that can
occur in the setup of interest. It is described in the table below,
\begin{equation} \label{eq_HW_other}
    \begin{tabular}{|c|| >{\centering}p{4mm} | >{\centering}p{4mm} | >{\centering}p{4mm} | 
    >{\centering}p{4mm}
    | >{\centering}p{4mm} |
    >{\centering}p{4mm}
    |
    >{\centering}p{4mm}
    |
    >{\centering\arraybackslash}p{24mm}
    |}
       \hline 
       \rule{0pt}{9pt}Brane & $x^0$ & $x^1$ &$x^2$ &$x^3$ & $x^4$ & $x^5$ & $x^6$ & $\RP^4$    \\\hline \hline
       \rule{0pt}{10.pt}M5  & $\mathsf X$ &$\mathsf X$  &$\mathsf X$ &  &   &   &   & $\RP^3$   \\   \hline
       \rule{0pt}{10.pt} M5'  &   &   &   &$\mathsf X$ & $\mathsf X$&$\mathsf X$ &  & $\RP^3$    \\ \hline
       \rule{0pt}{10.pt}M2  &   &   &  &    &  & & $\mathsf X$ & $\RP^2$ \\ \hline
    \end{tabular}
\end{equation}
This effect can be derived using the same logic as above.
The key point is ${\rm PD}_{\RP^4}[\RP^3] = t_1$ and that $t_1^2  = t_2 = {\rm PD}_{\RP^4}[\RP^2]$. This is why the M2-brane that gets generated (third line in the table)
wraps $\RP^2$. Making use of Table \ref{table:summary},
we translate \eqref{eq_HW_other} into a property of the topological operators of the 7d TQFT $\mathscr T_{\mathrm{7d}}^{D_N}$,
\begin{equation} \label{eq_7d_moves_other}
    \begin{tabular}{|c|| >{\centering}p{4mm} | >{\centering}p{4mm} | >{\centering}p{4mm} | 
    >{\centering}p{4mm}
    | >{\centering}p{4mm} |
    >{\centering}p{4mm}
    |
    >{\centering\arraybackslash}p{4mm}
    |}
       \hline 
       \rule{0pt}{9pt}Top.~op. & $x^0$ & $x^1$ &$x^2$ &$x^3$ & $x^4$ & $x^5$ & $x^6$    \\\hline \hline
       \rule{0pt}{10.pt} \rule[-2mm]{0mm}{6.5mm} $\widehat{\bm Q}_3$  & $\mathsf X$ &$\mathsf X$  &$\mathsf X$ &  &   &   &      \\   \hline
       \rule{0pt}{10.pt} \rule[-2mm]{0mm}{6.5mm} $\widehat{\bm Q}_3$  &   &   &   &$\mathsf X$ & $\mathsf X$&$\mathsf X$ &       \\ \hline
       \rule{0pt}{10.pt} \rule[-2mm]{0mm}{6.5mm} ${\bm Q}_1$  &   &   &  &    &  & & $\mathsf X$   \\ \hline
    \end{tabular}
\end{equation}
In other words, if we change the linking of two topological operators of type
$\widehat{\bm Q}_3$ in 7d spacetime,
then an operator
${\bm Q}_1$ is created/annihilated, stretching between them.

Notice that in \eqref{eq_7d_moves} and \eqref{eq_7d_moves_other}
we have not yet made a choice of which direction is identified with the radial/interval
directions in the sandwich construction. Indeed,
as emphasized in \cite{Apruzzi:2023uma}, the same physical effect
\eqref{eq_7d_moves} or \eqref{eq_7d_moves_other} in the SymTFT bulk can lead to different phenomena regarding global symmetries depending on which direction is the interval direction and on the choice of gapped boundary conditions (symmetry boundary) in the sandwich construction. A more explicit discussion of this point can be found below in Section \ref{sec:application}.

\subsubsection{Some remarks on intersecting branes}
\label{sec_cubic_and_branes}

The same physics underlying the Hanany-Witten transitions
discussed in the previous section is also at play 
in setups with intersecting branes.
More precisely, in this section we analyze configurations in which
two M5-branes wrapping $\RP^3$ or $\RP^1$ are placed inside a common codimension-one subspace $M_6$ of external spacetime $M_7$.
The M5-branes intersect in $M_6$. We argue that consistency requires the presence of an M2-brane (wrapping $\widetilde {\mathrm{pt}}$ or 
$\RP^2$) that emanates from the intersection.
Similar setups have been studied in \cite{Apruzzi:2023uma}.

\begin{figure} \centering
\begin{tikzpicture}[      
        every node/.style={anchor=south west,inner sep=0pt},
        x=1mm, y=1mm,
      ]   
     \node (fig1) at (0,0)
       {\includegraphics[scale=0.6]{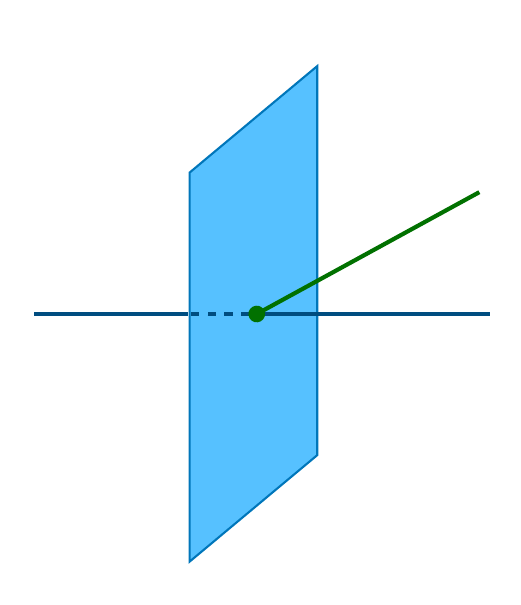}};
       \node (fig2) at (65,0)
       {\includegraphics[scale=0.6]{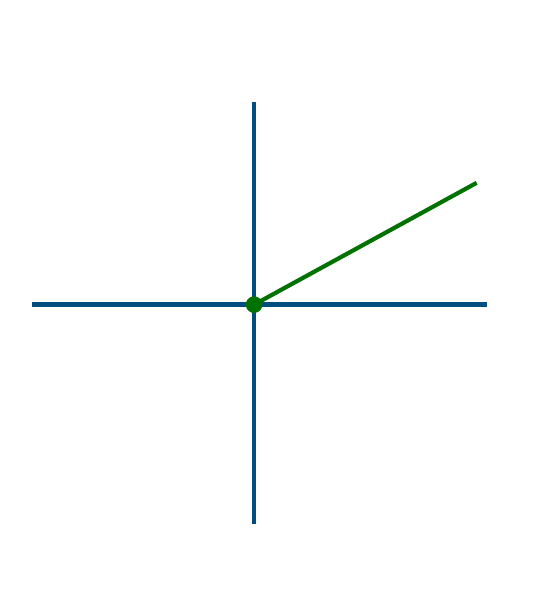}};
     \node   at (27,7) {$\small\text{M5 on $\Sigma_5 \times \RP^1$}$};
     \node   at (35,24) {$\small\text{M5 on $\widehat \Sigma_3 \times \RP^3$}$};
     \node   at (35,43) {$\small\text{M2 on $\Sigma_3 \times \widetilde{\rm pt}$}$};
     \node   at (65+32,24) {$\small\text{M5 on $\widehat \Sigma_3^{(1)} \times \RP^3$}$};
     \node   at (65+27,7) {$\small\text{M5 on $\widehat \Sigma_3^{(2)} \times \RP^3$}$};
     \node   at (65+32,43) {$\small\text{M2 on $\Sigma_1 \times \RP^2$}$};
\end{tikzpicture}
\caption{Some intersecting brane configurations. The supports $\Sigma_3$, $\widehat \Sigma_3$, $\Sigma_5$, $\widehat \Sigma_3^{(1,2)}$,
and $\Sigma_1$ are taken to lie inside a common 
codimension-one slice $M_6$ of external
spacetime $M_7$. On the left,
$\Sigma_5$ and $\widehat \Sigma_3$
generically intersect transversely along a
2d surface $\cS_2$ and $\partial \Sigma_3 = \cS_2$. On the right, $\widehat \Sigma_3^{(1)}$
and $\widehat \Sigma_3^{(2)}$ generically intersect transversely at a point $\cS_0$ and $\partial \Sigma_1 = \cS_0$.
}
\label{fig_branes}
\end{figure}
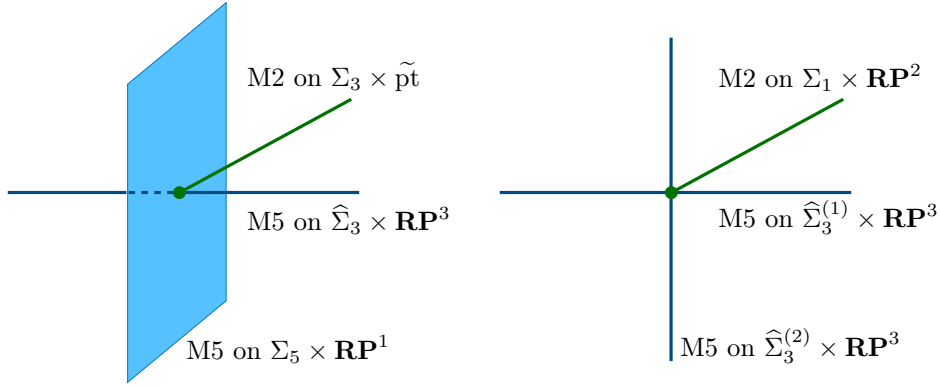

The first brane setup we consider is as follows.
Fix $M_6$ inside $M_7$ and place one M5-brane on $\Sigma_5 \times \RP^1$, and one M5-brane on $\widehat \Sigma_3 \times \RP^3$.
Here $\Sigma_5$, $\widehat \Sigma_3$ are submanifolds in $M_6$ in a generic configuration. Their intersection inside $M_6$ is a 2d surface $\cS_2$. This configuration is depicted in Figure \ref{fig_branes} on the left.

The M5-brane on $\Sigma_5 \times \RP^1$ sources a non-trivial profile for $G_4$,
\be 
dG_4 = \delta_5(\Sigma_5 \times \RP^1 \subset M_{11})
= \delta_1(\Sigma_5 \subset M_6) \wedge \delta_1(M_6 \subset M_7)
\wedge t_3 \ . 
\ee 
We have observed that the internal profile of the brane is
$\RP^1 \subset \RP^4$, which is Poincar\'e dual to $t_3$.
Within $M_6$, let us consider one of the half-spaces delimited by $\Sigma_5$. Let us denote it schematically as $\Sigma_5 \times \mathbb R_{\ge 0} \subset M_6$. We can write (up to closed terms)
\be \label{eq_G_flux_jump}
G_4 = \theta_0(\Sigma_5 \times \mathbb R_{\ge 0}  \subset M_6) \wedge \delta_1(M_6 \subset M_7)
\wedge t_3 \ ,
\ee 
where $\theta_0$ is a Heaviside-theta-like 0-form supported on
$\Sigma_5 \times \mathbb R_{\ge 0}$.
The relation 
\eqref{eq_G_flux_jump} shows that the M5-brane on $\Sigma_5 \times \RP^1$
induces a jump in $G_4$ flux within $M_6$.

The second M5-brane on $\widehat \Sigma_3 \times \RP^3$ detects
the jump in $G_4$ flux. This is because the M5-brane effective action includes a coupling of the form 
\be 
\int_{\widehat \Sigma_3 \times \RP^3} h_3 \wedge C_3 \ , 
\ee 
where $h_3$ denotes the field strength of the chiral 2-form on the M5-brane, satisfying
$dh_3 = G_4$ 
\cite{Townsend:1995af}, \cite[around eq.~(2.8)]{Witten:1995em}, \cite{Witten:1996hc}.
Consistency 
(charge conservation) requires
another effect to compensate the jump in flux. 
We now verify that this comes
precisely from an M2-brane
on $\Sigma_3\times \widetilde{\rm pt}$,
where $\Sigma_3$ is an open submanifold inside $M_6$ whose boundary is the intersection surface $\cS_2$ between the two M5-branes.
Indeed, 
recall that
the objects charged under the chiral 2-form with field strength $h_3$
are strings. 
They are identified with endpoints of M2-branes 
ending on the M5-brane.
The M2-brane 
on $\Sigma_3\times \widetilde{\rm pt}$ can thus induce a source in $h_3$ that compensates the jump in flux described above.
In order to reproduce the  internal profile 
on $\RP^4$ of the jump in flux \eqref{eq_G_flux_jump},
we take the 
 twisted point 
$\widetilde{\rm pt}$ to lie
inside the $\RP^3$ subspace of $\RP^4$ where the second M5-brane is located. In fact, 
the Poincar\'e dual of 
$\widetilde{\rm pt}$ inside
$\RP^3$
is $t^3$, matching
the internal profile of the flux jump.
In summary, we have argued that an M2-brane emanates from the
intersection of the two M5-branes within $M_6$, as depicted in Figure
\ref{fig_branes} on the left.

We now turn to a different brane configuration.
We still fix $M_6 \subset M_7$ and we place 
a first M5-brane on $\widehat \Sigma_3^{(1)} \times \RP^3$
and a second 
M5-brane on $\widehat \Sigma_3^{(2)} \times \RP^3$.
Here $\widehat \Sigma_3^{(1,2)}$ are 3d submanifolds
in $M_6$ at a generic configuration. They intersect at a point $\cS_0$.
See Figure \ref{fig_branes} on the right.

Using arguments analogous to the ones given above,
we can argue that an M2-brane must emerge from the intersection
of the two M5-branes. More precisely,
the M2-brane is placed on $\Sigma_1 \times \RP^2$,
where $\Sigma_1$ is an open line in $M_6$ with $\partial \Sigma_1 = \cS_0$.

\begin{figure}
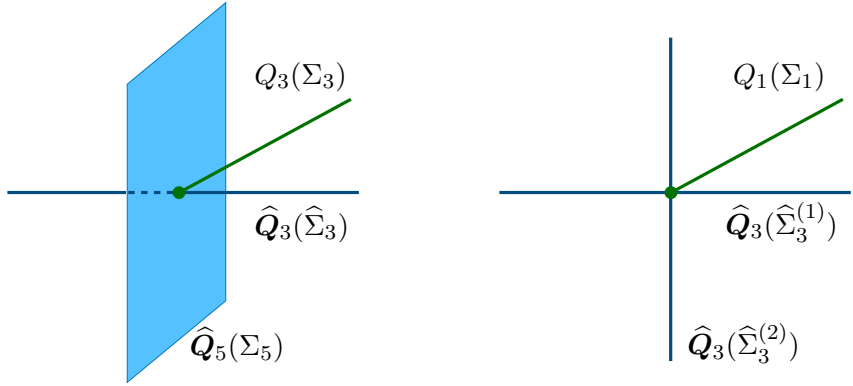
 \centering
\begin{tikzpicture}[      
        every node/.style={anchor=south west,inner sep=0pt},
        x=1mm, y=1mm,
      ]   
     \node (fig1) at (0,0)
       {\includegraphics[scale=0.6]{branefig1.pdf}};
       \node (fig2) at (65,0)
       {\includegraphics[scale=0.6]{branefig2.pdf}};
     \node   at (27.5,7) {$\small \widehat{\bm Q}_5(\Sigma_5)$};
     \node   at (35+1,24-1) {$\small \widehat{\bm Q}_3(\widehat \Sigma_3)  $};
     \node   at (35+1,43) {$\small  Q_3(\Sigma_3)    $};
     \node   at (65+32+1,24-1) {$\small \widehat{\bm Q}_3(\widehat \Sigma_3^{(1)})  $};
     \node   at (65+27+2-1,7) {$\small  \widehat{\bm Q}_3(\widehat \Sigma_3^{(2)})  $};
     \node   at (65+32+2,43) {$\small  Q_1(\Sigma_1)  $};
\end{tikzpicture}
\caption{The brane configurations of Figure \ref{fig_branes}
are translated into configurations involving the
topological operators of the 7d TQFT $\mathscr T_{\mathrm{7d}}^{D_N}$.
}
\label{fig_branes2}
\end{figure}

We close this section by noting that it is straightforward to translate
the brane configurations of Figure  \ref{fig_branes}
into configurations involving the topological operators
$\bm Q_3$, $\widehat{\bm Q}_3$, $\bm Q_1$,
$\widehat {\bm Q}_5$
of the 7d TQFT $\mathscr T_{\mathrm{7d}}^{D_N}$. This is done with the dictionary reported in Table \ref{table:summary}. We obtain the configurations depicted in Figure \ref{fig_branes2}.
In Section \ref{sec:application} we will see an application of these 
configurations of topological operators to one of the absolute global variants of the 6d (2,0) $D_N$ SCFT.

\section{Application: (non-)invertible symmetries of 6d SCFTs}\label{sec:application}

In this section we apply
the 7d TQFT $\mathscr T_{\mathrm{7d}}^{D_N}$
to the study of absolute 6d (2,0) SCFTs of type $D_N$ and their global symmetries.

For convenience, 
we  repeat here   the relevant terms in $\mathscr T_{\mathrm{7d}}^{D_N}$,
\be \label{eq_tot7d}
S_\text{7d TQFT} = 2\pi 
\int_{M_7} 
\bigg[  
\frac 12 \mathsf c_3 \cup \delta \mathsf b_3
- \frac N8 \mathsf c_3 \cup \delta \mathsf c_3
+ \frac 12 \mathsf a_1 \cup \delta \mathsf a_5
+  \frac 14 \mathsf a_1 \cup  \mathsf  c_3  \cup \mathsf c_3 
\bigg]  \ .
\ee 
Our strategy is to relate $\mathscr T_{\mathrm{7d}}^{D_N}$
to the analysis of 
global variants of the 6d (2,0) SCFT of type $D_N$ \cite{Gukov:2020btk}.
We extend such analysis by keeping track of the 0-form symmetry associated to $\mathsf a_1$. 
In particular, for each global variant we elucidate how the cubic coupling in $\mathscr T_{\mathrm{7d}}^{D_N}$ 
affects the global symmetries of the theory. In the process, we encounter non-trivial symmetry structures: mixed 't Hooft anomalies, higher-group symmetries, and non-invertible symmetries.

\subsection{The $SO(2N)$ variant}
\label{sec_SO_variant}
 
\begin{figure} \centering
\includegraphics[width=6cm]{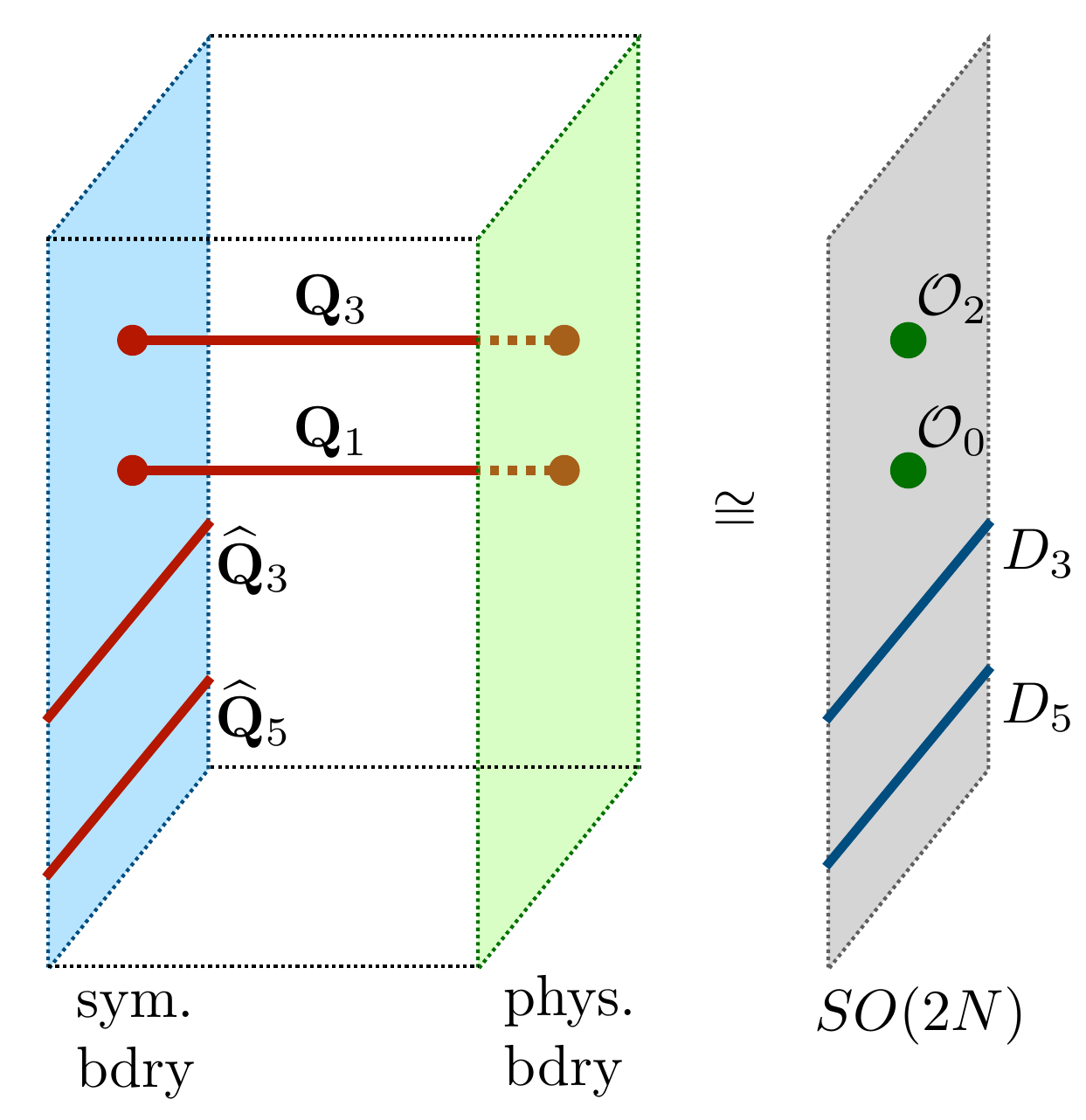}
\caption{Sandwich realization of the absolute global variant $SO(2N)$. The choice of (gapped) symmetry boundary is such that the operators $\bm Q_3$ and $\bm Q_1$ 
can end perpendicular to the boundary. Upon closing the sandwich, they yield non-topological operators $\cO_2$,
$\cO_0$. The operators $\widehat{\bm Q}_3$ and $\widehat{\bm Q}_5$  are projected parallel onto the symmetry boundary.
Upon closing the sandwich, they yield the topological operators
$D_3$ and $D_5$. $D_3$ implements the global $\mathbb Z_2$ 2-form symmetry of the $SO(2N)$ variant, while
$D_5$ implements the global $\mathbb Z_2$ 0-form symmetry.
Since $\bm Q_3$ and $\widehat {\bm Q}_3$ link in the 7d bulk,
$\cO_2$ is charged under $D_3$.
Since $\bm Q_1$ and $\widehat {\bm Q}_5$ link in the 7d bulk,
$\cO_0$ is charged under $D_5$.
}
\label{fig_SOops}
\end{figure}

  For any $N$, the topological action
\eqref{eq_tot7d} admits at least one gapped boundary condition, corresponding to the existence of an \emph{absolute}
6d (2,0) SCFTs.
(We recall that, in contrast, the 7d TQFT associated to SCFTs of type
of type $A_{N-1}$ admits
gapped boundary conditions
only if $N$ is a perfect square.)
Explicitly, 
we can consider the following gapped boundary condition for 
\eqref{eq_tot7d},
\be \label{eq_SO_bc}
\text{Dirichlet:} \;\; \mathsf c_3 \ , \ 
\mathsf a_1 \ ; 
\qquad 
\text{free:}  \;\; \mathsf b_3 \ , \ 
\mathsf a_5 \ . 
\ee 
These boundary conditions correspond to the absolute global variant of the SCFT of type $D_N$ denoted $SO(2N)$ in \cite{Gukov:2020btk}.
This absolute QFT has a global $\mathbb Z_2$ 0-form symmetry (with background field given by the boundary value of $\mathsf a_1$)
and a global $\mathbb Z_2$ 2-form symmetry (with background field given by the boundary value of $\mathsf c_3$).
The choice of topological boundary condition
\eqref{eq_SO_bc} determines which of the topological operators of the 7d TQFT $\mathscr T_{\mathrm{7d}}^{D_N}$ can end perpendicular to the boundary, and which can be projected parallel onto the boundary. This is summarized in Figure \ref{fig_SOops}.

We notice the appearance of a local operator $\cO_0$, which originates from a $\bm Q_1$ operator stretched along the interval direction in the sandwich construction.
Recall from Table \ref{table:summary} that
$\bm Q_1$ originates from an M2-brane wrapping $\RP^2 \subset \RP^4$. We can regard the local operator
$\cO_0$ as an analog of the Pfaffian operator in 4d $\cN = 4$ $SO(2N)$ super Yang-Mills. In terms of the holographic dual
$\AdS_5 \times \RP^5$ in Type IIB, the Pfaffian operator is realized as endpoint on the conformal boundary of a line operator in $\AdS_5$ obtained by wrapping a D3-brane on
$\RP^3 \subset \RP^5$ \cite{Witten:1998xy}.

With reference with Figure \ref{fig_SOops}, it is worth  remarking that the operators $\widehat {\bm Q}_3$ and $\widehat {\bm Q}_5$
are non-invertible operators in the 7d bulk, but they yield invertible operators $D_3$, $D_5$ upon closing the sandwich.
This is because the dressing factors $T_3$, $T_5$ in 
$\widehat {\bm Q}_3$, $\widehat {\bm Q}_5$, see \eqref{eq:T3T5dressing_def},
are decoupled from the rest of the system by the choice 
\eqref{eq_SO_bc} of boundary conditions.

From the couplings in 
\eqref{eq_tot7d} we infer 
the following 't Hooft anomalies of the $SO(2N)$ theory,
\be \label{eq_SO_anomaly}
\mathcal A_\text{6d SCFT $SO(2N)$} = \exp 2\pi i \int_{M_7} \bigg[ 
- \frac N4 \mathsf c_3 \cup \beta \mathsf c_3
+ \frac 14 \mathsf a_1 \cup \mathsf c_3 \cup \mathsf c_3
\bigg]  \ . 
\ee 
In the previous expression,
$\mathsf c_3$, $\mathsf a_1$
are understood as non-dynamical background fields.
The operation $\beta =\delta/2$ is the Bockstein homomorphism associated to the short exact sequence
$0 \rightarrow \mathbb Z_2 \rightarrow \mathbb Z_4 \rightarrow \mathbb Z_2 \rightarrow 0$, see e.g.~\cite{Cordova:2018acb}.

\begin{figure} \centering
\includegraphics[width=14.5cm]{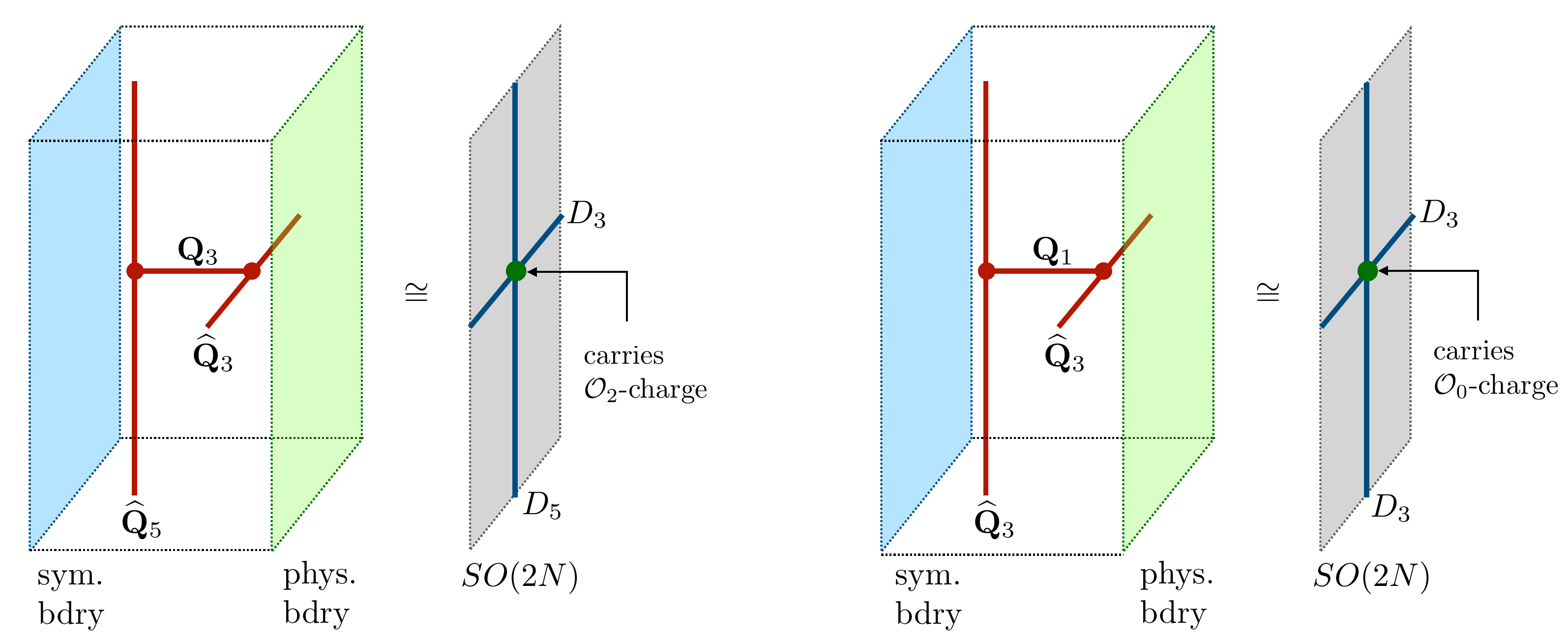}
\caption{On the left: the Hanany-Witten transition \eqref{eq_7d_moves} in the 7d bulk implies that the intersection of $D_3$ and $D_5$ defects in the $SO(2N)$ theory carries
$\cO_2$-charge.
On the right: the Hanany-Witten transition \eqref{eq_7d_moves_other} in the 7d bulk implies that the intersection of two $D_3$  defects in the $SO(2N)$ theory carries
$\cO_0$-charge.
Both these effects correspond to the cubic coupling
$\mathsf a_1 \mathsf c_3 \mathsf c_3$ in the anomaly theory
\eqref{eq_SO_anomaly} of the $SO(2N)$ variant.
}
\label{fig_SO_and_HW}
\end{figure}

The $\mathsf a_1 \mathsf c_3 \mathsf c_3$ term in \eqref{eq_SO_anomaly}
can also be seen at the level of the topological defects
$D_3$, $D_5$ defined in Figure \ref{fig_SOops}.
More precisely, the presence of this cubic anomaly is equivalent to stating that the intersection of a $D_3$ and a $D_5$ operator carries $\cO_2$-charge, namely it is charged under $D_3$. An equivalent point of view is that the intersection of two $D_3$ operators carries $\cO_0$-charge, namely is charged under $D_5$. Both these properties can be inferred using the Hanany-Witten transitions discussed in Section
\ref{sec_HW} \cite{Apruzzi:2023uma}.     This is depicted in Figure \ref{fig_SO_and_HW}.
For example, in order to realize an intersection of a $D_3$
and a $D_5$ defect, we start with a $\widehat{\bm Q_3}$
and a $\widehat{\bm Q}_5$ operator in the bulk of the sandwich construction, and we project them parallel onto the symmetry boundary one after the other. Here, however, we encounter a potential ambiguity: if we change the order in which they are projected parallel to the boundary, we create a $\bm Q_3$ operator stretched between them, see \eqref{eq_7d_moves}. This can link to $\widehat{\bm Q}_3$ in the 7d bulk, and thus carries $\cO_2$-charge. We infer that the intersection of $D_3$ and $D_5$ carries $\cO_2$-charge.
Similar remarks apply to the other process depicted in Figure
\ref{fig_SO_and_HW}, which is based on the Hanany-Witten
transition \eqref{eq_7d_moves_other}.

We expect that one could also detect the $\mathsf c_3 \beta \mathsf c_3$ anomaly in \eqref{eq_SO_anomaly} from branes, but we do not address this point in the present work.

\subsection{The $O(2N)$ variant and a higher-group structure}

\begin{figure} \centering
\includegraphics[width=6cm]{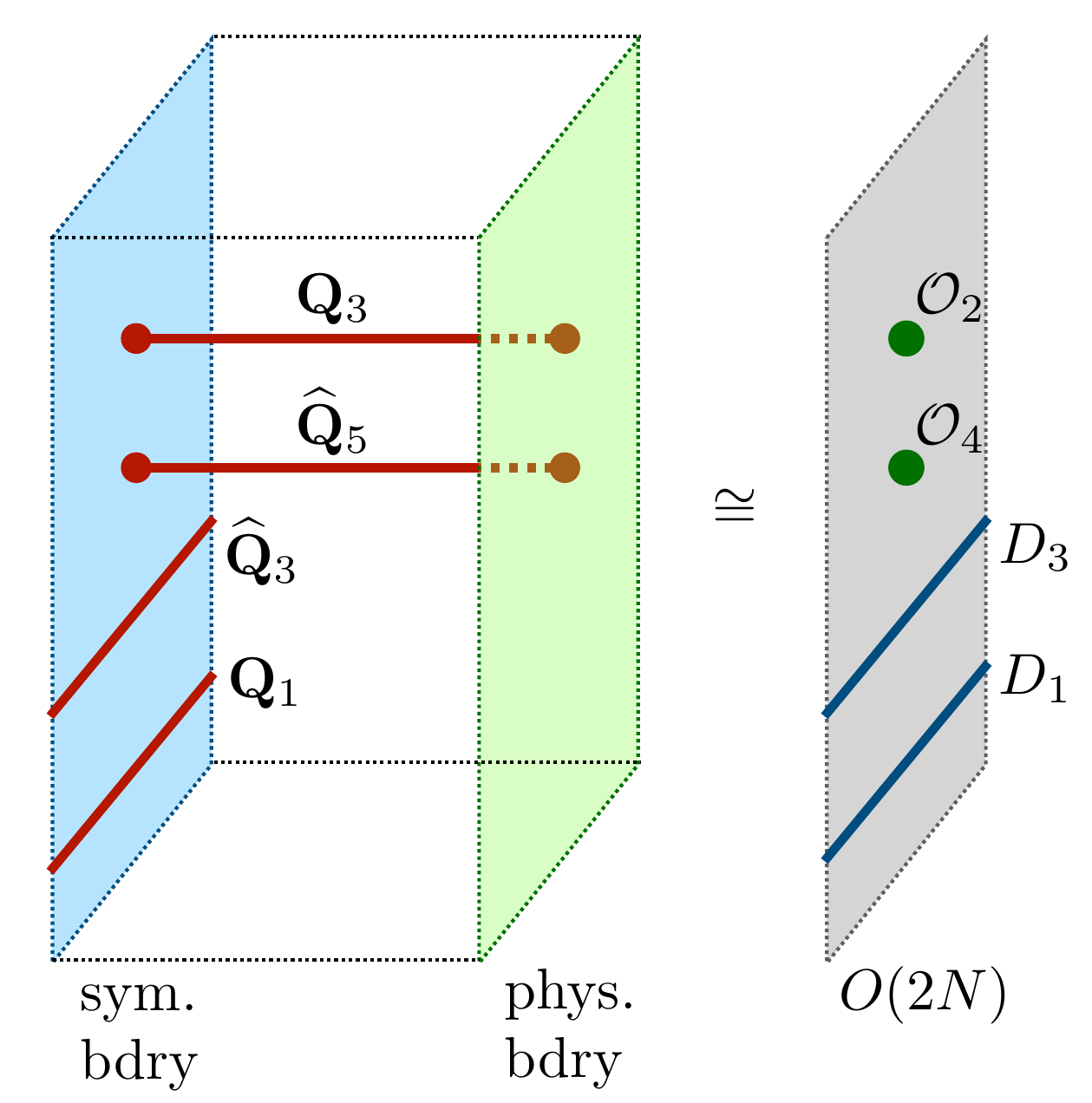}
\caption{Sandwich realization of the absolute global variant $O(2N)$. The choice of (gapped) symmetry boundary is such that the operators $\bm Q_3$ and $\widehat{\bm Q}_5$ 
can end perpendicular to the boundary. Upon closing the sandwich, they yield non-topological operators $\cO_2$,
$\cO_4$. The operators $\widehat{\bm Q}_3$ and $\bm Q_1$  are projected parallel onto the symmetry boundary.
Upon closing the sandwich, they yield the topological operators
$D_3$ and $D_1$. $D_3$ implements the global $\mathbb Z_2$ 2-form symmetry of the $O(2N)$ variant, while
$D_1$ implements the global $\mathbb Z_2$ 4-form symmetry.
}
\label{fig_Oops}
\end{figure}

For any $N$,
we can start from the $SO(2N)$ absolute global variant and gauge its $\mathbb Z_2$ 0-form symmetry. In terms of the sandwich constructions, 
this corresponds to 
the following gapped boundary condition of the 7d TQFT
$\mathscr T_{\mathrm{7d}}^{D_N}$,
\be \label{eq_O_bc}
\text{Dirichlet:} \;\; \mathsf c_3 \ , \ 
\mathsf a_5 \ ; 
\qquad 
\text{free:}  \;\; \mathsf b_3 \ , \ 
\mathsf a_1 \ . 
\ee 
We denote the resulting absolute global variant as $O(2N)$.
This QFT enjoys a global $\mathbb Z_2$ 4-form symmetry (with background field given by the boundary value of $\mathsf a_5$)
and a global $\mathbb Z_2$ 2-form symmetry (with background field given by the boundary value of $\mathsf c_3$).
The choice of topological boundary condition
\eqref{eq_O_bc} determines which of the topological operators of the 7d TQFT $\mathscr T_{\mathrm{7d}}^{D_N}$ can end perpendicular to the boundary, and which can be projected parallel onto the boundary. This is summarized in Figure \ref{fig_Oops}.

For the $O(2N)$ variant, the cubic $\mathsf a_1 \mathsf c_3 \mathsf c_3$ term in the 7d action does not translate into a mixed 't Hooft anomaly, but rather into a higher-group structure \cite{Cordova:2018cvg, Benini:2018reh}. Equivalently, per the general analysis of \cite{Tachikawa:2017gyf}, the mixed 't Hooft anomaly $\mathsf a_1 \mathsf c_3 \mathsf c_3$ of the $SO(2N)$ variant
induces a higher-group when we gauge the 0-form symmetry associated to $\mathsf a_1$. The higher-group structure 
in question is encoded in the relation
\be \label{eq_higher_group_rel}
\delta \mathsf a_5 = - \frac 12 \mathsf c_3 \cup \mathsf c_3 \ ,
\ee 
where $\mathsf a_5$ and $\mathsf c_3$ are   understood as non-dynamical background fields in the $O(2N)$ theory.

We can also see the higher-group structure in terms
of the topological operators of the theory.
Upon closing the sandwich, the operators $D_3$ and $D_1$
have a non-trivial relation: if two $D_3$ operators intersect
transversely at a point,
a $D_1$ operator emanates from the intersection \cite{Benini:2018reh}.
This can be seen from \eqref{eq_higher_group_rel} applying Poincar\'e duality.
In terms of the sandwich picture, this phenomenon originates from
the analogous phenomenon involving operators $\widehat{\bm Q}_3$ and
$\bm Q_1$, all projected parallel onto the symmetry boundary.
In Section \ref{sec_cubic_and_branes} we have described the brane origin of this effect in M-theory, see Figure \ref{fig_branes}
and Figure \ref{fig_branes2}.

\subsection{The $Sc(8k)$ and $Ss(8k)$ variants  and non-invertible symmetries}

Let us  start once more
from the absolute global variant $SO(2N)$ and specialize to $N=4k$.
The $\mathsf c_3 \cup \beta \mathsf c_3$ term in the anomaly \eqref{eq_SO_anomaly}
trivializes. As a result, 
we may now gauge the $\mathbb Z_2$ 2-form symmetry associated to $\mathsf c_3$. In performing this gauging, we have the option to include discrete torsion, 
namely stack with an SPT before gauging.
The SPT in question 
is the Arf-Kervaire invariant \cite{Hsin:2021qiy,Gukov:2020btk}.
The resulting absolute global variants are denoted $Sc(8k)$, $Ss(8k)$.
Following the conventions of Figure 13 in \cite{Gukov:2020btk}, 
gauging the $SO(8k)$ variant without torsion yields the $Sc(8k)$ variant, while gauging it with torsion yields $Ss(8k)$.

\begin{figure}
\centering
\begin{tikzpicture}[every text node part/.style={align=left}]
\node[draw, rectangle] (node1) at (4.4,-2) {\small
    $SO(8k)$ variant 
\\ \small
anomaly
$ \tfrac 14 \mathsf a_1 \mathsf c_3 \mathsf c_3$
};

\node[draw, rectangle] (node2) at (10.,-2) {
\small
    $Sc(8k)$ or $Ss(8k)$ variant 
\\   \small
non-inv.~0-form symm.
};

\node[draw, rectangle] (node4) at (-0.5,-2) { \small
    $O(8k)$ variant 
\\   \small
higher-group \\ \small 
$\delta \mathsf a_5 = \tfrac 12 \mathsf c_3 \mathsf c_3$
};

\draw[->] (node1) to (node2);

\draw[->] (node1) to (node4);

\node[] (node13) at (6.9,-2+0.2) {\footnotesize gauge $\mathsf c_3$};

\node (node13) at (1.9,-2+0.2) {\footnotesize gauge $\mathsf a_1$};

\end{tikzpicture}
\caption{
Absolute variants of the 6d (2,0) SCFT of type $D_N$ with $N=4k$.  All these variants share the same 7d TQFT \eqref{eq_tot7d} as SymTFT. 
The $SO(8k)$ variant exhibits a mixed 't Hooft anomaly. 
By ``gauge $\mathsf a_1$'' we mean gauging the 0-form symmetry of $SO(8k)$
associated to $\mathsf a_1$,
and similarly by ``gauge $\mathsf c_3$''
we mean gauging the 2-form symmetry
of $SO(8k)$ associated to $\mathsf c_3$.
We reach the $Sc(8k)$ or the $Ss(8k)$
variant depending on whether we
perform the gauging of the 2-form symmetry with a discrete torsion
(associated to the Arf-Kervaire SPT), or not. 
\label{fig_6d_variants_simpler}}
\end{figure}
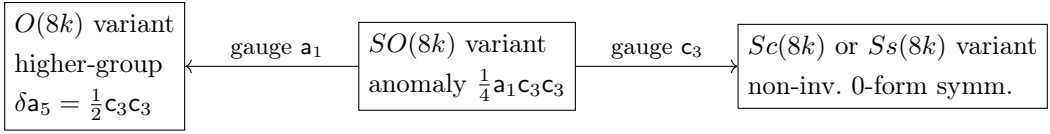

According to the general arguments of 
\cite{Tachikawa:2017gyf,Kaidi:2021xfk},
if we start from the $SO(8k)$ variant
with a mixed $\mathsf a_1 \mathsf c_3 \mathsf c_3$ 't Hooft anomaly,
and we gauge the 2-form symmetry associated to $\mathsf c_3$, the 0-form symmetry associated to $\mathsf a_1$
becomes a non-invertible symmetry.
The corresponding symmetry category, loosely speaking, can be 
thought of as
 a higher-dimensional version of a  $\mathbb Z_2$ Tambara-Yamagami fusion category.\footnote{ \ 
 Duality defects in 4d QFTs often form analogous higher versions of Tambara-Yamagami fusion categories,
 see e.g.~\cite{Choi:2021kmx,
 Kaidi:2021xfk,Cordova:2023bja,Antinucci:2023ezl}.
 We refer the reader to \cite{Decoppet:2023bay}
 for recent developments on a mathematically rigorous notion of
 Tambara-Yamagami fusion 2-category.
 }  
We conclude that the $Sc(8k)$,
$Ss(8k)$ global variants enjoy such non-invertible symmetry.
A schematic summary of the global variants discussed so far 
is given in Figure~\ref{fig_6d_variants_simpler}.

\begin{figure} \centering
\includegraphics[width=6cm]{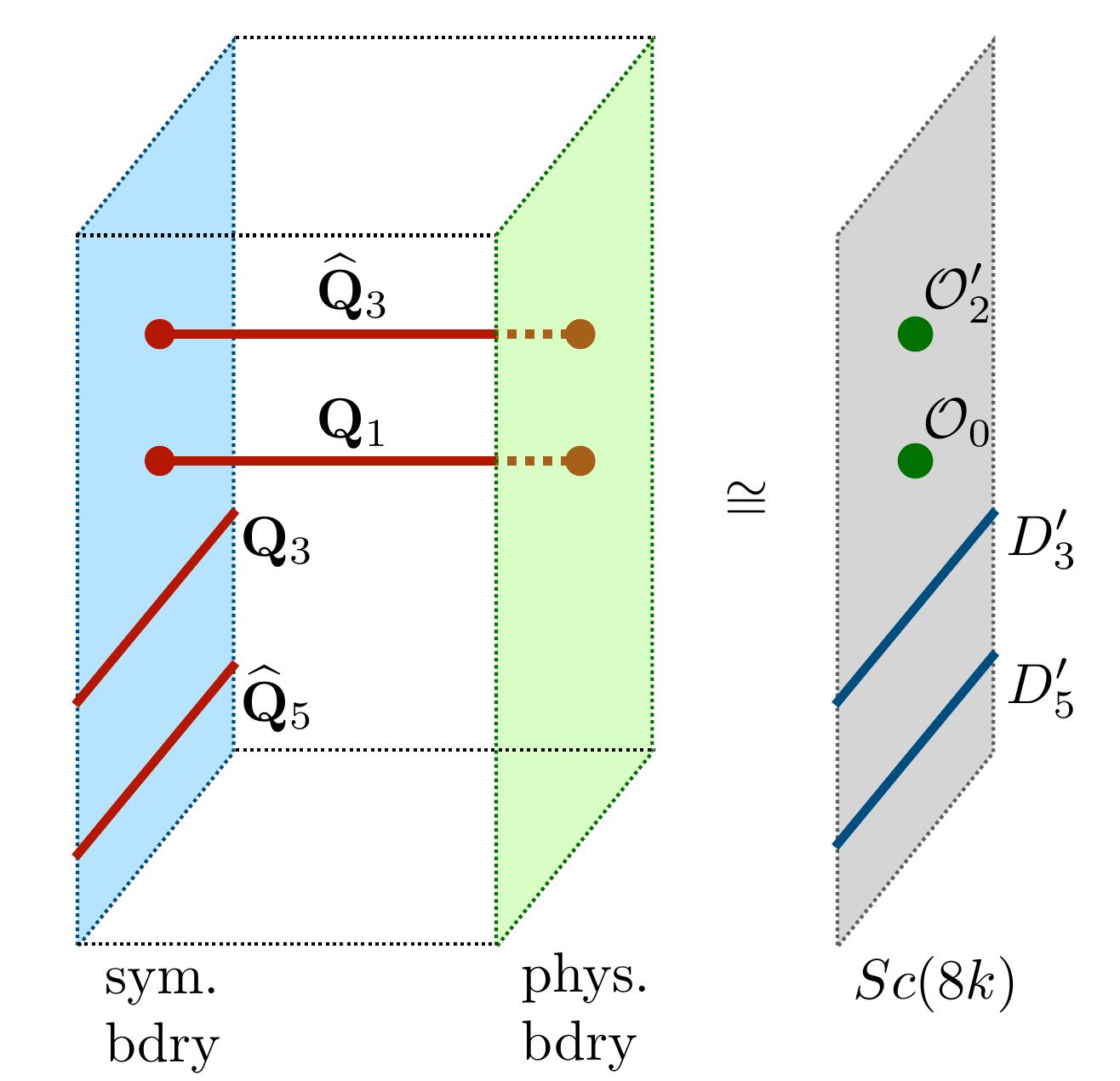}
\caption{Sandwich realization of the absolute global variant $Sc(2N)$. The choice of (gapped) symmetry boundary is such that the operators $\widehat{\bm Q}_3$ and $\bm Q_1$ 
can end perpendicular to the boundary. Upon closing the sandwich, they yield non-topological operators $\cO_2'$,
$\cO_0$. (The latter is the same as in the $SO(8k)$ theory.) The operators ${\bm Q}_3$ and $\widehat{\bm Q}_5$  are projected parallel onto the symmetry boundary.
Upon closing the sandwich, 
${\bm Q}_3$ yields an invertible topological operator
$D_3'$ implementing a global $\mathbb Z_2$ 2-form symmetry.
(The latter is the quantum symmetry associated to the gauging of the 2-form symmetry of the $SO(8k)$ theory.)
The operator $\widehat{\bm Q}_5$ yields instead a topological operator $D_5'$ that implements
a non-invertible 0-form symmetry of the $Sc(8k)$ theory.
}
\label{fig_Scops}
\end{figure}

In terms of gapped boundary conditions of the 7d TQFT
$\mathscr T_{\mathrm{7d}}^{D_N}$, we expect that 
the $Sc(8k)$, $Ss(8k)$ variants
 correspond to Dirichlet boundary conditions for $\mathsf b_3$
and for $\mathsf b_3 + \mathsf c_3$, respectively.
(In both cases, we assign Dirichlet boundary conditions to $\mathsf a_1$, and free boundary conditions to $\mathsf a_5$, $\mathsf c_3$.)
For definiteness, let us focus on $Sc(8k)$.
The relation between topological operators in the 7d bulk and operators in the 6d absolute SCFT
are summarized in Figure \ref{fig_Scops}. 
In particular, upon projecting $\widehat{\bm Q}_5$
parallel onto the symmetry boundary, we get a non-invertible operator $D_5'$. On the symmetry boundary, $\mathsf a_1$ is a $c$-number, due to the Dirichlet boundary conditions,
but $\mathsf c_3$ fluctuates. As a result, the $T_5$ dressing factor remains non-trivial: the non-invertible operator
$\widehat{\bm Q}_5$ in the 7d bulk yields
a non-invertible operator $D_5'$ in 6d.

\begin{figure} \centering
\includegraphics[width=6cm]{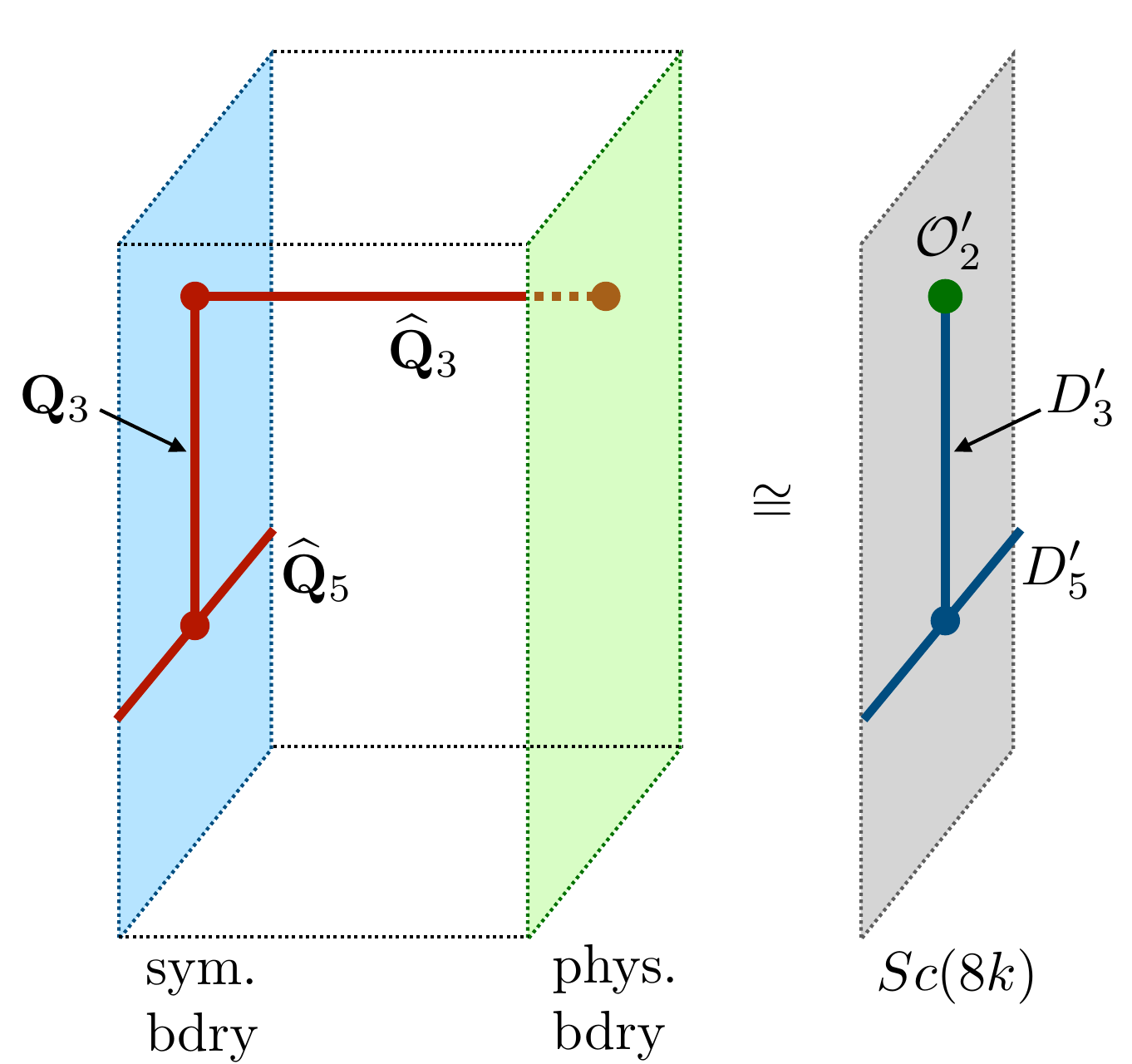}
\caption{The Hanany-Witten transition \eqref{eq_7d_moves}
implies a non-trivial action of the non-invertible 0-form symmetry defect $D_5'$ on the extended operator $\cO_2'$
in the $Sc(8k)$ variant.
If we sweep the defect $D_5'$ across $\cO_2'$,
when $\cO_2'$ reemerges on the other side of $D_5'$ it comes attached with a topological defect $D_3'$.
}
\label{fig_Sc_HW}
\end{figure}

The action of the non-invertible operator
$D_5'$ on the operators $\cO_0$, $\cO_2'$
can be inferred from the properties of the topological operators in the 7d bulk.
Firstly, we know that $\widehat {\bm Q}_5$ and
$\widehat Q_1$ link in 7d. When acting on $\cO_0$,
the symmetry defects $D_5'$ acts as an ordinary invertible
$\mathbb Z_2$ symmetry defects.
The action of $D_5'$ on $\cO_2'$, however, is non-trivial.
This can be inferred from the Hanany-Witten effect \eqref{eq_7d_moves}, as depicted in Figure \ref{fig_Sc_HW}.
More precisely, we find that, 
if we sweep the defect $D_5'$ across $\cO_2'$,
when $\cO_2'$ reemerges on the other side of $D_5'$ it comes attached with a topological defect $D_3'$.
This is similar to the action of the Kramers-Wannier duality defect on the spin operator of the Ising 2d CFT and to many other higher-dimensional analogs encountered in the study of non-invertible symmetries, see e.g.~the lecture notes \cite{Shao:2023gho}
and references therein.

\subsection{Further remarks on $SO(8k)$, $Sc(8k)$, and $Ss(8k)$}

Let us focus on the case $N=4k$ and consider in greater detail the relationships among  the global variants $SO(8k)$, $Sc(8k)$, $Ss(8k)$.
As discussed in \cite{Gukov:2020btk},
these variants fit into a commutative diagram of the form depicted below,
\be \label{eq_boson_diagram6d}
\begin{tikzcd}[sep=1.5cm]
\mathcal T_1 \arrow[shift left=0]{r}{\phi} \arrow[shift right=.7, swap]{d}{g}& \mathcal S_1 \arrow[shift right=.7, swap]{d}{h}
\\
\mathcal T_2 \arrow[shift left=0]{r}{\phi} \arrow[shift right=.7, swap]{u}{g^\vee} & \mathcal S_2 \arrow[shift right=.7, swap]{u}{h}
\end{tikzcd} \quad  \quad
\begin{array}{r l}
\small
g \equiv  &\text{gauging a non-anom. $\mathbb Z_2$ 2-form symmetry of $\cT_1$,} \\
\small
g^\vee \equiv  & \text{gauging the associated quantum symmetry of   $\cT_2$,} \\
h \equiv  & \text{stacking with the Arf-Kervaire SPT,} \\
\phi \equiv  & \text{analog of fermionization map in 2d.} 
\end{array}
\ee 
Some comments are in order.
The map $g$ is realized in the standard way at the level of partition functions 
\be 
Z_{g \cT_1}[B_3] = \frac{1}{|H^3(M_6;\mathbb Z_2)|^{1/2}} \sum_{b_3} Z_{\cT_1}[b_3] (-1)^{\int b_3 B_3} \ , 
\ee 
where the sum here and in what follows is over $H^3(M_6;\mathbb Z_2)$, the integral is over $M_6$, and we omit cup products. As usual, $Z_{\cT_1}[b_3]$ denotes the partition function of theory $\cT_1$ coupled to the background field $b_3$ for its global $\mathbb Z_2$ 2-form symmetry, while $B_3$ is the background field for the quantum symmetry of $ g\cT_1$.
The map $g^\vee$ is realized similarly.

The map $\phi$
is given as \cite{Hsin:2021qiy}\footnote{ \ We restrict ourselves to orientable $M_6$. We refer the reader to \cite{Hsin:2021qiy} for a discussion covering also non-orientable $M_6$.}
\be \label{eq_phi_6d_def}
Z_{\phi \cT_1}[\rho_3, C_3]
= \frac{1}{|H^3(M_6;\mathbb Z_2)|^{1/2}} \sum_{b_3} Z_{\cT_1}[b_3]  (-1)^{q_{\rho_3}(b_3)}  \ . 
\ee 
Here, the quantity $\rho_3$ is a trivialization of the fourth Wu class $\nu_4$ of $M_6$, $\nu_4 = d\rho_3$.
We refer to $\rho_3$ as a Wu structure.
Wu structures form a torsor over $H^3(M_6;\mathbb Z_2)$.
A choice of  $\rho_3$ determines  a
 $\mathbb Z_2$-valued quadratic refinement of the cup product in $H^3(M_6;\mathbb Z_2)$,
which we denote
 $ q_{\rho_3}: H^3(M_6;\mathbb Z_2) \rightarrow \mathbb Z_2$.
 (We refer the reader to Appendix \ref{app_refinement} for more details and references.)
The Arf-Kervaire SPT is the invertible topological theory with  partition function  
\be 
(-1)^{\mathrm{AK}(\rho_3)} = \frac{1}{|H^3(M_6 ; \mathbb Z_2)|^{1/2}} \sum_{b_3} (-1)^{q_{\rho_3}(b_3) } \ . 
\ee 
Notice that it depends on a choice of $\rho_3$.

As already alluded to above,
the map $\phi$ can be regarded as analogous to the fermionization map of 2d QFTs.
Bosonization and fermionization   have been studied extensively in recent years \cite{Karch:2019lnn,Gaiotto:2015zta,Bhardwaj:2016clt,Kapustin:2017jrc,Thorngren:2018bhj,Ji:2019ugf}. The fermionization map in 2d is given 
by an expression similar to 
\eqref{eq_phi_6d_def},
with the role of $\rho_3$ played by $\rho_1$, a spin structure (i.e.~trivialization of the second Stiefel-Whitney class
$w_2(M_6)$)
\cite{Ji:2019ugf}. The 6d commutative diagram \eqref{eq_boson_diagram6d} generalizes
a similar diagram for 2d theories. In the 2d case, the role of the Arf-Kervaire SPT is played by the Arf SPT,
which is the (spin structure dependent) invertible field theory whose partition function on an orientable 2-manifold is the Arf invariant.

After these preliminaries
we can go back to the problem at hand.
The 6d global variants $SO(8k)$,
$Sc(8k)$, $Ss(8k)$ fit into a diagram of the form
\eqref{eq_boson_diagram6d}
with 
$\cT_1 = Sc(8k)$,
$\cT_2 = Ss(8k)$,
$\cS_1 = \cS_2 = SO(8k)$,
\be \label{eq_SO_boson}
\begin{tikzcd}[sep=1.5cm]
Sc(8k) \arrow[shift left=0]{r}{\phi} \arrow[shift right=.7, swap]{d}{g}& SO(8k) \arrow[shift right=.7, swap]{d}{h}
\\
Ss(8k) \arrow[shift left=0]{r}{\phi} \arrow[shift right=.7, swap]{u}{g^\vee} & SO(8k) \arrow[shift right=.7, swap]{u}{h}
\end{tikzcd} \ . 
\ee 
This diagram encodes the equivalence of the following  statements: the theories
$Sc(8k)$ and $Ss(8k)$ 
are isomorphic (thus
$Sc(8k) \cong Ss(8k)$
is self-dual under gauging its
$\mathbb Z_2$ 2-form symmetry); the theory
$SO(8k)$ is invariant under stacking with the Arf-Kervaire SPT. 
In Section \ref{sec_SO_variant} we have argued that
the cubic coupling
$\mathsf a_1 \mathsf c_3 \mathsf c_3$ in the 7d bulk
implies that the global variant
$SO(8k)$ exhibits a 't Hooft anomaly between its global $\mathbb Z_2$ 0-form and 2-form symmetries.
In reference to the diagram
\eqref{eq_SO_boson}, we claim that
this mixed anomaly is equivalent to the fact that
$SO(8k)$ is invariant under stacking with the Arf-Kervaire
SPT. Accordingly, 
$Sc(8k)$ and $Ss(8k)$ are dual. This makes it possible to construct non-invertible duality defects for these global variants.
Indeed, such defects have been studied in 
\cite{Lawrie:2023tdz,Apruzzi:2024cty} for $k=1$.
Here we offer an alternative perspective on these duality defects, valid for any $k \ge 1$,   leveraging the mixed 't Hooft anomaly of the $SO(8k)$ variant.\footnote{ \ Since the map $\phi$ is not a mere (bosonic) gauging of a $\mathbb Z_2$ 2-form symmetry,
but rather the more complicated operation 
\eqref{eq_phi_6d_def}, we do \emph{not} claim that the non-invertible duality symmetries of the $Sc(8k)$,
$Ss(8k)$ variants are group-like
(also known as non-intrinsic).
}

We can elucidate further the connection between the mixed anomaly $\mathsf a_1 \mathsf c_3 \mathsf c_3$ of the $SO(8k)$ variant and the considerations of this subsection as follows.
We build on the arguments of 
\cite{Kaidi:2021xfk}.
They key idea is to use descent to relate 
the 7d anomaly term $ \mathsf a_1 \mathsf c_3 \mathsf c_3$
to the anomalous variation
of the 6d partition function
under a gauge transformation of the background field
$\mathsf a_1$ of the $\mathbb Z_2$ 0-form symmetry.
Upon turning off $\mathsf a_1$, we obtain a relation
that encodes the ability of the partition function to ``absorb'' a phase factor constructed from $\mathsf c_3$,
\be 
\text{$\mathsf a_1 \mathsf c_3 \mathsf c_3$ anomaly} 
\quad \rightarrow 
\quad 
Z_{SO(8k)}[\mathsf c_3] = 
Z_{SO(8k)}[\mathsf c_3] e^{2\pi i \frac 14 \widetilde q^{M_6}_{\rho_3}(\mathsf c_3)}
=
Z_{SO(8k)}[\mathsf c_3] (-1)^{q^{M_6}_{\rho_3}(\mathsf c_3)} \ . 
\ee 
We have used the fact that the naive expression
$\int_{M_6} \tfrac 14 \mathsf c_3 \cup \mathsf c_3$ is
best understood in terms of the $\mathbb Z_4$-valued quadratic refinement
$\widetilde q^{M_6}_{\rho_3}(\mathsf c_3)$ of Section \ref{sec_this_is_F7} and Appendix \ref{app_refinement}.
(Here $M_6$ is the spacetime where the 6d
$SO(8k)$ SCFT is formulated.)
Since we assume that $M_6$ is oriented,
we can trade the 
 $\mathbb Z_4$-valued
 refinement $\widetilde q_{\rho_3}^{M_6}$
for the $\mathbb Z_2$-valued
refinement 
$ q_{\rho_3}^{M_6}$,
see Appendix \ref{app_refinement}.

\subsection{Toroidal reduction} \label{sec_T2_reduction}

In this section we consider the reduction of the 7d topological theory \eqref{eq_tot7d} on $T^2$,
namely we set $M_7= M_5 \times T^2$ and integrate on the second factor.
The resulting theory on $M_5$ is compared against known terms \cite{Bergman:2022otk,Etheredge:2023ler} in the 5d topological action associated to 4d $\mathcal N = 4$ super Yang-Mills with gauge algebra $\mathfrak{so}(2N)$, thus providing a check of our findings.
We find it convenient to discuss separately the quadratic terms and the cubic term in \eqref{eq_tot7d}.

The reduction of 
the quadratic terms 
is readily performed in the continuum formulation:
all fields in this paragraph are differential forms.
The standard $A$, $B$ 1-cycles on $T^2$ are Poincar\'e dual to
two harmonic 1-forms, which we denote
$\omega^{(A)}_1$, $\omega^{(B)}_1$. The  reduction ansatz reads
\begin{align} 
\label{eq:torus_KK}
c_3^{\rm 7d} &= c_3 + c_2 \wedge \omega^{(A)}_1 +  c_2' \wedge \omega^{(B)}_1
+ c_1 \wedge \omega^{(A)}_1 \wedge \omega^{(B)}_1
\ , \\ 
b_3^{\rm 7d} &= b_3 + b_2 \wedge \omega^{(A)}_1 +  b_2' \wedge \omega^{(B)}_1
+ b_1 \wedge \omega^{(A)}_1 \wedge \omega^{(B)}_1
\ , \\ 
a_1^{\rm 7d} & = a_1 + a_0 \omega^{(A)}_1 
+  a_0' \omega^{(B)}_1  \ , \\ 
a_5^{\rm 7d} &= a_5 + a_4 \wedge \omega^{(A)}_1 +  a_4' \wedge \omega^{(B)}_1
+ a_3 \wedge \omega^{(A)}_1 \wedge \omega_1^{(B)}
\ .
\end{align}
The fields on the LHSs
are the 7d fields appearing in the original 7d action.
All fields on the RHSs live on
$M_5$. We are using the same symbol for some 7d and 5d fields; the distinction
should be clear from context.
A straightforward computation gives 
the 5d action 
\begin{align}
S_\text{5d TQFT}^\text{quadratic} &= \int_{M_5}\bigg[
-  N  c_2' \wedge d c_2
-2 c_2 \wedge d  b_2'
+ 2  c_2' \wedge db_2
+ 2 a_1 \wedge da_3
\nn \\ 
& -  N   c_1 \wedge d c_3
+2 c_3 \wedge db_1
+2 c_1 \wedge db_3
- a_0 d  a_4' +  a_0' da_4
\bigg] \ ,
\label{eq:5d_quad}
\end{align}
where we have used
\be  \label{eq:ABintersection}
\int_{T^2} \omega_A \wedge \omega_B = 1 \ .
\ee

Let us now turn to the reduction of the cubic coupling in \eqref{eq_tot7d}.
In this paragraph we work at the level of $\mathbb Z_2$ cohomology classes.
We  use the analog of  \eqref{eq:torus_KK},
\begin{align}
\mathsf c_3^{\rm 7d} &= \mathsf c_3  + \mathsf c_2 \cup \mathsf v^{(A)}_1 +  \mathsf c_2' \cup \mathsf v^{(B)}_1
+ \mathsf c_1 \cup \mathsf v^{(A)}_1 \cup \mathsf v^{(B)}_1 
\ , \label{eq_c37d_ansatz} \\ 
\mathsf a_1^{\rm 7d} & = \mathsf a_1 + \mathsf a_0 \cup \mathsf v^{(A)}_1 
+  \mathsf a_0' \cup  \mathsf v^{(B)}_1  \  .
\label{eq_a17d_ansatz}
\end{align}
Here $\mathsf v^{(A)}_1$, $\mathsf v^{(B)}_1$ are integral cohomology classes of degree 1,
generators of $H^1(T^2;\mathbb Z)$,
corresponding to the de Rham
classes of $\omega_A$, $\omega_B$ and satisfying
\be \label{eq_torus_intersection}
\int_{T^2} \mathsf v^{(A)}_1 \cup \mathsf v^{(B)}_1 = 1 \ . 
\ee 
Integration   on $T^2$ gives the 5d couplings 
\be 
\label{eq:5d_cubic}
S_\text{5d TQFT}^\text{cubic}  = \int_{M_5}
\bigg[
 \frac 12 \mathsf a_1 \cup \mathsf c_2'
\cup \mathsf c_2
+ \frac 12 \mathsf a_1 \cup \mathsf c_1 \cup \mathsf c_3
+ \frac 12 \mathsf a_0 \cup
\mathsf c_2' \cup \mathsf c_3
+ \frac 12 \mathsf a_0' \cup \mathsf c_2 \cup \mathsf c_3 
\bigg] \ . 
\ee 
To justify this claim,
we recall from Section 
\ref{sec_this_is_F7} that $\frac 14 \mathsf a_1 \cup \mathsf c_3 \cup \mathsf c_3$ is a
schematic notation.
More precisely, a suitable quadratic refinement is understood.
To make sense of $\tfrac 14 \mathsf a_1 \cup \mathsf c_3 \cup \mathsf c_3$ we use Poincar\'e duality to 
convert $\mathsf a_1$ into a subspace $X_6$
and we compute 
$\widetilde q_{\rho_3}^{X_6}(\mathsf c_3^{\rm 7d})$,
in 
 the notation 
of
Section \ref{sec_this_is_F7},
see also Appendix
\ref{app_refinement}.
The task at hand
is to plug 
\eqref{eq_c37d_ansatz}
into $\widetilde q_{\rho_3}^{X_6}(\mathsf c_3^{\rm 7d})$ and expand.
This is achieved by making use of the key property \eqref{eq_refinement_key} of the
quadratic refinement, repeated here for convenience, 
\be 
\widetilde q_{\rho_3}^{X_6}(\mathsf x_3 + \mathsf y_3)  = \widetilde q_{\rho_3}^{X_6}(\mathsf x_3 )
+ \widetilde q_{\rho_3}^{X_6}( \mathsf y_3)
+ \theta \int_{X_{6}} \mathsf x_3 \cup \mathsf y_3 \mod 4 \ .
\ee  
With the benefit of hindsight, we only display those terms that will eventually give a non-zero contribution when we integrate over $T^2$.
The relevant terms are 
\be 
\ba 
q_{\rho_3}^{X_6}(\mathsf c_3^{\rm 7d}) \supset \theta \int_{X_6} & \bigg[
\mathsf c_3 \cup \mathsf c_1 \cup \mathsf v^{(A)}_1 \cup 
\mathsf v^{(B)}_1
+ \mathsf c_2 \cup \mathsf c_2' \cup \mathsf v^{(A)}_1 \cup 
\mathsf v^{(B)}_1
\\ 
& +  \mathsf c_3 \cup \mathsf c_2 \cup \mathsf v^{(A)}_1
+ \mathsf c_3 \cup \mathsf c_2' \cup \mathsf v^{(B)}_1
\bigg] \ . 
\ea 
\ee 
We stress that all these terms originate from cross terms in the expansion of $q_{\rho_3}^{X_6}(\mathsf c_3^{\rm 7d})$.
This means that they can be written in a simpler
way in terms of the ordinary cup product.
In other words, 
the reduction of 
$\tfrac 14 \mathsf a_1 \cup \mathsf c_3 \cup \mathsf c_3$ yields 
\be 
\ba 
& \int_{M_7}\frac 14 \mathsf a_1^{\rm 7d} \cup \mathsf c_3^{\rm 7d} \cup \mathsf c_3^{\rm 7d}
\rightarrow 
\int_{M_5 \times T^2}
\frac 12 \Big(
\mathsf a_1 + \mathsf a_0 \cup \mathsf v^{(A)}_1 
+  \mathsf a_0' \cup  \mathsf v^{(B)}_1
\Big) \cup 
\\
& \cup \bigg[ 
 \mathsf c_3 \cup \mathsf c_1 \cup \mathsf v^{(A)}_1 \cup 
\mathsf v^{(B)}_1
 + \mathsf c_2 \cup \mathsf c_2' \cup \mathsf v^{(A)}_1 \cup 
\mathsf v^{(B)}_1
+  \mathsf c_3 \cup \mathsf c_2 \cup \mathsf v^{(A)}_1
+ \mathsf c_3 \cup \mathsf c_2' \cup \mathsf v^{(B)}_1
\bigg]  \ . 
\ea 
\ee 
We have made use of \eqref{eq_a17d_ansatz} for $\mathsf a_1^{\rm 7d}$.
Finally, we get the result
\eqref{eq:5d_cubic} by collecting terms with exactly one $\mathsf v_1^{(A)}$ and one 
$\mathsf v_1^{(B)}$
factor, and using 
\eqref{eq_torus_intersection}.

The terms on the first line of \eqref{eq:5d_quad}, together with the first term in \eqref{eq:5d_cubic}, match
the 5d SymTFT for 4d $\mathcal N = 4$ SYM with gauge algebra $\mathfrak{so}(2N)$ \cite{Bergman:2022otk,Etheredge:2023ler}.
More precisely, let us make contact with the notation used in \cite{Etheredge:2023ler}.
With the identifications
\be 
c_2 ' = \mathbf B_{\rm F1} \ , \quad 
c_2 = \mathbf B_{\rm D1} \ ,  \quad 
b_2 = - \mathbf B_{\rm NS5} \ , \quad 
b_2' = \mathsf B_{\rm D5} \ , \quad 
a_1 = \mathbf A_1 \ , \quad 
c_3 = \mathbf A_3 \ ,
\ee 
the first line of \eqref{eq:5d_quad} yields
\be 
\int_{M_5} \bigg[ 
- N \mathbf B_{\rm D1}  \wedge d
\mathbf B_{\rm F1}
- 2 \mathbf B_{\rm F1} \wedge d \mathbf B_{\rm NS5}
- 2 \mathbf B_{\rm D1} \wedge d \mathbf B_{\rm D5}
+ 2 \mathbf A_1 \wedge d\mathbf A_3
\bigg]  \ .
\ee 
Similarly, at the level of $\mathbb Z_2$ discrete gauge fields, we use the identifications
\be 
\mathsf a_1 = A_1 \ , \quad 
\mathsf c_2 = B_{\rm D1} \ ,  \quad 
\mathsf c_2' = B_{\rm F1} \ ,
\ee 
to get the cubic coupling 
\be 
\int_{M_5} \frac 12 A_1 \cup B_{\rm F1} \cup B_{\rm D1} \ .
\ee 
We have reproduced (2.50),
(2.60), (2.68)
in \cite{Etheredge:2023ler}.

\subsection{Comments on reduction on a Riemann surface}
\label{sec_Sigma}

The computations of the previous subsection
are readily adapted to the case in which the 7d spacetime is of the form $M_7 = M_5 \times \Sigma_g$, with $\Sigma_g$ a smooth genus-$g$ Riemann surface.
The analog of \eqref{eq:torus_KK}
reads
\be \label{eq_Sigmag_KK}
\ba 
c_3^{\rm 7d} & = \sum_{i=1}^{g}\bigg[ c_2^{(i)} \wedge \omega_1^{(A,i)}
+ c_2^{\prime(i)} \wedge \omega_1^{(B,i)}
\bigg] + \dots \ , \\ 
b_3^{\rm 7d} & = \sum_{i=1}^{g}\bigg[ b_2^{(i)} \wedge \omega_1^{(A,i)}
+ b_2^{\prime(i)} \wedge \omega_1^{(B,i)}
\bigg] + \dots  \ , \\
a_1^{\rm 7d} & = a_1 + \dots \ , \\
a_5^{\rm 7d} & = a_5 \wedge {\rm vol}_{\Sigma_g} + \dots \ . 
\ea 
\ee 
The harmonic 1-forms
$\omega_1^{(A,i)}$,
$\omega_1^{(B,i)}$, $i=1,\dots,g$, represent cohomology classes dual to a
standard basis of 1-cycles on $\Sigma_g$, with 
\be 
\int_{\Sigma_g} \omega_1^{(A,i)} \wedge \omega_1^{(B,j)} = \delta^{ij} \ , \qquad 
\int_{\Sigma_g} \omega_1^{(A,i)} \wedge \omega_1^{(A,j)} = 0 =
\int_{\Sigma_g} \omega_1^{(B,i)} \wedge \omega_1^{(B,j)}  \ . 
\ee 
The quantity ${\rm vol}_{\Sigma_g}$ is the volume form on $\Sigma_g$, normalized to integrate to 1.
The ellipsis in \eqref{eq_Sigmag_KK}
stand for additional terms that are not relevant for obtaining the 5d couplings we are focusing on.
Upon reducing the 7d quadratic terms  in 
\eqref{eq_tot7d}, we obtain the following quadratic terms in 5d,
\be \label{eq_quadratic_Sigma}
S^{\rm quadratic}_\text{5d TQFT} = \int_{M_5} \bigg[ 
2 a_1 \wedge da_3 + \sum_{i=1}^g \bigg( 
- N c_2'^{(i)} \wedge dc_2^{(i)}
- 2 c_2^{(i)} \wedge db_2'^{(i)}
+ 2 c_2'^{(i)} \wedge db_2^{(i)}
\bigg)
\bigg]  \ . 
\ee 
The reduction of the 7d cubic term in \eqref{eq_tot7d} can also be performed repeating the steps of the previous subsection. The relevant terms are
\be \label{eq_cubic_Sigma}
S^{\rm cubic}_\text{5d TQFT} 
 = \int_{M_5} \sum_{i=1}^g
 \frac 12 \mathsf a_1 \cup \mathsf c_2'^{(i)} \cup \mathsf c_2^{(i)} \ . 
\ee 
As in the previous subsection,
we find it convenient to write the quadratic terms in the 5d TQFT action using a continuum formulation, while
we write the cubic term using
$\mathbb Z_2$ cochains 
 $\mathsf a_1$,
$\mathsf c_2^{(i)}$,
$\mathsf c_2'^{(i)}$.

We expect that the 5d topological actions
\eqref{eq_quadratic_Sigma}, \eqref{eq_cubic_Sigma}
capture the discrete 
0-form and 1-form symmetry sectors of 4d $\cN = 2$ SCFTs of class $\cS$ 
\cite{Gaiotto:2009we,Gaiotto:2009hg}
obtained from reducing the 6d (2,0) SCFT of type $D_N$ on the Riemann surface $\Sigma_g$
(for $g \ge 2$). In particular, the cubic couplings
\eqref{eq_cubic_Sigma} indicate that
there exist global variants of these class $\cS$ theories that
can exhibit non-invertible 0-form symmetries, by the same mechanism at play in the global variants $Sc(4N)$
and $Ss(4N)$ of 4d $\cN = 4$ SYM with gauge algebra $\mathfrak{so}(4N)$ \cite{Bhardwaj:2022yxj}.
A different class of non-invertible 0-form symmetries in 4d $\cN = 2$ class $\cS$ theories of type $A$ has been studied in \cite{Bashmakov:2022jtl,Bashmakov:2022uek} -- see also \cite{Garding:2023unh} for an analysis from the perspective of BPS quivers. 
A detailed analysis of 
class $\cS$ theories, however, is beyond the scope of this work.

\section{Anomaly polynomial
of the 6d (2,0) $D_N$ SCFT 
from holography}\label{sec:DtypeANOMAL}

In this section we study 
the gravitational and R-symmetry
't Hooft anomalies of the 6d (2,0) SCFT of type $D_N$ from a holographic
point of view. 
Our strategy is to analyze the topological couplings of M-theory
reduced to seven dimensions on $\RP^4$, 
following the general program of \cite{Bah:2019rgq,Bah:2020uev}, building on \cite{Witten:1998qj,Freed:1998tg,Harvey:1998bx}.
By keeping track 
of both the $C_3 G_4 G_4$ two-derivative coupling,
as well as the $C_3 X_8$ higher-derivative correction,
we are able to reproduce
the \emph{exact}
anomaly polynomial of the 6d SCFT,
and not only the leading
terms in the large $N$ expansion.
Our analysis mirrors
the probe brane/anomaly inflow
analysis of \cite{Yi:2001bz}.
The anomaly polynomial for 6d (2,0) SCFTs of type $D$ was first conjectured in \cite{Intriligator:2000eq}, and later
obtained with field theory
arguments in \cite{Ohmori:2014kda}.

In order to probe the gravitational
and R-symmetry 't  Hooft anomalies
of the 6d (2,0) SCFT of type $D_N$,
we have to turn on the corresponding
background fields. The latter are mapped
on the gravity side to the metric
on external 7d spacetime,
and to the Kaluza-Klein gauge fields
associated to the $SO(5)$ isometry
of $\RP^4$.
Since we are studying perturbative
anomalies for continuous symmetries,
it is convenient to use
the formalism of the anomaly polynomial.
The object of interest is
thus the anomaly polynomial
of the 6d (2,0) SCFT of type
$D_N$, denoted $I_8^{\rm SCFT}$.

Our starting point is the 12-form
\be 
\mathcal I_{12} =  - \frac 16 G_4 \wedge G_4 \wedge G_4 - G_4 \wedge X_8 \ ,  
\ee 
which encodes  the topological
couplings in the 11d M-theory effective action. 
The 8-form is constructed
from Pontryagin classes of the tangent bundle, see \eqref{eq:X8def} repeated here for convenience,
\be \label{eq:X8defbis}
X_8 = \frac{1}{192} \bigg[
p_1(TM_{11})^2 - 4 p_2(TM_{11})
\bigg] \ ,
\ee 
We consider a fiducial 12d spacetime $M_{12}$ of the form
\be  \label{eq:M12}
\RP^4 \hookrightarrow
M_{12} \rightarrow M_8 \ . 
\ee 
The base $M_8$ of the fibration is a fiducial 8d spacetime where the
anomaly polynomial $I_8^{\rm SCFT}$ lives.
In the fibration \eqref{eq:M12}
we allow for a non-trivial
twisting of the fiber $\RP^4$ by $SO(5)$ rotations.

Our aim is to compute the fiber
integration
\be 
\mathcal I_8  := \int_{\RP^4} \mathcal I_{12} \  .
\ee 
The quantity $\mathcal I_8$ is
expected to be equal and opposite
to the SCFT anomaly polynomial
$I_8^{\rm SCFT}$. This expectation
turns out to be correct, up to 
a subtler point addressed below.

\subsection{Global angular form}

In the total space $M_{12}$
of the fibration \eqref{eq:M12},
the volume form $v_4$ on $\RP^4$ is not a globally
defined object, due to the twisting
of the $\RP^4$ base
over $M_8$.
There exists a systematic way
to promote $v_4$ to a globally defined
form on $M_{12}$, while at the same
time preserving its closure.
For the problem at hand, it is useful to
regard $\RP^4$
as the quotient of $\sphere^4$ by antipodal
identification, and to map 
the task of completing $v_4$
to the analogous task for an $\sphere^4$
fibration. The latter can be addressed
using global angular forms
\cite{Freed:1998tg,Harvey:1998bx}, see also the textbook \cite{bott1982differential}.
(Systematic tools
to address arbitrary fibrations
are developed in \cite{Bah:2019rgq}, where
a connection to equivariant cohomology is established.)

We can realize the fibration \eqref{eq:M12} starting from
an auxiliary sphere fibration,
\be 
\sphere^4 \hookrightarrow \widetilde M_{12} \rightarrow M_8 \ ,
\ee 
and performing a fiberwise antipodal
identification on $\sphere^4$.
In the auxiliary space $\widetilde M_{12}$ a \emph{globally defined}
4-form $\widetilde e_4$ exists, 
dubbed global angular form, 
with the following properties:
\begin{itemize}
\item $\widetilde e_4$ is closed;
its de Rham cohomology class is the image in $H^4(\widetilde M_{12};\mathbb R)$ of an integral cohomology class
$H^4(\widetilde M_{12};\mathbb Z)$;
\item $\widetilde e_4$, restricted to
a generic fiber, is proportional to the volume form on the fiber, and satisfies
\be  
\int_{\sphere^4} \widetilde e_4 = 2 \ .
\ee 
\end{itemize}
We can furnish an explicit local
formula for $\widetilde e_4$ \cite{Harvey:1998bx},
\be \label{eq:explicite4}
\widetilde e_4 = \frac{1}{32 \pi^2} \epsilon_{ABCDE}\bigg[ 
y^A Dy^B Dy^C Dy^D Dy^E
+ 2  y^A F^{BC} Dy^D Dy^E
+ y^A F^{BC} F^{DE}
\bigg] \ . 
\ee 
Our notation is as follows.
The fiber $\sphere^4$ is parametrized 
by five real constrained coordinates
$y^{A}$, $A =1,\dots,5$, satisfying
$\delta_{AB} y^A y^B = 1$.
The 1-form $Dy^A$ describes
the twisting of the $\sphere^4$ fiber over the base space
and is defined as $Dy^A = dy^A + A^{AB} y_B$. Here, $A^{AB}$ is the $SO(5)$
Kaluza-Klein gauge field;
its field strength is
$F^{AB} = dA^{AB} + A^{AC} A_C{}^B$.

The form $\widetilde e_4$
satisfies the Bott-Cattaneo formula
\cite{BottCattaneo},
\be \label{eq:bottcattaneo}
\int_{\sphere^4} \widetilde e_4^{2s+2} = 0 \ , \quad 
\int_{\sphere^4} \widetilde e_4^{2s+1} = 2 \big[ p_2(\mathcal N) \big]^s \ , \quad
s = 0,1,2,\dots
\ee 
where $p_2(\mathcal N)$ denotes
the second Pontryagin form
of the $SO(5)$ bundle $\cN$
that describes the twisting of the $\sphere^4$ fiber. Explicitly, 
 $p_2(\mathcal N)$
can be written in terms of the
field strength $F^{AB}$
of the $SO(5)$ Kaluza-Klein gauge field,
\be
p_2(\cN) = \frac{1}{(2\pi)^4} \bigg[ 
\frac 18 ({\rm tr} F^2)^2
- \frac 14 {\rm tr} F^4
\bigg]  \ , 
\ee 
where ${\rm tr} F^2 = F^{AB} F_{BA}$, and similarly for
${\rm tr}F^4$.

From the explicit expression \eqref{eq:explicite4}
we observe that:
(i) if we turn off the
$SO(5)$ gauge fields,
$\widetilde e_4$
reduces to the volume form
on $\sphere^4$, normalized to integrate to $2$;
(ii) under the antipodal identification $y^A \mapsto - y^A$,
$\widetilde e_4$ transforms as 
$\widetilde e_4 \mapsto - \widetilde e_4$.
This means that, after modding out by the antipodal identification,
$\widetilde e_4$ 
yields a  twisted 4-form on the fibration 
$M_{12}$, which reduces to $v_4$ on each $\RP^4$ fiber.
We denote this twisted
form $e_4$.
It satisfies a version
of the Bott-Cattaneo formula,
\be \label{eq:bottcattaneoBIS}
\int_{\RP^4}  e_4^{2s+2} = 0 \ , \quad 
\int_{\RP^4}  e_4^{2s+1} =  \big[ p_2(\cN) \big]^s \ , \quad
s = 0,1,2,\dots
\ee 
We observe that
in the second equality, where we get a non-zero quantity,
the integrand is an odd power of $e_4$,
hence a twisted form,
as is required 
because the fiber $\RP^4$ is not orientable.
The above relations are motivated as follows.
The original $\widetilde e_4$ integrates to 2 on each $\sphere^4$ fiber.
As usual, 
going from $\sphere^4$ to $\RP^4$
brings a factor 1/2 to each integral.
Thus, $e_4$ integrates
to 1 on each $\RP^4$ fibers,
which is the same normalization as $v_4$,
see \eqref{eq:v4normalization}.
This means that we do not have to rescale $\widetilde e_4$
to achieve the desired normalization after antipodal identification.
In particular, the only 
modification in the second equality 
in \eqref{eq:bottcattaneo}
in going from $\sphere^4$ to $\RP^4$
is an overall 1/2 factor.

\subsection{Extracting $I_8^{\rm SCFT}$}

Let us first address the $G_4^3$ term in 
$\mathcal I_{12}$. We set
\be
G_4 = \left( N - \frac 12 \right) e_4 \ , 
\ee 
where $e_4$ was introduced in the above subsection, and the prefactor
is needed to reproduce
\eqref{eq:flux_bk} upon switching off
the $SO(5)$ gauge fields.
Making use of \eqref{eq:bottcattaneoBIS}
we compute 
\begin{align}
 \frac 16 \int_{\RP^4} G_4^3 & = \frac 16  \left( N - \frac 12 \right)^3 \int_{\RP^4} e_4^3 = \frac 12 (2N-1)^3 \frac{p_2(\cN)}{24} \ . 
\end{align}
Let us now turn to
the $G_4 X_8$ term.
We need the definition \eqref{eq:X8defbis}
of $X_8$
in terms of Pontryagin classes of the 
total spacetime. The tangent
bundle to the fiducial total space
$M_{12}$ in \eqref{eq:M12} splits
as $TM_8 \oplus \mathcal N$,
where $\mathcal N$ is the $SO(5)$
bundle encoding the twisting
of the $\RP^4$ fiber.
We make use of the identities\footnote{ \ These relations hold modulo torsion; they are enough for the purposes of this section.}
\be 
p_1 (TM_8 \oplus \mathcal N )= p_1(TM_8) + p_1(\mathcal N) \ , \quad 
p_2 (TM_8 \oplus \mathcal N )
= p_2(TM_8) + p_2(\mathcal N)
+p_1(TM_8)  p_1(\mathcal N) \ . 
\ee 
We get
\be 
X_8 = J_8 - \frac{p_2(\mathcal N)}{24}
\ ,  \qquad 
J_8 = \frac{1}{48} 
\bigg[ 
p_2(\cN) - p_2(TM_8)
+ \frac{(p_1(TM_8) - p_1(\cN))^2}{4}
\bigg] \ , 
\ee 
where we have isolated the
combination $J_8$,
which is equal to the anomaly
polynomial of a free 6d $(2,0)$ tensor
multiplet.
In the integral of $G_4 X_8$
over $\RP^4$,
the integral is saturated by
$G_4$, yielding a factor $N-1/2$.
In total, we arrive at
\be 
- \cI_8 = -\int_{\RP^4} \cI_{12} = 
\left( N - \frac 12 \right) J_8
+ N(2N-1) (2N-2)\frac{p_2(\cN)}{24} \ . 
\ee

The final step is the extraction of
the sought-for
$I_8^{\rm SCFT}$ from $\cI_8$.
The naive relation
$I_8^{\rm SCFT} \stackrel{?}{=} - \cI_8$ must be replaced by 
\begin{align} \label{eq:subtraction}
I_8^{\rm SCFT} & = \Big(-\cI_8 \Big) - \Big(- \cI_8 \Big|_{N \rightarrow 0} \Big) 
 = 
N   J_8
+ N(2N-1) (2N-2)\frac{p_2(\cN)}{24} \ . 
\end{align}
This 
matches the general formula for the anomaly polynomial of the 6d (2,0) SCFT of type~$\mathfrak g$ \cite{Intriligator:2000eq,Ohmori:2014kda}
\be 
I_8^{{\rm SCFT},  \mathfrak{g}}
= r_{\mathfrak g} J_8
+ d_{\mathfrak g} h^\vee_{\mathfrak g} \frac{p_2(\cN)}{24} \ , 
\ee 
where $r_{\mathfrak g}$,
$d_{\mathfrak g}$,
$h^\vee_{\mathfrak g}$ denote the rank, dimension, and dual Coxeter number of $\mathfrak g$, respectively
Indeed, for $\mathfrak g = D_{N} = \mathfrak{so}(2N)$,
we have 
$r_\mathfrak g =N$,
$d_{\mathfrak g} = N(2N-1)$,
$h^\vee_{\mathfrak g}
= 2N-2$.

The origin of the subtraction
of $(- \cI_8|_{N \rightarrow 0})$
lies in the consistency 
of the 11d low-energy M-theory
effective action \cite{Witten:1996md}.
We are considering a spacetime in which
the flux quantization of $G_4$ is shifted to be half-integral,
\be
G_4 =  a_4 - \frac 12 v_4 \ , 
\ee 
where $a_4$ is a closed 4-form
with integral periods.
This shift affects the integrality properties of the 12-form
$\mathcal I_{12}$,
which are in turn pivotal
for the consistency of the effective action.
In order to account
for such effects,
one must consider
the combination \cite{Witten:1999vg} (eq.~(5.21))\footnote{ \ Here $v_4$ should be understood in general
as the twisted integral lift of
$w_4$ that specifies the $m_{\rm c}$ structure.
In \cite{Witten:1999vg} the orientable setting is considered, and accordingly $v_4$
there is replaced with $\lambda$, which provides a canonical integral lift of $w_4$.
}
\be  \label{eq:Msubtraction}
\int_{M_{12}} \bigg\{ \bigg[ 
- \frac 16 \left( a_4 - \frac 12 v_4\right)^3 
- \left( a_4 - \frac 12 v_4\right) X_8 \bigg] - (a_4 \rightarrow 0)
\bigg\}  \ , 
\ee 
where $(a_4 \rightarrow 0)$
is a shorthand notation
for the  previous two terms,
with $a_4$ set to zero.
On a spacetime where $G_4$ is half-integrally quantized,
the topological terms in the action
yield a phase factor in the path integrand that is defined up to a sign. This amibuigity disappears once an analogous ambiguity in the
Pfaffian of the  
Rarita-Schwinger operator is taken into account (see \cite{Witten:1996md}
for the orientable setting
and \cite{Freed:2019sco}
for the extension to the
non-orientable case).
We may then regard
the subtraction of $(a_4 \rightarrow 0)$ as a proxy
for the effects of the gravitino
measure.

When we consider M-theory on $\AdS_7 \times \RP^4$,
the general formula 
\eqref{eq:Msubtraction} can be specialized.
Indeed, in this case $a_4 = N e_4$,
and subtracting $(a_4 \rightarrow 0)$ is equivalent to subtracting
$(N \rightarrow 0)$.
We reproduce the prescription
\eqref{eq:subtraction}.

We observe that the rule
\eqref{eq:Msubtraction} is universal,
i.e.~can also be applied to
setups in which $G_4$ is 
integrally quantized. In that case 
we should take $v_4 = 0$,
and subtracting
$(a_4 \rightarrow 0)$ has no effect at all. 
For example, if we consider M-theory on $\AdS_7 \times \sphere^4$,
we can use the same
prescription as in \eqref{eq:subtraction}
to extract the SCFT anomaly
from $\cI_8$;
the subtracted term 
$(- \cI_8|_{N \rightarrow 0})$ is simply zero in this case. This subtraction
should not be confused with
the subtraction of the anomaly
of a free 6d (2,0) tensor multiplet, which is necessary
to isolate the anomaly of the interacting 6d (2,0) $A_{N-1}$ theory from the total anomaly
of the M5-brane stack.
In the case of 6d (2,0) SCFTs of type $D$, there are no center-of-mass degrees of freedom
to subtract.

\section{Conclusion and outlook}\label{sec:conclusion}

In this work we have revisited
6d (2,0) SCFTs of type $D_N$
and their holographic realization via M-theory on
$\AdS_7 \times \RP^4$.
Our main focus has been on absolute   variants of these SCFTs and their global discrete 0- and 2-form symmetries. 
We have derived
the 7d SymTFT capturing these global symmetries from M-theory, both from the point of view of the low-energy supergravity description,
and from branes. In the process, we have highlighted several subtle points related to the non-orientability of $\RP^4$, the half-integral $G_4$-flux that threads it,
and the non-commutativity of fluxes. We have also made contact with brane dynamics
(fat branes and Hanany-Witten effect).
The  7d SymTFT derived from M-theory exhibits a  cubic coupling which is the origin of non-trivial symmetry structures of the various absolute global variants:
a mixed 't Hooft anomaly for 
the $SO(2N)$ variant;
a higher-group structure for
the $O(2N)$ variant;
a non-invertible 0-form symmetry
for the $Sc(8k)$, $Ss(8k)$  variants.
Finally, we have given a purely holographic derivation of the anomaly polynomial of 6d (2,0) SCFTs of type $D_N$,
first studied in \cite{Intriligator:2000eq,Yi:2001bz}, by 
 reducing the 11d topological couplings of M-theory on $\RP^4$. 
This offers a concrete realization 
\eqref{eq:subtraction}
of the general
subtraction prescription
\eqref{eq:Msubtraction}
discussed 
in \cite{Witten:1996hc,Witten:1999vg,Freed:2019sco}.

Our findings suggest several possible directions for future investigation. We comment on some of them below.

Firstly, it would be interesting to revisit the bulk-boundary description of those 6d (2,0) SCFTs of type $A_{N-1}$ that can have 0-form symmetries along the lines we encountered for theories of D-type (and analogous cubic terms). In particular, 
there are absolute variants 
of type $SU(M^2)/\mathbb Z_M$ for $N = M^2$
\cite{Gukov:2020btk}, where we expect our methods can lead to an improved understanding of global discrete symmetries by uncovering possible structures related
to the outer automorphism of the 
$A_{N-1}$ Dynkin diagram via M-theory realizations in terms of M5-brane stacks and their holographic duals.\footnote{ \ The isomorphism $A_3 \cong D_3$, combined with the findings of this work, shows that the absolute variant $SU(4)/\mathbb Z_2$ does exhibit both  0-form and 2-form symmetries
with a non-trivial interplay.
More interesting are the cases 
$SU(M^2)/\mathbb Z_M$ with $M \ge 3$.
}

Secondly,  the parallelism between 
bosonization/fermionization in 2d \cite{Karch:2019lnn,Gaiotto:2015zta,Bhardwaj:2016clt,Kapustin:2017jrc,Thorngren:2018bhj,Ji:2019ugf}
and the web of relations \eqref{eq_boson_diagram6d} among absolute global variants of 6d (2,0) SCFTs of type $D_{N = 4k}$
\cite{Gukov:2020btk} raises an intriguing possibility of a deeper connection.
In this paper we focused on a 7d 
SymTFT description of such relations and we highlighted the role played by the cubic coupling in the 
7d topological action. A natural direction to explore is to consider possible ``fermionized'' versions of 6d theories (along the lines of \cite{Hsin:2021qiy}), and to study the interplay of our results with 3d/2d topological-bulk/boundary interactions capturing the 2d analog of \eqref{eq_boson_diagram6d}.\footnote{ \ A famous example of bosonization/fermionization in 2d is provided by the relationship between 
the Ising CFT and the Majorana fermion.
See \cite{Freed:2018cec,Ji:2019ugf,Kaidi:2022cpf,Bhardwaj:2024ydc} for a 3d SymTFT perspective.}

Next, it is interesting to investigate potential implications of our findings for lower-dimensional QFTs obtained by compactification of 6d (2,0) $D_N$ SCFTs.
As already alluded to in Section \ref{sec_Sigma},
our results can have applications to 4d $\cN = 2$ SCFTs of class $\cS$ \cite{Gaiotto:2009we,Gaiotto:2009hg}, obtained by reducing 
6d (2,0) $D_N$ SCFTs on a Riemann surface. More precisely, the results of this paper point toward a strategy for constructing novel non-invertible 0-form symmetries in
class $\cS$ theories of type $D$, building  
on couplings such as \eqref{eq_cubic_Sigma} in their 5d SymTFTs.\footnote{ \ Non-invertible symmetries of class $\cS$ theories of type $A_{p-1}$ with $p$ prime have been constructed in \cite{Bashmakov:2022jtl,Bashmakov:2022uek}
by tuning the complex structure moduli of $\Sigma_g$
to special points of enhanced symmetries
(analogously to tuning $\tau_{\rm YM} = i$ to study duality defects in 4d $\cN = 4$ SYM). The non-invertible symmetries of class $\cS$ theories of type $D$
that may originate from couplings such as 
\eqref{eq_cubic_Sigma} would be of a different kind, as they would not hinge on tuning
the  complex structure moduli of $\Sigma_g$
to special values (and so we expect them to be ubiquitous). In particular, we expect generalizations of the arguments of \cite{Argurio:2024kdr} in the context of orthosymplectic quivers.
}
This strategy also extends to 4d $\cN = 1$ theories obtained from 6d (2,0) SCFTs \cite{Maruyoshi:2009uk,Benini:2009mz,Bah:2011je,Bah:2012dg}.

Furthermore, it 
would also be interesting
to consider reductions of the 6d (2,0) $D_N$ SCFT to 2d and 3d theories.
A concrete case study could be provided by
2d (0,2) SCFTs obtained by compactification on 
a negatively-curved K\"ahler-Einstein four-manifold $\Sigma_4$
\cite{Benini:2013cda}, thus extending the analysis of \cite{Bashmakov:2023kwo,Chen:2023qnv} to 6d (2,0) theories of type $D_N$.
The 7d SymTFT \eqref{eq_tot7d}
reduced on $\Sigma_4$ yields
the 3d SymTFT for the  2d (0,2) SCFT. We expect the latter to contain cubic couplings of the schematic forms
$a_1b_1c_1$ as well as 
$a_1 c_1 c_1$.
(In contrast, the 5d SymTFT \eqref{eq_cubic_Sigma} only contains
trilinear cubic couplings of the form
$a_1 b_2 c_2$.)
Studying these systems might also 
shed light on
analogies and differences between 2d bosonization/fermionization and its 6d counterpart.

Building on the previous two paragraphs, it would be 
beneficial to clarify some geometric aspects of
the holographic description
of compactifications of 6d (2,0) SCFTs of type $D$.
On general grounds, if the 6d SCFT is reduced on a compact space $X_m$ with a partial topological twist to yield an
SCFT dual to a supersymmetric $\AdS_{7-m}$ solution, we expect the internal space to
be a non-trivial
$\RP^4$ fibration over
$X_m$. M-theory requires the total 11d spacetime to admit
an $m_{\rm c}$ structure
(thus in particular a Pin$^+$ structure). It would be useful to understand concretely how this is realized, and how to describe the internal
Killing spinors (or rather,
pinors) associated to 
preserved supersymmetry upon reduction.

Finally, in this work we have mainly focused on couplings for discrete gauge fields that originate from flux non-commutativity
or from the two-derivative
$C_3G_4G_4$ term in the 11d supergravity effective action.
It is natural to explore whether the higher-derivative term $C_3 X_8$ can also be a source of couplings for discrete gauge fields, and to investigate its interplay with the corresponding SymTFT for continuous symmetries, extending the analysis of  \cite{Antinucci:2024zjp,Brennan:2024fgj,Bonetti:2024cjk,Apruzzi:2024htg}.

\acknowledgments

We thank 
Guillaume Bossard,
Thomas Dumitrescu, Kenneth Intriligator,
Justin Kaidi,
Edward  Mazenc, Kantaro Ohmori, Piljin Yi
for useful discussions and correspondence.
RM is supported in
part by ERC Grant 787320-QBH Structure and by ERC Grant 772408-Stringlandscape. The work of MDZ is supported by the European Research Council (ERC) under the European Union’s Horizon 2020 research and innovation program (grant agreement No. 851931) and by the VR project grant No. 2023-05590. MDZ also acknowledges the VR Centre for Geometry and Physics (VR grant No. 2022-06593). 
FB is supported by 
the Program ``Saavedra Fajardo'' 22400/SF/23.
MDZ and FB also acknowledge support from the Simons Foundation Grants \#888984  (Simons Collaboration on Global Categorical Symmetries).

\appendix
\section{Differential cochains
and the $C_3$ field}
\label{app:diff_cochains}

The $C_3$ field of 11d supergravity
can be modeled using differential
cochains \cite{Hopkins:2002rd}.
In the orientable case, 
an alternative approach is the $E_8$ model of \cite{Diaconescu:2000wy,Diaconescu:2003bm};
see \cite{Moore:2004jv} for comments on its extension to the non-orientable case.

Let us review some useful notions from
\cite{Hopkins:2002rd},
adapted to the case of a 
not necessarily orientable
spacetime manifold $M$.
A twisted differential cochain of degree $n$ on a $M$ is a triple 
\be 
(c,h,\omega) \in C^n(M; \widetilde{\mathbb Z}) \times 
C^{n-1}(M; \widetilde{\mathbb R}) \times \widetilde \Omega^n(M) \ ,
\ee 
where the integral cochain $c$,
real cochain $h$, and differential form $\omega$ are all twisted by the orientation bundle of $M$. We use the notation $\breve C^n(M;\widetilde {\mathbb Z})$
for the set of twisted 
differential cochains of degree $n$, while we write 
$\breve C^n_{\rm flat}(M;\widetilde {\mathbb Z})$
for those that are flat, \emph{i.e.}~have $\omega = 0$.
The differential $d$ acting 
on twisted differential cochains is defined as
\be
d(c,h,\omega) = (\delta c, \omega - c - \delta h , d \omega) \ . 
\ee
We define the twisted differential
cohomology group $\breve H^n(M;\widetilde {\mathbb Z})$
as
\be 
\breve H^n(M;\widetilde {\mathbb Z}) = \frac{\breve C^n(M; \widetilde{\mathbb Z})}{d \breve C^{n-1}_{\rm flat}(M; \widetilde{\mathbb Z})} \ . 
\ee 
This is the same object
as in the commutative diagram
\eqref{eq:diff_coho_hexagon}.

Next, let $u_4$ be
a twisted $\mathbb R/\mathbb Z$
cocycle on $M$,
\be
u_4 \in C^4(M ; \widetilde{\mathbb R / \mathbb Z}) \ , \qquad 
\delta u_4 = 0 \ .
\ee 
We define the category 
$\breve {\mathcal H}^4_{u_4}(M;\widetilde{\mathbb Z})$
of twisted
differential cocycles shifted
by $u_4$ as follows. 
The objects are triples 
\be 
(c,h,\omega) \in C^4(M; \widetilde{\mathbb R}) \times 
C^{3}(M; \widetilde{\mathbb R}) \times \widetilde \Omega^4(M) \ ,
\ee 
where
\be 
c = u_4 \mod \mathbb Z \ , \qquad 
\delta c = 0 \ , \qquad 
d\omega = 0 \ , \qquad \delta h = \omega-c \ .
\ee 
The morphisms from $(c^{(1)}, h^{(1)}, \omega^{(1)})$ to 
$(c^{(2)}, h^{(2)}, \omega^{(2)})$
are equivalence classes of pairs 
\be
(b,k) \in C^3(M ; \widetilde{\mathbb Z}) 
\times C^2(M ; \widetilde {\mathbb R}) \ , 
\ee 
satisfying
\be 
c^{(2)} = c^{(1)} - \delta b \ , \qquad 
h^{(2)} = h^{(1)} + b + \delta k \ , \qquad
\omega^{(2)} = \omega^{(1)} \ . 
\ee 
The equivalence relation is generated by
\be
(b,k) \sim (b - \delta b' , k + \delta k' + b') \ , \qquad 
(b',k') \in C^2(M;\widetilde {\mathbb Z}) \times 
C^1(M ; \widetilde {\mathbb R}) \ .
\ee 
(In particular, if 
$\omega^{(2)} \neq \omega^{(1)}$,
there are no morphisms
from $(c^{(1)}, h^{(1)}, \omega^{(1)})$ to 
$(c^{(2)}, h^{(2)}, \omega^{(2)})$.)

The physically interesting case
is the case in which $M$ is an $m_{\rm c}$
manifold, so that 
the fourth Stiefel-Whitney class
$w_4(TM_4) \in H^4(M;\mathbb Z_2)$ admits a twisted integral lift
$v_4 \in H^4(M ; \widetilde{\mathbb Z})$.
Let $ v_4' \in C^4(M ; \widetilde{\mathbb Z})$,
$\delta v_4'=0$ be a twisted integral
cocycle representing $v_4$.
We then set 
\be 
u_4 = \frac 12 v_4' \mod \mathbb Z \  . 
\ee 
The objects in the category
$\breve {\mathcal H}^4_{u_4}(M;\widetilde{\mathbb Z})$
model 
3-form gauge fields configurations.
A morphism from $(c^{(1)}, h^{(1)}, \omega^{(1)})$ to 
$(c^{(2)}, h^{(2)}, \omega^{(2)})$
models a gauge transformation
from one configuration to another.
The set of isomorphisms classes of objects
in $\breve {\mathcal H}^4_{u_4}(M;\widetilde{\mathbb Z})$
is a torsor for $\breve H^n(M;\widetilde {\mathbb Z})$.
It is in this sense that
$\breve H^n(M;\widetilde {\mathbb Z})$
models the gauge equivalence class
of the difference between
two 3-form configurations,
as stated in the main text.

\section{Remarks on the quadratic refinement}
\label{app_refinement}
 
In this appendix we review the salient properties of the quadratic refinement $q_{\rho_n}$ of the intersection pairing in the middle $\mathbb Z_2$ cohomology of a $2n$-dimensional manifold.
The cases relevant for this paper are $n=1$ and $n=3$.
We follow the expositions in the physics papers 
\cite[Appendix B]{Gukov:2020btk}, \cite[Appendix B]{Hsin:2021qiy}.
For $n=1$ see also \cite{Debray:2018wfz}.

Let $X_{2n}$ be a (non-necessarily orientable) compact manifold of even dimension $2n$. The $(n+1)$-th Wu class $\nu_{n+1}$ of $X_{2n}$
vanishes in cohomology, because $\nu_{n+1}$ vanishes on any manifold of dimension $\le 2n+1$.\footnote{ \ The Wu class $\nu_{p}$ represents the Steenrod square operation
${\rm Sq}^{p}: H^*(X;\mathbb Z_2) \rightarrow H^{*+p}(X;\mathbb Z_2)$, in the sense that
${\rm Sp}^p(x) = \nu_p \cup x$ for $\deg (x) = \dim X-p$.
One of the axioms of 
${\rm Sq}^{p}$, however, is 
${\rm Sq}^{p}(x)=0$
if $\deg(x) < p$.
Thus, if $\dim X \le 2p-1$,
${\rm Sq}^{p}(x)$ vanishes
for any $x$
(due to the aforementioned axiom
if $\deg(x) < p$,
for dimensional reasons if
$\deg(x) \ge p$).
}
Thus, there exist a trivialization $\rho_n$ of $\nu_{n+1}$,
$d\rho_n = \nu_{n+1}$. We refer to $\rho_n$ as a Wu structure of degree $n$.\footnote{ \ For $n=1$, when $X_2$ is two-dimensional, 
a Wu structure is the same as a Pin$^-$ structure,
because the second Wu class is given as $\nu_2 = w_2 + w_1^2$.} Inequivalent Wu structures from a torsor
over $H^n(X_{2n};\mathbb Z_2)$. We write $\rho_n+z_n$
for the Wu structure obtained from $\rho_n$ by acting with the element
$z_n \in H^n(X_{2n}; \mathbb Z_2)$. 

It is known \cite{brown_math, browder_math} that Wu structures $\rho_n$ are in 1-to-1
correspondence with $\mathbb Z_4$-valued refinements of the
intersection pairing in $H^n(X_{2n} ; \mathbb Z_2)$.
Let us write $\widetilde q_{\rho_n}$ for the refinement
associated to $\rho_n$. It is an operation
\be 
\widetilde q_{\rho_n}: H^n(X_{2n}  ; \mathbb Z_2) \rightarrow \mathbb Z_4 \ . 
\ee 
It satisfies the properties
\be 
\ba 
\widetilde q_{\rho_n}(a_n + b_n) & = \widetilde q_{\rho_n}(a_n )
+ \widetilde q_{\rho_n}( b_n)
+ \theta \int_{X_{2n}}a_n \cup b_n \mod 4 \ , \\ 
\widetilde q_{\rho_n + z_n} (a_n) & = \widetilde q_{\rho_n}(a_n) + \theta \int_{X_{2n}} z_n \cup a_n \mod 4 \ . 
\ea 
\ee 
In the above expressions $a_n$, $b_n$, $z_n \in H^n(X_{2n};\mathbb Z_2)$ and $\theta: H^*(X_{2n} ; \mathbb Z_2) \rightarrow H^*(X_{2n} ; \mathbb Z_4)$
is the cohomology  map induced by the inclusion
$\mathbb Z_2 \hookrightarrow \mathbb Z_4$
that sends (1 mod 2) to (2 mod 4).

We now proceed assuming that the manifold $X_{2n}$ has trivial
$n$-th Wu class,
\be 
\nu_n(TX_{2n}) = 0 \ . 
\ee 
In particular, this will be the case if $n$ is odd and $X_{2n}$ is orientable.\footnote{ \ Indeed, it is known that all odd Wu classes vanish on an orientable manifold. This can also be seen explicitly from the identity \cite{yoshida1980wu}
\be 
\nu_{2k+1} = \sum_{i\ge 1} (w_1)^{2^i-1} \nu_{2k+2-2^i} \ . 
\ee }
Recall the property $a_n \cup a_n = a_n \cup \nu_n$ mod 2.
It implies that, if $\nu_{n}$ vanishes, the $\mathbb Z_4$-valued
operation $\widetilde q_{\rho_n}$ takes even values.
As a result, we can define a $\mathbb Z_2$--valued quadratic refinement.
It is an operation
\be 
 q_{\rho_n}: H^n(X_{2n}  ; \mathbb Z_2) \rightarrow \mathbb Z_2 \ . 
\ee 
It satisfies the properties
\be 
\ba 
 q_{\rho_n}(a_n + b_n) & =  q_{\rho_n}(a_n )
+  q_{\rho_n}( b_n)
+  \int_{X_{2n}}a_n \cup b_n \mod 2 \ , \\ 
 q_{\rho_n + z_n} (a_n) & =  q_{\rho_n}(a_n) +  \int_{X_{2n}} z_n \cup a_n  \mod 2 \ . 
\ea 
\ee 
We can recover the $\mathbb Z_4$-valued quadratic function $\widetilde q_{\rho_n}$ via 
\be 
\widetilde q_{\rho_n}(a_n) = \theta q_{\rho_n}(a_n) \mod 4  \ . 
\ee

As an aside, we observe that the Pontryagin square operation
$\mathfrak P : H^n(X_{2n};\mathbb Z_2) \rightarrow H^n(X_{2n};\mathbb Z_4)$ is not of help in constructing $\widetilde q_{\rho_n}$ for odd $n$ (the main case of interest for this paper).
This can be seen from 
\cite[eq.~(4.6)]{whitehead1949simply}
\be 
\mathfrak P(\mathsf x_p + \mathsf y_p) = 
\mathfrak P(\mathsf x_p )
+
\mathfrak P( \mathsf y_p)
+ \big[ 1 + (-1)^p \big]
\mathsf x_p \cup \mathsf y_p
+ \text{(exact terms)} \ , 
\quad 
\text{as $\mathbb Z_4$-cochains} \ .
\ee 
For $p$ odd, the mixed term drops, and we cannot obtain a refinement of $\mathsf x_p \cup \mathsf y_p$.
In fact, for $p$ odd $\mathfrak P(\mathsf x_p)$ is not an independent cohomology operation, and can be written in terms of Steenrod squares and Bockstein homomorphisms
\cite[eq.~(4.2)]{massey1969pontryagin}.

Using the quadratic function $\widetilde q_{\rho_n}$ we write the
Arf-Brown-Kervaire invariant  as \cite{Arf_math, Kervaire_math, browder_math, brown_math} 
\be 
Z_{\rm ABK}[\rho_n] = 
 \frac{1}{|H^n(X_{2n} ; \mathbb Z_2)|^{1/2}} \sum_{a_n \in H^n(X_{2n};\mathbb Z_2)} e^{2\pi i \frac 14 \widetilde q_{\rho_n}(a_n) }   \ . 
\ee 
In general this is an 8th root of unity ($\mathbb Z_8$ invariant).
In the case $\nu_n = 0$, we have instead a $\mathbb Z_2$ invariant,
which we refer to as Arf-Kervaire invariant.
We adopt the notation
\be 
(-1)^{\mathrm{AK}(\rho_n)} = \frac{1}{|H^n(X_{2n} ; \mathbb Z_2)|^{1/2}} \sum_{a_n \in H^n(X_{2n};\mathbb Z_2)} (-1)^{q_{\rho_n}(a_n) } \ . 
\ee 
For $n=1$ this is the Arf invariant of a spin structure
(here we consider $\nu_1 = w_1 = 0$ hence $\nu_2 = w_2$ and a trivialization of $\nu_2$ is a spin structure).

\section{Killing spinors on $\sphere^4$ and antipodal identification}
\label{app:killing}

In this appendix we study the behavior
of the Killing spinors on $\sphere^4$
under the antipodal identification
that yields $\RP^4$.

As a warm-up, let us briefly revisit the Ho\v{r}ava-Witten setup \cite{Horava:1995qa,Horava:1996ma}, i.e.~M-theory on $\sphere^1$ (with coordinate
$x^{10} \sim x^{10} + 2\pi R$) further
modded out by the $\mathbb Z_2$
action $\sigma: x^{10} \mapsto - x^{10}$,
which turns $\sphere^1$ into an interval.
M-theory on $\mathbb R^{1,9}\times \sphere^1$
is invariant under supersymmetry generated by an arbitrary constant spinor $\epsilon$. The reflection
$\sigma$ induces an action on $\epsilon$, $\sigma  \epsilon = \gamma^{10} \epsilon$.
The invariance condition $\sigma  \epsilon = \epsilon$ selects those
constant spinors that satisfy $\gamma^{10} \epsilon = \epsilon$:
supersymmetry is reduced by half.

Let us now perform an analogous analysis
for Killing spinors on $\sphere^4$
with $\sigma: \sphere^4 \rightarrow \sphere^4$ the antipodal map.
It is actually no harder to consider
a general $n$-dimensional sphere.
We work in stereographic coordinates, see \emph{e.g.}~\cite{Metlitski:2015yqa}. The same conclusions
can alternatively be reached
using the explicit
expressions for Killing spinors on spheres  in angular coordinates~\cite{Lu:1998nu}.

\subsubsection*{Stereographic coordinates
and Killing spinors}

We can describe $\sphere^n$ in terms of $n+1$
constrained coordinates $y^A$ satisfying
$\delta_{AB} y^A y^B = 1$.
We introduce stereographic coordinates
$x^i$, $i=1,\dots,n$ defined by 
\be 
y^i = \frac{2 x^i}{1 + x \cdot x} \ , \qquad 
y^{n+1} = \frac{x \cdot x -1}{1 + x \cdot x} \ , 
\ee 
where $x \cdot x = \delta_{ij} x^i x^j$.
These coordinates cover all $\sphere^n$
expect an arbitrarily small neighborhood
of the pole at $y^i=0$, $y^{n+1}=1$.\footnote{ \ Our  analysis in this section
is a local analysis restricted to this coordinate patch. Below, we implicitly study the action of antipodal identification
on $\sphere^n \setminus \{\text{poles at $y^i =0$, $y^{n+1} =\pm 1$}\}$.
}
The round metric on $\sphere^n$ reads
\be 
ds^2(\sphere^n) = \frac{4 dx \cdot dx}{(1 + x\cdot x)^2} \ .
\ee 
We use the vielbein
\be 
e^{\underline i} = \frac{2 dx^i }{1 + x\cdot x} \ , 
\ee 
where underlined indices are flat  indices on $\sphere^n$.
The Clifford algebra reads
$\{ \gamma^{\underline i} , \gamma^{\underline j} \} = 2 \delta^{\underline i \underline j} \mathbb I$.

The Killing spinor equation on $\sphere^n$ reads
\be  \label{eq:Killing}
\quad D_i \epsilon = \frac  i2   \gamma_i \epsilon
   \ . 
\ee 
The gamma matrix on the RHS  
has a curved index ($\gamma_i = \gamma_{\underline j} e^{\underline j}{}_i$),
and $D_i \epsilon dx^i = d\epsilon
+ \tfrac 14 \omega^{\underline j \underline k} \gamma_{\underline j \underline k} \epsilon$ is the standard
spinor covariant derivative,
written in terms of the spin connection
1-form $\omega^{\underline j \underline k}$. 
The equation \eqref{eq:Killing}
admits the following explicit solutions for our choice of coordinates and vielbein,
\be  \label{eq:explicit_Killing}
\quad \epsilon(x) =
\frac{1}{\sqrt{1 + x\cdot x}} (\mathbb I +i x \cdot \gamma ) \epsilon_0
\  \ , 
\ee 
where $x \cdot \gamma  = x^i \gamma_{\underline i}$
and $\epsilon_0$ is a constant spinor.

\subsubsection*{Antipodal involution}

The antipodal involution $\sigma$
acts on the embedding coordinates
$y^A$ as $\sigma: y^A \mapsto - y^A$.
This action corresponds to
\be 
\sigma: x^i \mapsto - \frac{x^i}{x \cdot x}  \ , 
\ee 
in  stereographic coordinates.
We adopt an ``active'' point of view
on the transformation $\sigma$.
Let $x' = \sigma(x)$ denote the transformed point. The Jacobian of the transformation reads
\be 
\frac{\partial x'^i}{\partial x^j} = - \frac{1}{x \cdot x} M^i{}_j(x) \ , \qquad 
M^i{}_j(x) := \delta^i_j  - \frac{2 x^i x_j}{x \cdot x} \ . 
\ee 
The matrix $M(x)$ satisfies $M(x)^2 = \mathbb I$,
$M^i{}_j(x) M^k{}_\ell (x) \delta_{ik} =\delta_{j\ell}$, $\det M(x) = -1$.
Thus $M(x) \in O(n)$
but $M(x) \not \in SO(n)$. 
Let 
$(\sigma e)^{\underline i}{}_j$ denote the transformed vielbein
under the map $\sigma$. It
is defined  by the relation
\be
(\sigma e)^{\underline i}{}_j(x') = L^{\underline i}{}_{\underline k}(x)
\frac{\partial x^\ell}{\partial x'^{j}}
e^{\underline k}{}_\ell(x) \ .
\ee 
Here $L^{\underline i}{}_{\underline k}(x)$ denotes a compensating $x$-dependent $O(n)$ transformation.
It is determined by demanding that
the transformed vielbein, evaluated at $x$,
be equal to the original vielbein
at the same point $x$,
$(\sigma e)^{\underline i}{}_j(x) \stackrel{!}{=} e^{\underline i}{}_j(x)$.
(This condition
is motivated by the fact that $\sigma$
is an isometry.)
We find
\be 
L^{\underline i}{}_{\underline j}(x) = - M^i{}_j(x) \ . 
\ee 
If $n$ is odd, $L \in SO(n)$;
if $n$ is even,  $L(x) \in O(n)$
but $L(x) \not \in SO(n)$. 
The transformation $L(x)$
induces an element $S(x)$ of the Clifford
algebra via
\be 
S(x)^{-1} \gamma^{\underline i} S(x) =   L^{\underline i}{}_{\underline j} (x)\gamma^{\underline j} \ , \qquad 
S(x) = \frac{-i \, x \cdot \gamma}{\sqrt{x \cdot x}}  
\ . 
\ee 
The expression for $S(x)$
is valid for $x \neq 0$,
normalized 
for later convenience,
cfr.~\eqref{eq:squares_to_id}.

Given a spinor field $\psi(x)$ on $\sphere^n$
(a section of the spinor bundle),
the transformed spinor
$(\sigma \psi)(x')$
under the action of $\sigma$ is written in terms of
$S(x)$,
\be 
(\sigma \psi)(x') = S(x) \psi(x) \ , 
\qquad \text{or equivalenty} 
\qquad 
(\sigma  \psi)(x) = S(\sigma(x)) \psi(\sigma(x)) \ . 
\ee 
Applying the $\sigma$ transformation twice leaves any spinor invariant,
\be \label{eq:squares_to_id}
(\sigma \sigma \psi)(x) = S(\sigma(x)) S(x) \psi(x) = \psi(x) \ , \qquad \text{since }
S(\sigma(x)) S(x) = \mathbb I \,,
\ee 
thanks to our normalization for $S(x)$.
If we specialize to the case $\psi(x)=\epsilon(x)$,
the explicit Killing spinor
given in \eqref{eq:explicit_Killing},
we verify
\be 
(\sigma \epsilon)(x) =  \epsilon(x)
\   \ . 
\ee

Let us now focus on the case $n=4$
of relevance for this paper.
If $\psi(x)$ is a section of the spinor bundle on $\sphere^4$,
it descends to a section of
a Pin$^+$ bundle on 
$\RP^4$ if it satisfies the condition
\be  \label{eq:pin_sections}
(\sigma \psi)(x)  = \pm \psi(x) \ ,
\ee 
where the choice of sign
corresponds to the two possible
Pin$^+$ structures on $\RP^4$ \cite{Witten:2015aba}. 
We then see that \emph{all} Killing spinors on $\sphere^4$ automatically
descend to sections of the 
Pin$^+$ bundle associated to the plus sign in \eqref{eq:pin_sections}.

Let us conclude this section with some
remarks for general $n$.
Let $\alpha$ denote the generator
of $H^1(\RP^n;\mathbb Z_2) \cong \mathbb Z_2$. Then,
\begin{align}
    n & = 0 \mod 4 \ : &
    w_1(T\RP^n) & = \alpha \ , &
    w_2(T\RP^n) & =0 \ , & 
    &(\text{Pin}^+) \\ 
    n & = 1 \mod 4 \ : &
    w_1(T\RP^n) & = 0 \ , &
    w_2(T\RP^n) &  = \alpha^2 \ , & 
    &(\text{Spin$^c$}) \\ 
    n & = 2 \mod 4 \ : &
    w_1(T\RP^n) & = \alpha \ , &
    w_2(T\RP^n) &  = \alpha^2 \ , & 
    &(\text{Pin}^-) \\ 
    n & = 3 \mod 4 \ : &
    w_1(T\RP^n) & = 0 \ , &
    w_2(T\RP^n) &  = 0 \ . & 
    &(\text{Spin})   
\end{align}
For a proof that $\RP^n$
with $n=1$ mod 4 admits a Spin$^c$ structure
see for instance \cite{Sati:2010mj}.\footnote{ \ Here we assume $n=1$ mod 4 and $n\ge 5$. The case $n=1$
yields $\RP^1$, isomorphic
with $\sphere^1$ and thus endowed with two Spin structures.}
Recall that (isomorphism classes of)
Spin and Pin$^\pm$ structures on a manifold $M$
(if they exist)
are in 1-to-1 correspondence
with elements of $H^1(M;\mathbb Z_2)$,
while (isomorphism classes of) Spin$^c$
structures on a manifold $M$
(if they exist)
are in 1-to-1 correspondence
with elements of $H^2(M;\mathbb Z)$.
Since $H^1(\RP^n ; \mathbb Z_2) \cong \mathbb Z_2$
for any $n$, and 
$H^2(\RP^n;\mathbb Z) \cong \mathbb Z_2$
for $n\ge 2$,
we see that, in each of the four cases in the list above,
we find two structures of the type indicated
(Spin, Spin$^c$, Pin$^\pm$).
In the double cover $\sphere^n$,
the two structures correspond
to the conditions $(\sigma \psi)(x) = \pm \psi(x)$ on the Dirac
spinor $\psi(x)$.

\bibliographystyle{JHEP}
\bibliography{refs}

\end{document}